\documentclass[article]{JHEP3}

\usepackage{amsmath,amssymb}
\usepackage[dvips]{graphics}
\usepackage{epsfig}
\usepackage{psfrag,comment}

\newcommand{\be}{\begin{equation}}
\newcommand{\ee}{\end{equation}}
\newcommand{\bea}{\begin{eqnarray}}
\newcommand{\eea}{\end{eqnarray}}
\newcommand{\bean}{\begin{eqnarray*}}
\newcommand{\eean}{\end{eqnarray*}}
\newcommand{\ba}{\begin{aligned}}
\newcommand{\ea}{\end{aligned}}

\newcommand{\sectiono}[1]{\section{#1}\setcounter{equation}{0}}

\newcommand{\CA}{{\cal A}}

\newcommand{\CC}{{\cal C}}

\newcommand{\CF}{{\cal F}}
\newcommand{\CG}{{\cal G}}
\newcommand{\CH}{{\cal H}}
\newcommand{\CI}{{\cal I}}

\newcommand{\CN}{{\cal N}}
\newcommand{\CO}{{\cal O}}
\newcommand{\CP}{{\cal P}}

\newcommand{\CS}{{\cal S}}

\def\BZ{{\mathbb Z}}

\def\BC{{\mathbb C}}
\def\BP{{\mathbb P}}

\def\r{\right\rangle}

\def\1{\mathbf{1}}
\def\0{|\1\r}

\newcommand{\tr}{{\rm Tr}}
\newcommand{\rme}{{\rm e}}
\newcommand{\rmi}{{\rm i}}
\newcommand{\rmd}{{\rm d}}
\newcommand{\re}{{\rm e}}
\newcommand{\ri}{{\rm i}}
\newcommand{\rd}{{\rm d}}
\newcommand{\figref}[1]{Fig.~\protect\ref{#1}}

\newdimen\tableauside\tableauside=1.0ex
\newdimen\tableaurule\tableaurule=0.4pt
\newdimen\tableaustep
\def\phantomhrule#1{\hbox{\vbox to0pt{\hrule height\tableaurule width#1\vss}}}
\def\phantomvrule#1{\vbox{\hbox to0pt{\vrule width\tableaurule height#1\hss}}}
\def\sqr{\vbox{%
  \phantomhrule\tableaustep
  \hbox{\phantomvrule\tableaustep\kern\tableaustep\phantomvrule\tableaustep}%
  \hbox{\vbox{\phantomhrule\tableauside}\kern-\tableaurule}}}
\def\squares#1{\hbox{\count0=#1\noindent\loop\sqr
  \advance\count0 by-1 \ifnum\count0>0\repeat}}
\def\tableau#1{\vcenter{\offinterlineskip
  \tableaustep=\tableauside\advance\tableaustep by-\tableaurule
  \kern\normallineskip\hbox
    {\kern\normallineskip\vbox
      {\gettableau#1 0 }%
     \kern\normallineskip\kern\tableaurule}%
  \kern\normallineskip\kern\tableaurule}}
\def\gettableau#1{\ifnum#1=0\let\next=\null\else
\squares{#1}\let\next=\gettableau\fi\next}
\preprint{
{\small{\textsf{CERN-PH-TH/2007-218}}}}

\title{Nonperturbative Effects and the Large--Order Behavior of Matrix Models and Topological Strings}

\author{Marcos Mari\~no$^1$, Ricardo Schiappa$^1$ and Marlene Weiss$^{1,2}$
\\
$^1$Theory Division, Department of Physics, CERN,\\
CH--1211 Gen\`eve 23, Switzerland\\
\\
$^2$ITP, ETH Z\"urich,\\
CH--8093 Z\"urich, Switzerland\\
\\
\email{marcos@mail.cern.ch}, \quad
\email{ricardos@mail.cern.ch}, \quad
\email{marlene.weiss@cern.ch}
}
\abstract{
This work addresses nonperturbative effects in both matrix models and topological strings, and their relation with the large--order behavior of the 
$1/N$ expansion. We study instanton configurations in generic one--cut matrix models, obtaining explicit results for the one--instanton amplitude at both one and two loops. The holographic description of topological strings in terms of matrix models implies that our nonperturbative results also apply to topological strings on toric Calabi--Yau manifolds. This yields very precise predictions for the large--order behavior of the perturbative genus expansion, both in conventional matrix models and in topological string theory. We test these predictions in detail in various examples, including the quartic matrix model, topological strings on the local curve, and Hurwitz theory. In all these cases we provide extensive numerical checks which heavily support our nonperturbative analytical results. Moreover, since all these models have a critical point describing two--dimensional gravity, we also obtain in this way the large--order asymptotics of the relevant solution to the Painlev\'e I equation, including corrections in inverse genus. From a mathematical point of view, our results predict the large--genus asymptotics of simple Hurwitz numbers and of local Gromov--Witten invariants.}


\begin{document}



\vfill

\eject

\sectiono{Introduction}

A well--known result in field theory states that the $1/N$ expansion of gauge theories has nonperturbative corrections which behave as $\re^{-N}$ \cite{w79, shenker}. Physically, these corrections are due to instantons in the collective field theory which describes the large $N$ limit. In cases where the gauge theory has a string theory dual, they typically correspond to D--brane instanton effects. Based on rather general field theoretic arguments, one should expect that these $\re^{-N}$ effects are further related to the large--order behavior of the $1/N$ expansion, as is familiar in standard perturbation theory \cite{lgzj}.

Perhaps the simplest class of large $N$ gauge theories with string theory duals are matrix models. In spite of their apparent simplicity, matrix models hide a great deal of nontrivial information, as there are two different classes of string theories which can be described with these models. The first class of examples are the so--called noncritical or minimal string theories, defined as two--dimensional gravity coupled to conformal matter with central charge $c<1$. To be precise, these theories are described by matrix models in the double--scaling limit, \textit{i.e.}, near critical points (see \cite{dfgz} for an excellent review). The second class of examples which are described by matrix models are topological strings: as it was shown in \cite{dv}, the genus expansion of the topological B--model, on certain noncompact Calabi--Yau (CY) backgrounds, is described by the $1/N$ expansion of a certain type of matrix models. Furthermore, it has been recently shown that the mirrors of toric manifolds can also be holographically described via matrix model technology \cite{mm,bkmp}. In this set of examples there is no need to go near a critical point in order to have a string dual; a generic CY background is here described by a matrix model off--criticality, whose couplings precisely correspond to the moduli of the CY in question. If one further tunes the matrix model couplings to a critical point, the matrix model will describe a particular CY background at, say, a given fixed value of the K\"ahler parameter.

One of the main goals in our present work is to understand nonperturbative phenomena, as described by instantons, in both matrix models and topological strings (where, in the latter, we always have in mind their dual gauge theoretic description via matrix models). For the case of matrix models, the instanton configurations in the $1/N$ expansion have been identified long ago in terms of eigenvalue tunneling \cite{shenker, david1, david2}, and they have been studied in great detail in the double--scaling limit. In \cite{david1, david2}, David considered the action of an instanton configuration, which is obtained by analyzing the tunneling of a single matrix eigenvalue across the unstable effective potential---in this context this corresponds to a one--instanton effect; tunneling of several eigenvalues would correspond to multi--instanton effects. David explicitly showed that, near the critical point, this one--instanton action precisely agrees with the large--genus behavior of the free energy, which is in turn obtained from the matrix model in the double--scaling limit, via a solution to the so--called string equation. In the dual string theory, these effects were later identified as D--instanton effects \cite{m03,kms03} due to the so--called ZZ branes \cite{zz01}, and it was shown in \cite{m03,akk} that a direct D--brane calculation reproduces the instanton action obtained from the double--scaled matrix model. This line of research thus made precise the connection between D--instantons in string theory and eigenvalue tunneling in the matrix model dual.

Quantum fluctuations around this one--instanton configuration, again restricted to the double--scaling limit, were further analyzed in \cite{davidtwo}, and more recently in \cite{lvm,iy}, but the connection to the large--order behavior of perturbation theory was never explicitly addressed in any of those papers. In fact, it it surprising that to this date there has been no detailed study of instanton configurations in the matrix model \textit{per se}, \textit{i.e.}, off--criticality, nor of their connection to the large--order behavior of the $1/N$ expansion. In \cite{lvm,iy} a general setting for this study was presented but, unfortunately, the results of these papers, albeit written in terms of general matrix model data, are incorrect once we move away from the critical point. 

In this paper, we study in detail the perturbative expansion around a one--instanton configuration in a generic, one--cut matrix model. In particular, we shall give explicit formulae for both the one and two--loop contributions, and we shall write them in terms of geometric data which only depend on the spectral curve associated to the matrix model. This is a critical aspect of our analysis as it makes it possible to apply our results, not only in the realm of  conventional one--cut matrix models, but also to more general theories which are defined by geometric constructions based on a spectral curve, as in \cite{eo}. In particular, and of special interest to us, this is the case of topological strings on certain toric backgrounds \cite{mm}, and our general formulae make it possible to compute instanton effects in these models as well.

Indeed, an important motivation for this paper is to use the dual matrix model description of topological strings on local CY manifolds as a \textit{nonperturbative definition}, which then makes it possible to compute instanton effects in these theories for the first time. This is very similar to the nonperturbative holographic definition of noncritical string theories by double--scaled matrix models. By using this description, we deduce that the nonperturbative completion of the topological string theories considered in this paper includes an infinite number of nontrivial topological sectors, corresponding to the different instanton sectors of the matrix model. Geometrically, we interpret these nonperturbative effects as due to domain walls interpolating between D--brane configurations, as it had already been anticipated in \cite{dv}. 

A rather important aspect of all our nonperturbative computations is that they are \textit{testable} via their connection to the large--order behavior of perturbation theory. Since we compute instanton effects up to (and including) two loops, we can determine the large--order behavior of the genus $g$ free energy up to (and including) the $1/g$ correction. There are various examples where one can compute the $1/N$ expansion to high order, and by making use of standard numerical techniques which extract the asymptotic behavior of a perturbative series, we find an impressive agreement between the large--order numerical data and our theoretical instanton predictions. We will analyze in detail two types of examples. The first example concerns the standard hermitian quartic matrix model, studied for example in \cite{biz}. The second class of examples deals with topological string theory on local curves, which was extensively studied in \cite{cgmps,mm}. We shall confirm and improve the predictions of \cite{mm} about the large--order behavior of these models, and we will also consider a special limit of topological strings on local curves which describes simple Hurwitz numbers (studied in \cite{ksw}). All of these models have a critical point, describing pure 2d gravity, which is controlled by the Painlev\'e I equation. The double--scaling limit of our instanton calculations provides results for the large--order behavior of 2d gravity which refine those obtained \cite{gzj,eyzj} and agree with the analysis of the asymptotics in \cite{jk}. In fact, with the help of the Painlev\'e I equation one can derive the full perturbative expansion around the one--instanton sector, and in this way we provide a further check of our explicit two--loop calculation. 

Mathematically, our results are highly nontrivial predictions for the asymptotics of the $1/N$ expansion of a one--cut matrix model, and they provide some clues concerning the analytic structure of the total free energy of topological string theory, as a function of the string coupling constant. In the case of topological strings, our tests of large--order behavior provide a further check of the conjecture in \cite{mm}, as well as new conjectures about asymptotic properties of enumerative invariants that have not been explored so far. 

This paper is organized as follows. We begin in section 2 by presenting a short review of instanton effects and their connection to the large--order behavior of perturbation theory. We review a simple quantum mechanical example and further provide an extension of the main ideas to the $1/N$ expansion and string theory. These ideas are then explicitly applied in the analysis of one--instanton effects in matrix models in section 3. Here, we shall follow the general strategy put forward in \cite{lvm,iy}, but we shall both simplify and considerably improve their results. In particular, we shall present complete formulae for both the one--loop and the two--loop corrections around the one--instanton configuration, in generic one--cut matrix models. Applications of these results are then considered, starting in section 4 where we consider the quartic matrix model both off--criticality and in the double--scaling limit where it becomes pure 2d gravity. We further present numerical tests of the predictions given by the instanton calculation, by analyzing the large--order behavior of both the quartic matrix model and the Painlev\'e I equation. We then proceed to consider applications in topological string theory. In section 5 we shall consider topological string theories on local curves, verifying and extending the predictions of \cite{mm}, and we shall discuss the spacetime interpretation of the instanton effects in terms of domain walls. Then, in section 6, we analyze in detail the large--order behavior of the generating functionals for simple Hurwitz numbers as a further example of our formalism. In all cases, we find impressive agreement between theoretical and numerical results. A concluding section presents a list of open problems raised 
by our work. Finally, we also collect some explicit formulae for the free energies of both the quartic matrix model and Hurwitz theory, at high genera, in an appendix.

\sectiono{Instantons and Large--Order Behavior}

In this section we shall review the connection between instantons and the large--order behavior of perturbation theory. Good references on this subject include \cite{lgzj,zj,zjbook,dunne}.

\subsection{Field Theory Models}

Let us start by considering a quantum mechanical or field theoretical model which depends on a coupling constant, $g$, in such a way that for $g>0$ the theory has an unstable vacuum and that this vacuum gets stabilized for $g<0$. A simple example of such a situation is the familiar quartic anharmonic oscillator with potential
\FIGURE[ht]{\label{qpot}
    \centering
    \psfrag{g}{$g<0$}
    \psfrag{x}{$g>0$}
    \epsfxsize=0.3\textwidth
    \leavevmode
    \epsfbox{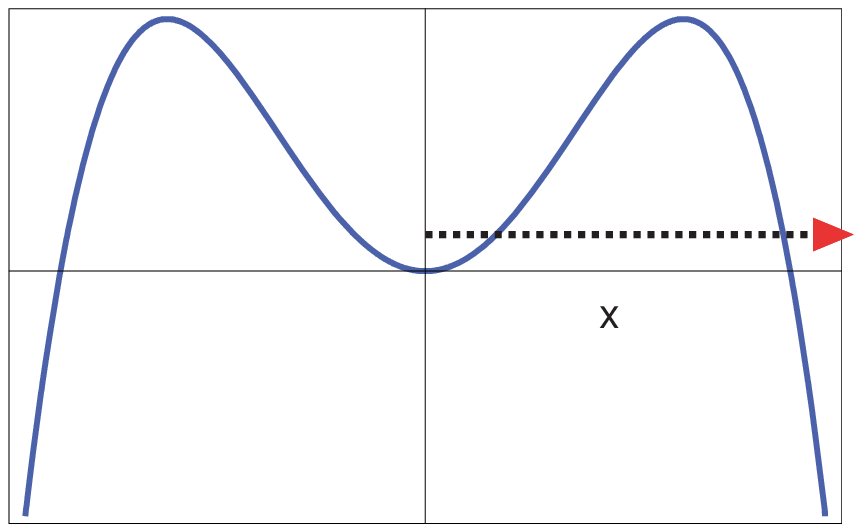} $\qquad$ $\qquad$
    \epsfbox{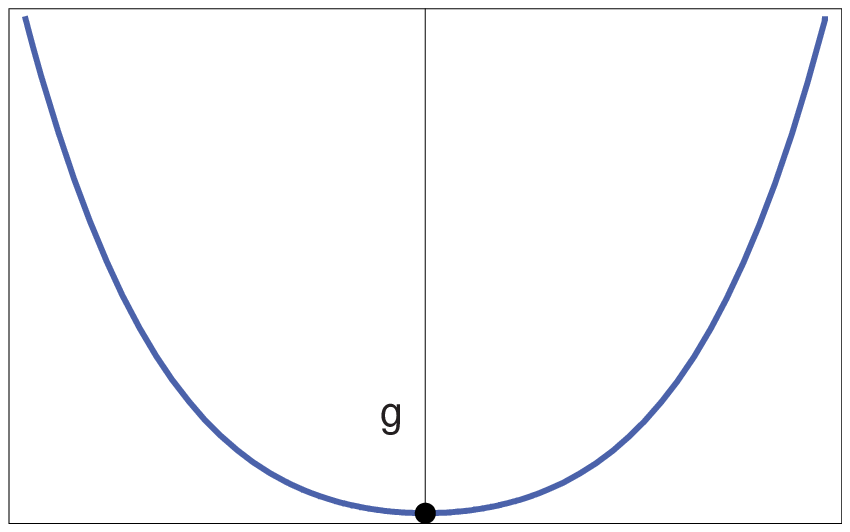}
\caption{The potential for the quartic anharmonic oscillator. When $g>0$ the theory has an unstable vacuum at the origin, which decays via instanton tunneling. This vacuum gets stabilized when $g<0$.}
}
\be
V = \frac{1}{2} x^2 - g x^4. 
\ee
\noindent
Due to the instability, there will be instanton solutions (sometimes called \textit{bounces} in this context) which mediate the decay of the false vacuum. This is illustrated in \figref{qpot}. As one analytically continues the coupling constant to the full complex plane, one finds that the partition function will have a branch cut along the real, positive $g$ axis, with a discontinuity which is purely imaginary. In particular, one may write for the full partition function \cite{cs}
\be
Z(g \pm \ri \epsilon) = Z^{(0)} (g) \pm {1\over 2}\, \mathrm{disc}\, Z (g), 
\ee
\noindent
defining both $Z^{(0)}$ and the discontinuity across the branch cut $\mathrm{disc}\, Z(g) = Z(g + \ri \epsilon) - Z(g - \ri \epsilon)$. A careful analysis of the physics of this problem, in the particular example of the anharmonic oscillator \cite{cs,cc,zj}, shows that $Z^{(0)}$ is given by the path integral around the perturbative vacuum (or zero--instanton configuration), while the leading contribution to $\mathrm{disc}\, Z (g)$ turns out to be given by the path integral calculated around the one--instanton configuration, i.e. the instanton configuration with the lowest action in absolute value. We shall denote this path integral by $Z^{(1)}(g)$.

Let us be slightly more precise on this point. If we want the partition function to remain meaningful, as one performs the analytical continuation in the coupling constant from the stable to the unstable case, it is required that the contour of integration is also rotated, in a compensating way \cite{cs} (\textit{e.g.}, in the quartic oscillator as one continues $-g$ to $-g \exp(\pm \ri \pi)$ one must rotate $x$ to $x \exp(\mp \ri\pi/4)$). The rotated integration contours are illustrated in \figref{contour}. What the analysis in \cite{cs,cc,zj} shows is that $Z^{(0)}$ is computed as the integral over the \textit{sum} of both contours, $\CC^+ + \CC^-$. In particular, if one is to compute the path integral in a saddle--point approximation, the contribution to $Z^{(0)}$ arises from the saddle--point at the origin. On the other hand, the discontinuity $\mathrm{disc}\, Z (g)$ is computed on the \textit{difference} of the two rotated contours, $\CC^+ - \CC^-$. This immediately implies that the saddle--point at the origin cancels, between the two contours. One thus needs to consider the sub--leading saddle--points, which correspond to the one--instanton configuration. These sub--leading saddle--point contributions are also illustrated in \figref{contour}. In particular, notice that
\FIGURE[ht]{\label{contour}
    \centering
    \psfrag{Cp}{$\CC^+$}
    \psfrag{Cm}{$\CC^-$}
    \psfrag{S1}{$\CS_1$}
    \psfrag{S2}{$\CS_2$}
    \epsfxsize=0.35\textwidth
    \leavevmode
    \hspace{3cm}\epsfbox{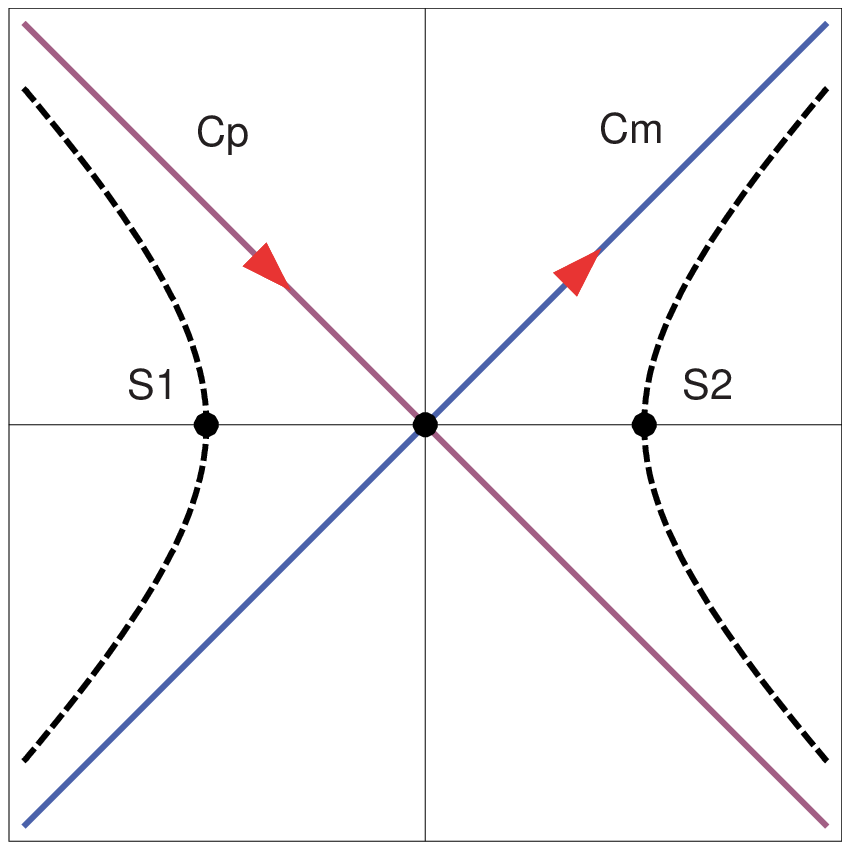}\hspace{3cm}
\caption{The complex plane for the functional integration. Here, $\CC^+$ and $\CC^-$ are the rotated contours one needs to consider for $g>0$. Their sum may be evaluated by the contribution of the saddle--point at the origin. Their difference is evaluated by the contribution of the sub--leading saddle--points, here denoted as $\CS_1$ and $\CS_2$.}
}
\be
\label{supp}
Z^{(1)}(g) \sim \re^{-1/g}
\ee
\noindent
and it is exponentially suppressed for small $g$ as compared to $Z^{(0)}$. This is exactly as one should expect from the discussion above. If we now consider the free energy, defined by $F=\log Z$, we similarly have
\be\label{fdisc}
F(g \pm \ri \epsilon) = F^{(0)} (g) \pm  {1\over 2} {\rm disc}\, F (g), 
\ee
\noindent
where $F^{(0)} (g) = \log Z^{(0)}(g)$ and
\be\label{mfdisc}
\mathrm{disc}\, F (g) = \log \frac{Z(g + \ri \epsilon)}{Z(g - \ri \epsilon)} = \frac{Z^{(1)}(g)}{Z^{(0)}(g)} + \cdots, 
\ee
\noindent
at leading order in $\re^{-1/g_s}$. We will denote by
\be
F^{(1)}(g) = \frac{Z^{(1)}(g)}{Z^{(0)}(g)}
\ee
\noindent
the one--instanton contribution to the discontinuity. The zero--instanton sector has a perturbative expansion around $g=0$ given by
\be
F^{(0)}(g) = \sum_{k=0}^{+\infty} a_k g^k,
\ee
\noindent
while the contribution from the one--instanton sector to the discontinuity $\mathrm{disc}\, F(g)$ turns out to have an expansion of the form 
\be\label{oneinst}
F^{(1)}(g) = \ri g^{-b} \re^{-A/g} \sum_{n=0}^{+\infty} c_n g^n.
\ee
\noindent
In this equation, $A$ is the action of the single instanton, $b$ is a characteristic exponent, and $c_n$ is the $(n+1)$--loop contribution around the instanton configuration.

If one now assumes analyticity of $F(g)$ in the $g$--plane, except for the branch cut along the positive real axis which we alluded to before, as well as some suitable conditions on the $g\rightarrow \infty$ behavior, one can deduce the following relation between the coefficients of the perturbative expansion around the zero--instanton sector and the discontinuity across the cut
\be\label{disp}
a_k ={1\over 2\pi \ri} \int_0^{\infty} \rd z\, {F^{(1)} (z) \over z^{k+1}}. 
\ee
\noindent
Plugging the expansion for $F^{(1)}$ (\ref{oneinst}) in the above formula (\ref{disp}) we find an asymptotic expansion for large $k$, 
\be\label{loinst}
a_k \sim {1\over 2 \pi} \sum_{n=0}^{+\infty} c_n A^{-k-b+n} \Gamma(k+b-n).
\ee
\noindent
This can be equivalently written as
\be\label{loinst2}
a_k \sim {A^{-b-k} \over 2 \pi}\, \Gamma(k+b) \left[c_0 + {c_1 A \over k+b-1} + {c_2 A^2 \over (k+b-2)(k+b-1)} + \cdots \right].
\ee
\noindent
What one learns from this analysis is that the computation of the one--instanton partition function, at one--loop, determines the leading order of the asymptotic expansion for the perturbative coefficients of the zero--instanton partition function, while higher--loop corrections yield the $1/k$ corrections. Notice that instanton configurations with an action $A'>A$ (in particular, multi--instanton configurations with action $n A$, $n\ge 2$) give corrections to the asymptotics of $a_k$ which are exponentially suppressed in $k$, and will not be considered in here. The relation between a nonperturbative instanton computation and the large--order behavior of perturbation theory was first implemented by Bender and Wu in the case of the quartic anharmonic oscillator in quantum mechanics \cite{bw}. They used the WKB method in order to perform a two--loop computation around the bounce, and thus obtain precise numerical values for $c_0$ and $c_1$. Furthermore, they performed accurate numerical tests of their prediction (\ref{disp}) for the large--order behavior of the $a_k$ coefficients. Their results were later reproduced in path integral language \cite{cs}.

In this quantum mechanical example the analyticity conditions for the free energy can be justified rigorously (see \cite{simon} for a review). In more general situations (such as in quantum field theory) one cannot justify these same assumptions; however the relation (\ref{disp}) can be tested in a number of examples with surprising numerical precision (see, \textit{e.g.}, \cite{zj, zjbook} for a review of these tests).

\subsection{The $1/N$ Expansion and String Theory}\label{sec:expansion}

The existence of a connection between instantons and large--order behavior has also been addressed in the context of the $1/N$ expansion; for example in \cite{hb} where one considers vector models in low dimension. In the case of matrix models and their double--scaling limit, such a connection was used in \cite{david1, david2, gzj} in order to infer on the large--order behavior of pure 2d gravity, by computing the instanton action directly in the matrix model (see \cite{dfgz} for a review). However, precise tests at one--loop or higher (the $c_n$ coefficients in the expressions above) have not been performed to date, and we shall fill such a gap in the present work. In order to proceed to loop--level, one first needs a generalization of both the dispersion relation (\ref{disp}) and the expression for the perturbation theory asymptotics (\ref{loinst}), to the present setting.

We shall proceed in a heuristic way. Let us first consider the perturbative series in the zero--instanton sector of a closed string theory or its matrix model dual,
\be
F^{(0)} (g_s) = \sum_{g=0}^{+\infty} F_g(t)\, g_s^{2g-2}.
\ee
\noindent
In this equation the sum is over all genera, $g_s$ is the string coupling constant and $t$ is the 't~Hooft coupling $t = g_s N$ in the context of matrix models, or a geometric modulus in string theory. Observe that while in the previous case of the anharmonic oscillator one wanted to study the asymptotics of a standard numerical series, one now wants to address the asymptotics of a series of functions, naturally enlarging the complexity of the problem \cite{hb}. In order to have a perturbative series with standard structure, we shall consider instead
\be
\CF (g_s) = g_s^2\, F(g_s).
\ee
\noindent
In this case, the one--instanton path integral yields a series of the form 
\be\label{curlfa}
\CF^{(1)} (z) = \ri z^{-b/2} \re^{-\frac{A}{\sqrt{z}}} \sum_{n=0}^{+\infty} c_n z^{n/2},
\ee
\noindent
where $z=g_s^2$. This is an important feature distinguishing matrix models and string theory from field theory: the action of an instanton goes like $1/\sqrt{z}$, and not as $1/z$. Similarly, the perturbation series around the instanton sector is a series in powers of $\sqrt{z}$, and not a series in powers of $z$. As such, we may now write
\be
\CF^{(0)} (z)=\sum_{g=0}^{+\infty} F_g(t)\, z^g.
\ee
\noindent
Our basic assumption is that a dispersion relation of the form (\ref{disp}) holds in here, as it did in field theory. In this case, one finds 
\be\label{lostringl}
F_g = {1\over 2\pi} \int_0^{\infty} {\rd z \over z^{g+1}} z^{-b/2} \re^{-\frac{A}{\sqrt{z}}} \sum_{n=0}^{+\infty} c_n z^{n/2} \sim {1\over \pi} \sum_{n=0}^{+\infty} c_n A^{-2g-b+n} \Gamma(2g+b-n),
\ee
\noindent
which may be explicitly written as
\be\label{lostring}
F_g \sim {A^{-2g-b} \over \pi}\, \Gamma(2g+b)\, \mu_1 \left[1 + {\mu_2 A \over 2g+b-1} + {\mu_3 A^2 \over (2g+b-2)(2g+b-1)} + \cdots \right],
\ee
where we have introduced for later convenience
\be
\mu_1=c_0, \qquad \mu_{i+1}={c_i\over c_0}, \quad  i\geq 1.
\ee
The series inside the brackets in (\ref{lostring}) must be understood as an asymptotic expansion in powers of $1/g$, therefore up to two loops we can write it as 
\be
F_g \sim {A^{-2g-b} \over \pi}\, \Gamma(2g+b)\, \mu_1 \left[1 + {\mu_2 A \over 2g} + \cdots \right].
\ee
\noindent
Justifying that the dispersion relation (\ref{disp}) holds in the present context is more delicate. The underlying reason is that $g_s^2$ or $1/N^2$ appear naturally as coupling constants only in a collective field treatment of the problem (or, equivalently, in a formulation in terms of a closed string field theory). In spite of this, one could still present a heuristic derivation of (\ref{lostringl}) by making use of the Lipatov approach to the large--order behavior, and applying it within the context of collective/string field theory. In this approach one does not use the analyticity properties of the free energy, but instead performs a saddle--point evaluation in both field space and coupling space \cite{lipatov}. Another heuristic derivation of (\ref{lostringl}) can be done by using Borel transforms \cite{dfgz}. Instead of trying to provide a more rigorous foundation for (\ref{lostringl}), we shall proceed to test it in various examples, also in the spirit of the many tests performed in field theory.

In writing (\ref{lostring}) we have implicitly assumed that there is a single instanton solution that contributes to the asymptotic behavior. In general there might be various instanton configurations in the system, with the same action in absolute value, and in this case $F^{(1)}$ will denote the sum of all these contributions. For example, in the quartic matrix model, which we will analyze in section 4, due to the symmetry of the potential there are two instantons which contribute equally. It is also common to have complex instanton solutions which give complex conjugate contributions to $F^{(1)}$, and in this case the asymptotic behavior of $F_g$ is again obtained by adding their contributions \cite{blgzj}. If we write
\be\label{thetas}
A =|A| \re^{\ri \theta_A}, \qquad \mu_1=|\mu_1|\re^{\ri \theta_{\mu_1}}, 
\ee
\noindent
the leading asymptotics will read in this case
\be\label{acos}
F_g \sim {|A|^{-2g-b}  \over \pi}\, \Gamma(2g+b)\, |\mu_1|\,  \cos \bigl( (2g+b) \theta_A + \theta_{\mu_1}\bigr).
\ee
\noindent
We shall also find examples of this situation in the models studied in this paper.

\subsection{Numerical Methods and Richardson Transforms}\label{sec:num}

The instanton computations we perform in this work yield predictions for the quantities $A$, $b$, $\mu_1$ and $\mu_2$ appearing in (\ref{lostring}) above. In order to test these predictions, one has to extract these quantities from the asymptotics of the sequence $\{ F_g\}_{g\ge 0}$. However, computation of the amplitudes $F_g$ is, in most cases, rather involved and therefore they will typically only be available at low genus, of order $g<20$. This will also be the case for our examples, apart from 2d gravity where the Painlev\'e I equation allows for a computation to arbitrarily high genus. We shall therefore use a standard numerical technique known as Richardson extrapolation (see, \textit{e.g.}, \cite{bo}), in order to be able to extract the asymptotic behavior more accurately from the very first terms of the series. This method removes the first terms of the subleading tail and thus accelerates convergence towards the leading asymptotics.

The basic idea of Richardson extrapolation is as follows. Given a sequence
\be\label{srich}
S(g)=s_0+{s_1\over g}+{s_2\over g^2}+\cdots,
\ee
\noindent
its Richardson transform is defined as
\be
A_S(g,N)=\sum_{k\geq 0}{S(g+k)(g+k)^N(-1)^{k+N}\over k!(N-k)!}.
\ee
\noindent
This cancels the sub--leading terms in $S(g)$ up to order $g^{-N}$. Indeed, one can show that if $S(g)$ truncates at order $g^{-N}$, the Richardson transform gives exactly the leading term $s_0$. 

The first quantity that one may extract from the sequence $\{ F_g\}_{g\ge 0}$, assuming it is of the form \eqref{lostring}, is the instanton action. In order to apply the Richardson method, we need a sequence with large $g$ asymptotics of the form (\ref{srich}). This is achieved by considering the sequence
\be\label{richainst}
Q_g={F_{g+1}\over 4g^2F_g}={1\over A^2}\left(1+{1+2b\over 2g}+\CO\left({1\over g^2} \right)\right).
\ee
\noindent
Once $A$ has been found, one can then simply extract the parameter $b$ from the new sequence
\be
2g\left(A^2{F_{g+1}\over 4g^2F_g}-1\right)=1+2b+\CO\left({1\over g}\right).
\ee
\noindent
Finally, one obtains the coefficients $\mu_1$ and $\mu_2$ from the sequences
\be\label{richc0}
{\pi A^{2g+b}F_g\over \Gamma(2g+b)}=\mu_1\left(1+{\mu_2 A\over 2g}+\CO\left({1\over g^2}\right)\right)
\ee
\noindent
and
\be\label{richc1}
{2g\over A}\left({\pi A^{2g+b}F_g\over \mu_1\Gamma(2g+b)}-1 \right)=\mu_2+\CO\left({1\over g}\right),
\ee
\noindent
whose asymptotics are already of the form (\ref{srich}), with leading terms $\mu_1$ and $\mu_2$, respectively. This is the basic picture behind most of our numerical work.

The situation is slightly more complicated when we have to deal with two complex conjugate instantons. In this case, the ansatz for $F_g$ is given by \eqref{acos}. If the absolute value of the instanton action is known, its phase $\theta_A$ can be checked using the sequence
\be
{|A|^{2g+2}F_{g+1}\over (2g+b+1)(2g+b)F_g}-{|A|^{2g-2}F_{g-1}(2g+b-2)(2g+b-1)\over F_g} = 2\cos(2\theta_A)\left(1+\CO\left({1\over g^2}\right)\right).
\ee

\sectiono{Instanton Calculus in Matrix Models}

We shall now perform a more systematic implementation of the ideas discussed in the previous section, in the context of generic, one--cut matrix models.

\subsection{Preliminary Results on Matrix Models}

We will consider a hermitian matrix model for an $N \times N$ matrix $M$, with 
generic potential $V(M)$. We shall use the normalizations of \cite{mmhouches}, so that the partition function will be defined by 
\be\label{matrix}
Z_N={1 \over {\rm vol}(U(N))}\int \rd M\, \re^{-{1\over g_s}\tr V(M)},
\ee
\noindent
where the factor ${\rm vol}(U(N))$ is the usual volume factor of the gauge group that arises after fixing the gauge. In terms of eigenvalues in the diagonal gauge, $Z_N$ reads
\be
Z_N={1\over N! (2\pi)^N} \int \prod_{i=1}^N \rd \lambda_i\, \Delta^2(\lambda)\, \re^{-{1 \over g_s} \sum_{i=1}^N V(\lambda_i)},
\ee
\noindent
where $\Delta(\lambda)=\prod_{i<j}(\lambda_i-\lambda_j)$ is the familiar  Vandermonde determinant. The normalized free energy of the matrix model is then defined by
\be
F = \log\, {Z_{N} \over Z^G_{N}},
\ee
\noindent
where $Z^G_N$ is the partition function of the Gaussian matrix model, defined by the potential $V(M)=\frac{1}{2} M^2$. The free energy has a perturbative genus expansion 
\be\label{oneovern}
F = \sum_{g=0}^{+\infty} F_g(t)\, g_s^{2g-2},
\ee
\noindent
where 
\be
t=g_s N
\ee
\noindent
is the 't~Hooft coupling. Another important set of quantities in a matrix model are the connected correlation functions
\be\label{wcor}
W_h (p_1, \ldots, p_h) = \left\langle \tr\, {1\over p_1-M} \cdots \tr\, {1\over p_h-M} \right\rangle_{(\mathrm{c})}, 
\ee
\noindent
where the subscript $(\mathrm{c})$ means connected. These correlation functions are generating functions for multi--trace correlators of the form 
\be\label{scor}
W_h (p_1, \ldots, p_h) = \sum_{n_i \ge 1} \frac{1}{p_1^{n_1+1} \cdots p_h^{n_h+1}}\, \left\langle \tr\, M^{n_1} \cdots \tr\, M^{n_h} \right\rangle_{(\mathrm{c})},
\ee
\noindent
and they further have a $g_s$ expansion of the form
\be
W_h (p_1, \ldots, p_h) = \sum_{g=0}^{+\infty} g_s^{2g+h-2} W_{g,h} (p_1, \ldots, p_h;t).
\ee
\noindent
In the literature one may find a great deal of work concerning the computation of the quantities $F_g(t)$ and $W_{g,h} (p_1, \ldots, p_h;t)$, starting with the seminal work of \cite{bipz} and culminating in the recent formulation of \cite{eo}. In the following, we shall focus on the so--called one--cut matrix models and present some well known results which will be required at a later stage in our computation.

At large $N$ the zero--instanton sector, or trivial saddle--point, of the matrix model is characterized by a density of eigenvalues $\rho(\lambda)$. In the one--cut case, this density has support on a single, connected interval $\CC=[a,b]$ in the complex plane. This density is completely determined by the condition that the effective potential on an eigenvalue, 
\be\label{veff}
V_{\rm eff}(\lambda) = V(\lambda) - 2t \int {\rm d} \lambda'\, \rho(\lambda') \log |\lambda -\lambda'|,
\ee
\noindent
has to be {\it constant}---at fixed 't~Hooft coupling---on the interval $\CC$:
\be\label{vconst}
V_{\rm eff}(\lambda) = t \xi(t), \qquad \lambda \in \CC. 
\ee
\noindent
A quantity which is closely related to the density of eigenvalues is the resolvent, defined by
\be\label{zeroresint}
\omega_0(p) =\int \rmd \lambda \,{\rho (\lambda)\over p -\lambda}.
\ee
\noindent
Once the resolvent is known, the eigenvalue density follows as
\be\label{rhow}
\rho(\lambda) = - {1 \over 2 \pi \ri} \bigl( \omega_0 (\lambda + \ri\epsilon) - \omega_0 (\lambda - \ri \epsilon) \bigr).
\ee
\noindent
It turns out that the resolvent may be written as
\be
\omega_0(p) ={1\over 2t} \bigl( V'(p) - y(p) \bigr), 
\ee
\noindent
where $y(p)$ is a function which has a branch cut along $\CC$, called the \textit{spectral curve} of the matrix model. It is explicitly given by
\be\label{scurve}
y(p) = M(p) {\sqrt{(p-a)(p-b)}}, 
\ee
\noindent
where $M(p)$, known as the \textit{moment function}, is given by 
\be\label{momentf}
M(p) = \oint_{\infty} {\rd z \over 2 \pi \ri}\, {V'(z) \over z-p}\, {1 \over{\sqrt{(z-a)(z-b)}}},
\ee
\noindent
with the contour of integration being around the point at $\infty$. The endpoints of the cut follow from the equations
\be\label{endpo}
\ba
\oint_{\cal C} {\rd z \over 2\pi \ri}\, {V'(z) \over {\sqrt{(z-a)(z-b)}}}&=0, \\
\oint_{\cal C}{\rd z \over 2\pi \ri}\, {z V'(z) \over {\sqrt{(z-a)(z-b)}}}&=2t.
\ea
\ee
\noindent
There is also a useful formula for the moments of the function $M(p)$, which are defined as
\be\label{moments}
M_{a,b}^{(k)} = {1\over (k-1)!} \left. {\rd^{k-1} \over \rd p^{k-1}} M(p) \right|_{p=a,b}, \qquad k\ge 1, 
\ee
\noindent
given in terms of contour integrals \cite{ackm}:
\be\label{momentsint}
M_{a}^{(k)} = \oint_{\CC} {\rd z \over 2 \pi \ri}\, {V'(z) \over (z-a)^{k+{1\over 2}} (z-b)^{1\over 2}}, \qquad 
M_{b}^{(k)} = \oint_{\CC} {\rd z \over 2 \pi \ri}\, {V'(z) \over (z-a)^{{1\over 2}} (z-b)^{k+{1\over 2}}}.
\ee

To make a long story short, it turns out that the quantities $F_g(t)$ and $W_{g,h} (p_1, \ldots, p_h;t)$ can be computed in terms of the spectral curve alone. More precisely, knowledge of the endpoints of the cut, $a$ and $b$, and of the moments (\ref{moments}), is all one needs in order to compute them. This was first made clear in \cite{ackm} and later culminated in the geometric formalism of \cite{eynard, ce,eo}. For example, one has for the genus--one free energy \cite{ackm} 
\be\label{fone}
F_1 = -{1\over 24} \log \left[ M(a) M(b) (a-b)^4 \right].
\ee
\noindent
The two and three--point correlators at genus zero are given by \cite{ajm}
\be\label{annmm}
\ba
W_{0,2} (p,q) &= {1 \over 2 (p-q)^2} \left( {p q - {1\over 2} (p+q) (a+b) + ab \over {\sqrt{(p-a)(p-b)(q-a)(q-b)}}} - 1 \right),\\
W_{0,3} (p,q,r) &= \frac{1}{8\sqrt{(p-a)(p-b)(q-a)(q-b)(r-a)(r-b)}}\, \cdot \\
& \cdot \left( \frac{a-b}{M(a)}\, \frac{1}{(p-a)(q-a)(r-a)} + \frac{b-a}{M(b)}\, \frac{1}{(p-b)(q-b)(r-b)} \right),
\ea
\ee
\noindent
while the one--point function at genus one is given by \cite{ackm}
\be
\ba
W_{1,1} (p) &= \frac{1}{16 M(a) (p-a) \sqrt{(p-a)(p-b)}} \left( \frac{2p+b-3a}{(p-a)(b-a)} - \frac{M'(a)}{M(a)} \right) + \nonumber \\
&  + \frac{1}{16 M(b) (p-b) \sqrt{(p-a)(p-b)}} \left( \frac{2p+a-3b}{(p-b)(a-b)} - \frac{M'(b)}{M(b)} \right).
\ea
\ee
\noindent
The only exceptions to this rule are the genus--zero free energy, $F_0(t)$, which is given by
\be\label{planarf0}
F_0(t) = -{t\over 2} \int_{\mathcal C} {\rm d}\lambda\, \rho (\lambda) V (\lambda) - {1\over 2} t^2 \xi(t),
\ee
\noindent
and the one--point function 
\be
W_{0,1} (p)= t \omega_0(p).
\ee
\noindent
It is clear from the expressions above that, for these two quantities, the spectral curve is not enough and one also needs to know the explicit form of the potential. As we shall soon unfold, the perturbative expansion around the one--instanton solution is again completely determined by the geometry of the spectral curve.

In the one--instanton computation that we shall perform in the next section, we will also need some results about the derivatives with respect to $t$ of various quantities that characterize the large $N$ solution. A result we need is (see \cite{dfgz})
\be\label{derom}
{\partial (t \omega_0(p)) \over \partial t}={1\over {\sqrt {(p-a)(p-b)}}}
\ee
\noindent
together with the following derivatives, which follow from the defining relations (\ref{endpo}) and (\ref{momentsint}), 
\be\label{abder}
\frac{\partial a}{\partial t} = \frac{4}{a-b}\, \frac{1}{M(a)}, \qquad
\frac{\partial b}{\partial t} = \frac{4}{b-a}\, \frac{1}{M(b)}.
\ee
\noindent
Using these formulae one finds,
\be
\ba
\partial_t y (z) &= - \frac{2}{\sqrt{(z-a)(z-b)}}, \\
\partial_t M (z) &= \frac{2}{(z-a)(z-b)} \left( \frac{(z-b) M(z)}{(a-b) M(a)} + \frac{(z-a) M(z)}{(b-a) M(b)} - 1 \right),
\ea
\ee
\noindent
as well as 
\be
\ba
\partial_t M (a) &= \frac{6}{a-b}\, \frac{M'(a)}{M(a)} + \frac{2}{(a-b)^2} \left( 1 - \frac{M(a)}{M(b)} \right), \\
\partial_t M (b) &= \frac{6}{b-a}\, \frac{M'(b)}{M(b)} + \frac{2}{(b-a)^2} \left( 1 - \frac{M(b)}{M(a)} \right).
\ea
\ee
\noindent
Finally, we will also need derivatives of the free energies. One finds\footnote{Notice that $\partial_t V_{\mathrm{eff}} (b)$ can be obtained by integrating $\partial_t y (z)$ from $-\infty$ to $b$, an integral which diverges logarithmically. Sensible results are obtained \cite{iy} by always considering its regulated version, where one simply drops the divergent $\log z$ term.} \cite{iy}
\be\label{fzeroders}
\ba
\partial_t F_0 (t) &= - t \xi(t) = - V_{\mathrm{eff}} (b), \\
\partial_t^2 F_0 (t) &= - \partial_t V_{\mathrm{eff}} (b) = 2 \log \left( b-a \right) - 2 \log 4, 
\ea
\ee
\noindent
while higher derivatives with respect to $t$ follow from (\ref{abder}). 

\subsection{The One--Instanton Sector of the Matrix Model}

We shall now compute the ``path'' integral around the one--instanton configuration of the matrix model. In this process, we will adopt the framework put forward in \cite{lvm}, but we shall use saddle--point technology rather than the approach based on orthogonal polynomials. Such a strategy has been considered before, as the approach of \cite{lvm} was first rephrased in terms of the saddle--point perspective in \cite{iy} (for a third point of view, based on collective field theory, see \cite{djr}). At this stage, it is important to point out that our calculation will improve on the calculations in \cite{lvm,iy} in three different ways. First of all, we shall fully exploit the saddle--point technology in order to present a much more succinct derivation of the final results. Secondly, we shall compute explicit formulae for the quantum expansion around the one--instanton solution up to two loops. Thirdly, and more importantly, we shall correct both the approach and the one--loop result in \cite{lvm,iy} which, as they stand, are incorrect once one moves away from criticality.
\FIGURE[ht]{\label{mmeffpot}
    \centering
    \epsfxsize=0.4\textwidth
    \leavevmode
    \hspace{3cm}\epsfbox{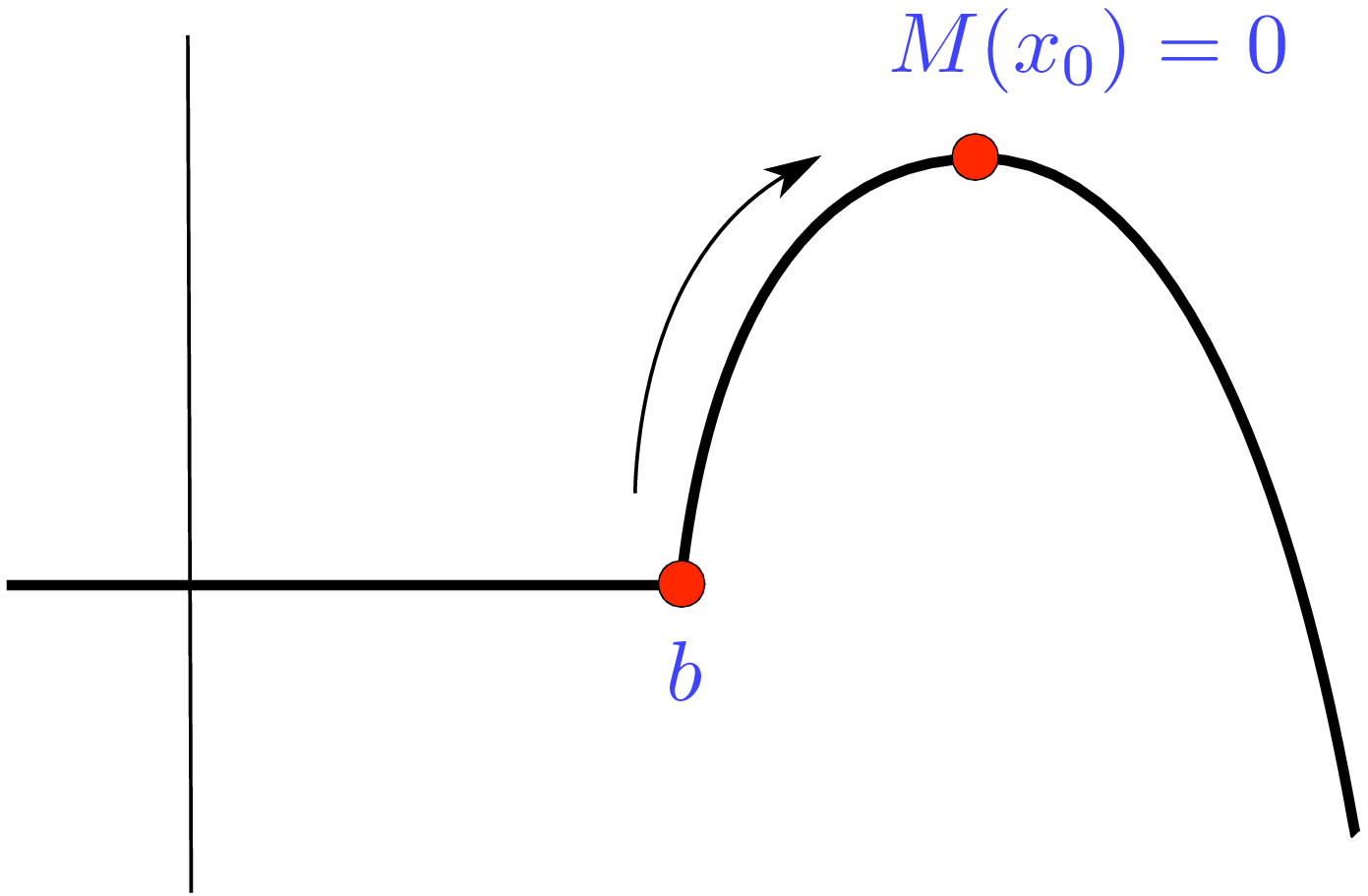}\hspace{3cm}
\caption{The effective potential for the matrix model eigenvalues.}
}

We thus consider a one--cut matrix model in which the effective potential has the form depicted in \figref{mmeffpot}. It is constant along the cut $\CC=[a,b]$, where there is a local, unstable minimum, and it has a local maximum at the point $x_0$. The standard $1/N$ expansion is computed by considering the saddle--point configuration in which all of the $N$ eigenvalues have support in the cut $\CC$. As was first pointed out in \cite{david1, david2, shenker}, a $k$--instanton configuration corresponds to a distinct saddle--point, in which $N-k$ of the eigenvalues remain with support in the interval $\CC$, while $k$ eigenvalues are placed at the local maximum $x_0$ (of course one still assumes that $k\ll N$).

The matrix integral for the one--instanton sector (this is the case where a single eigenvalue sits at $x_0$) is given by \cite{lvm}
\be\label{onemm}
Z_N^{(1)} = {N \over N! (2\pi)^N} \int_{x\in \CI} \rd x \, \re^{-{1\over g_s}  V (x)}
 \int_{\lambda \in \CI_0}  \prod_{i=1}^{N-1}\rd \lambda_i\, \Delta^2 (x, \lambda_1, \ldots, \lambda_{N-1})\, \re^{-{1\over g_s} \sum_{i=1}^{N-1} V (\lambda_i)},
\ee
\noindent
where the first integral in $x$ is over the nontrivial saddle--point contour, which we have denoted by $x\in \CI$, while the rest of the $N-1$ eigenvalues are integrated around the standard saddle--point contour $\CI_0$. At this stage, it might be useful to have in mind \figref{contour} and the discussion concerning the quartic oscillator in section 2. The overall factor of $N$ in front of the integral is a symmetry factor, counting the $N$ possible distinct ways of choosing one eigenvalue out of a set of $N$. One can easily write similar integrals for the $k$--instanton contribution (see \cite{lvm}) but we shall not consider them in here. Taking all normalization factors into careful account, one finds for (\ref{onemm}) \cite{lvm}
\be\label{oneinstex}
\ba
Z_N^{(1)} &= {N \over N! (2\pi)^N}\, (2\pi)^{N-1} (N-1)!\, Z^{(0)}_{N-1} \int_{x\in \CI} \rd x \left\langle \det (x {\bf 1} - M')^2 \right\rangle^{(0)}_{N-1}\, \re^{-{1\over g_s} V(x)} \\
&\equiv {1\over 2\pi}\, Z_{N-1}^{(0)} \int_{x \in \CI} \rd x\, f(x).
\ea
\ee
\noindent
The notation in this equation is as follows. $Z_N^{(0)}$ is the partition function evaluated around the standard saddle--point, and within the standard $1/N$ expansion. $M'$ is an $(N-1) \times (N-1)$ hermitian matrix, and all of its eigenvalues are still integrated around the standard saddle--point. $\langle \CO\rangle_N^{(0)}$ is the normalized vacuum expectation value of the gauge--invariant operator $\CO$, again computed around the standard saddle--point, 
\be
\left\langle \CO \right\rangle^{(0)}_N = {\int_{\lambda\in \CI_0} \prod_{i=1}^N \rd \lambda_i\, \Delta^2(\lambda)\, \CO(\lambda)\, \re^{-{1\over g_s} \sum_{i=1}^N V(\lambda_i)} \over \int_{\lambda\in \CI_0} \prod_{i=1}^N \rd \lambda_i\, \Delta^2(\lambda)\, \re^{-{1\over g_s} \sum_{i=1}^N V(\lambda_i)}}.
\ee
\noindent
Finally, we have also defined 
\begin{equation}\label{fx}
f(x) = \left\langle \det (x {\bf 1} - M')^2 \right\rangle^{(0)}_{N-1}\, \re^{-{1\over g_s} V(x)}.
\end{equation}
\noindent
As we have seen in (\ref{mfdisc}), the one--instanton contribution to the free energy may be expressed in terms of the partition function, at leading order in $\rme^{-1/g}$, by 
\begin{equation}\label{chempot}
 F^{(1)} = {Z_N^{(1)} \over Z_N^{(0)}} = {1\over 2\pi}\, { Z_{N-1}^{(0)} \over Z_N^{(0)}} \int_{x\in \CI} \rd x\, f(x).
\end{equation}
In the rest of this section we shall present a careful computation of this quantity. 

In order to calculate the instanton contribution (\ref{chempot}), we shall first compute $f(x)$, as defined above in (\ref{fx}). Making use of the familiar relation $\det (x {\bf 1} - M) = \exp\left({\rm tr}\,{\rm ln} (x {\bf 1}-M)\right)$ we obtain,   
\be\label{conex}
\left\langle \det (x {\bf 1} - M)^2 \right\rangle = \exp \left[ \sum_{s=1}^{+\infty} {2^s\over s!} \left\langle \left( {\rm tr}\, {\rm ln} (x {\bf 1} - M) \right)^s \right\rangle_{(\rm c)} \right],
\end{equation}
\noindent
which is written in terms of connected correlation functions (recall that the cumulant expansion precisely relates the generating functional of \textit{standard} correlation functions to the generating functional of \textit{connected} correlation functions as in this equality). The correlation functions appearing in (\ref{conex}) are nothing but integrated versions of the $W_{h}$ correlators in (\ref{wcor}), evaluated at coincident points. Let us define
\be\label{theas}
\ba
A_{g,h} (x;t) &= \left. \int^{x_1} \rd p_1 \cdots \int^{x_h} \rd p_h\, W_{g,h} (p_1,\cdots, p_h) \right|_{x_1=\cdots =x_h=x}, \\
\CA_n(x;t) &= \sum_{k=0}^{\left[\frac{n}{2}\right]} \frac{2^{n-2k+1}}{(n-2k+1)!}\, A_{k,n-2k+1} (x;t), \quad n\ge 1.
\ea
\ee
\noindent
In this notation, the general perturbative formula for the determinant follows as
\be
\left\langle \det ( x \mathbf{1} - M )^2 \right\rangle = \exp \left( \sum_{n=0}^{+\infty} g_s^{n-1} \CA_n (x;t) \right),
\ee
\noindent
where $\CA_n(x;t)$ is the $n$--loop contribution. We have, for example,  
\be
\ba
\CA_0(x;t) &= 2 A_{0,1} (x;t), \\
\CA_1(x;t) &= 2 A_{0,2} (x;t), \\
\CA_2(x;t) &= \frac{4}{3} A_{0,3} (x;t) + 2A_{1,1} (x;t), \\
\CA_3(x;t) &= \frac{2}{3}A_{0,4} (x;t) + 2A_{1,2} (x;t).
\ea
\ee
\noindent
One observes that in order to compute the determinant at $n$--loops, one would require analytic expressions for the $W_{g,h}$ with $(g,h) = (0,n+1), (1,n-1), (2,n-3), \ldots, (\frac{n}{2},1)$. Let us also point out that the integration constants involved in the integrations in (\ref{theas}) may be simply fixed by the large $x$ expansion of the correlators. Indeed, we have the expansion
\be
\left\langle ({\rm tr}\, {\rm ln} (x {\bf 1} - M))^s \right\rangle_{(\rm c)} =  \sum_{n_i\ge 1} {(-1)^s \over \prod_{i=1}^{s} n_i}\, \langle \tr\, M^{n_1} \cdots \tr\, M^{n_s} \rangle_{(\rm c)} \, x^{-\sum_{i=1}^s n_i}.
\ee
\noindent
Next, we define the \textit{holomorphic} effective potential, which combines the matrix model potential together with $\CA_0(x;t)$, as
\be\label{vheff}
V_{\rm h,eff}(x;t) = V(x) - 2t \int^x \rd p\, \omega_0 (p) = V(x) - 2t \int \rd p\, \rho(p) \log (x -p), 
\ee
\noindent
which satisfies
\be\label{dery}
V_{\rm h,eff}' (x;t) = y(x)
\ee
\noindent
as well as
\be
{\rm Re}\, V_{\rm h,eff}(x;t) = V_{\rm eff}(x),
\ee
\noindent
where $V_{\rm eff}(x)$ was earlier defined in (\ref{veff}). Altogether, one finally has for the integrand
\be\label{fint}
f(x) = \exp \left( - \frac{1}{g_s} V_{\rm h,eff} (x;t') + \sum_{n=1}^{+\infty} g_s^{n-1} \CA_n (x;t') \right),
\ee
\noindent
where
\be
t'=g_s(N-1)=t-g_s.
\ee
\noindent
This shift in the 't~Hooft parameter is due to the fact that the correlation function involved in (\ref{fx}) is computed in a matrix model with $N-1$ eigenvalues (recall we removed one eigenvalue from the single--cut). Since we are computing the one--instanton contribution in the theory with $N$ eigenvalues, we thus have to expand (\ref{fint}) around $t$. This gives further corrections in $g_s$, which we make explicit as
\be
f(x) = \exp \left( - \sum_{k=0}^{+\infty} g_s^{k-1}\, \frac{(-1)^k}{k!}\, \partial_t^k V_{\mathrm{h,eff}} (x;t) + \sum_{n=1}^{+\infty} \sum_{k=0}^{+\infty} g_s^{n+k-1}\, \frac{(-1)^k}{k!}\, \partial_t^k \CA_n (x;t) \right).
\ee
\noindent
We shall write this expression as
\be\label{fvphi}
f(x) = \exp \left( - \frac{1}{g_s} V_{\mathrm{h, eff}} (x) + \Phi (x) \right),
\ee
\noindent
where we define
\be
\Phi (x) \equiv \sum_{n=1}^{+\infty} g_s^{n-1}\, \Phi_{n} (x) \equiv \sum_{n=1}^{+\infty} g_s^{n-1} \left[ \frac{(-1)^{n-1}}{n!}\, \partial_t^n V_{\mathrm{h,eff}} (x) + \sum_{k=0}^{n-1} \frac{(-1)^k}{k!}\, \partial_t^k \CA_{n-k} (x) \right].
\ee
\noindent
One finds, for example,
\be
\ba
\Phi_1 (x) &= \CA_1 (x) + \partial_t V_{\mathrm{h,eff}} (x), \\
\Phi_2 (x) &= \CA_2 (x) - \partial_t \CA_1 (x) - \frac{1}{2!}\, \partial_t^2 V_{\mathrm{h,eff}} (x), \\
\Phi_3 (x) &= \CA_3 (x) - \partial_t \CA_2 (x) + \frac{1}{2!}\, \partial_t^2  \CA_1 (x) + \frac{1}{3!}\, \partial_t^3 V_{\mathrm{h,eff}} (x).
\ea
\ee
\noindent
In expression (\ref{fvphi}) all quantities now depend on the standard 't~Hooft parameter $t$ for the model with $N$ eigenvalues, and we have thus dropped the explicit dependence on $t$. The derivatives with respect to $t$ can be performed by using the formulae we presented at the end of the last subsection.

One may now proceed with the integration of $f(x)$, 
\be\label{fxint}
\int_{x \in \CI} \rd x\, \exp \left( - \frac{1}{g_s} V_{\mathrm{h,eff}} (x) + \Phi (x) \right).
\ee
\noindent
If we wish to evaluate this integral as a perturbative expansion around small string coupling, $g_s$, we can do it using a saddle--point evaluation \cite{lvm,iy}. The integration contour is over the nontrivial saddle characterizing the one--instanton sector, which is defined by the usual saddle--point requirement
\be
V'_{\mathrm{h,eff}} (x_0) = 0 \quad \Rightarrow \quad y(x_0)=0, 
\ee
\noindent
with $x_0$ located \textit{outside} of the cut. If we use the explicit form of the spectral curve (\ref{scurve}) we find the equivalent condition
\FIGURE[ht]{\label{riemanninst}
    \centering
    \epsfxsize=0.45\textwidth
    \leavevmode
    \hspace{3cm}\epsfbox{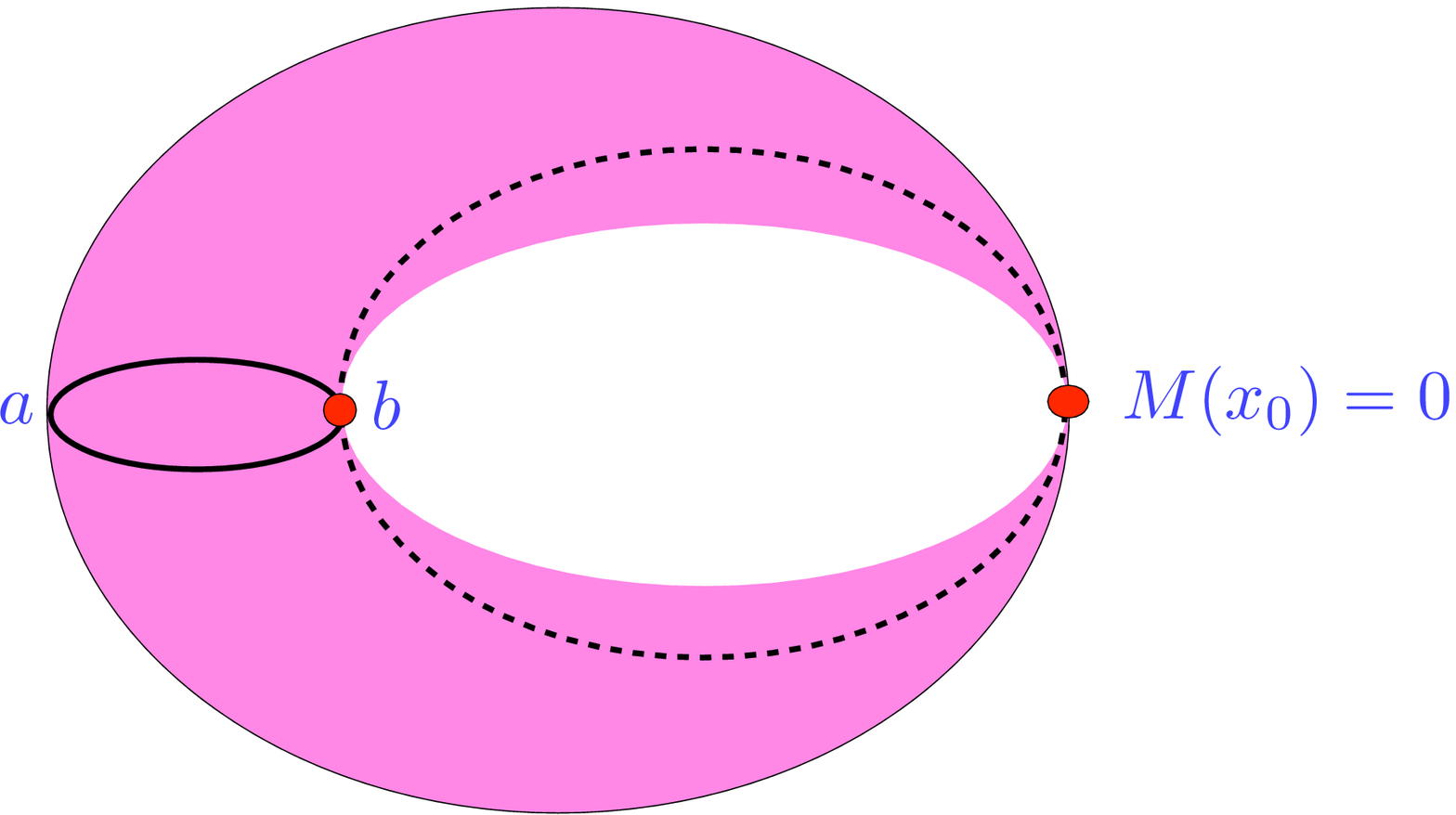}\hspace{3cm}
\caption{The spectral curve $y(x)$ has a singular point at the nontrivial saddle $x_0$.}
}
\be\label{mx}
M(x_0) = 0.
\ee
\noindent
The saddle--point $x_0$ is typically a local maximum of the effective potential, as depicted in \figref{mmeffpot}. Geometrically, the spectral curve is a curve of genus zero pinched at $x_0$, as shown in \figref{riemanninst}. This was observed in \cite{ss} in the context of spectral curves for double--scaled matrix models, and their relation with minimal strings. Of course, it can happen that there is more than one solution to (\ref{mx}). In this case, there will be various instantons and we will have to add up their contributions (the leading contribution arising from the instanton with the highest action, in absolute value). The calculation of (\ref{fxint}) is now completely standard, and it reduces to Gaussian integrations. The result is
\be\label{fexpansion}
\int_{x\in \CI} \rd x\, f(x) =  \sqrt{\frac{2 \pi g_s}{V''_{\mathrm{h,eff}} (x_0)}}\, \exp \left( - \frac{1}{g_s} V_{\mathrm{h,eff}} (x_0) + \Phi_1 (x_0) \right) \left( 1 + \sum_{n=2}^{+\infty} g_s^n\, f_n \right), 
\ee
\noindent
where the $f_n$ can be systematically computed in terms of the functions $\Phi_n(x)$ and their derivatives, evaluated at the saddle--point $x_0$, by making use of the Gaussian integral and the Gaussian moments. This is a long and tedious process, where one should be very careful with factors of $g_s$. In particular, one splits the integrand into the standard Gaussian integrand plus the rest, where the rest should be power--series expanded in order to produce Gaussian moments. This process is source to some extra factors of $g_s$ that must be properly considered. In any case, there are no conceptual difficulties in taking this calculation to arbitrary order. In order to find an explicit expression for the two--loop contribution to the one--instanton path integral, we shall later need
\be
\ba
f_2 &= \Phi_{2} (x_0) + \frac{1}{2 V''_{\mathrm{eff}} (x_0)} \left\{ \partial_x^2 \Phi_{1} (x_0) + \left( \partial_x \Phi_{1} (x_0) \right)^2 \right\} - \\
& - \frac{1}{2 \left( V''_{\mathrm{h,eff}} (x_0) \right)^2} \left\{ \frac{1}{4}\, \partial_x^4 V_{\mathrm{h,eff}} (x_0) + \partial_x^3 V_{\mathrm{h,eff}} (x_0) \partial_x \Phi_{1} (x_0) \right\} + \frac{5 \left( \partial_x^3 V_{\mathrm{h,eff}} (x_0) \right)^2}{24 \left( V''_{\mathrm{h,eff}} (x_0) \right)^3}.
\ea
\ee
\noindent
Observe that the required evaluation of derivatives at $x_0$, in the expression above, is a rather straightforward exercise as we are dealing in this case with rational functions.

The last ingredient needed to compute the one--instanton contribution is the quotient of partition functions in the expression for $\mathrm{disc}\, F$, (\ref{chempot}). This quotient can be written in terms of the standard, perturbative free energies, since
\be
\frac{Z^{(0)}_{N-1}}{Z^{(0)}_N} = \exp \left( F(t')-F(t) \right). 
\ee
\noindent
In the rest of this section both $F(t)$ and $F_g(t)$ shall denote the {\it unnormalized} free energies, \textit{i.e.}, $F=\log\, Z_N$. If one explicitly expands in $g_s$, by both writing the above expression in terms of the standard 't~Hooft parameter $t$ alone, and further expanding the free energy in its perturbative genus expansion (\ref{oneovern}), it follows
\be\label{quotex}
\frac{Z^{(0)}_{N-1}}{Z^{(0)}_N} = \exp \left( \sum_{n=0}^{+\infty} g_s^{n-1} \CG_n \right), \qquad
\CG_n \equiv \sum_{k=0}^{\left[ \frac{n}{2} \right]} \frac{(-1)^{n-2k+1}}{(n-2k+1)!}\, \partial_t^{n-2k+1} F_k (t).
\ee
\noindent
One has, for example,
\be\label{gex}
\ba
\CG_0 &=- \partial_t F_0 (t), \\
\CG_1 &= \frac{1}{2}\, \partial_t^2 F_0 (t), \\
\CG_2 &= -\frac{1}{3!}\, \partial_t^3 F_0 (t) - \partial_t F_1 (t).
\ea
\ee
\noindent
Putting together (\ref{fexpansion}) and (\ref{quotex}) above, we finally find that $F^{(1)}$ has the structure
\be\label{muex}
F^{(1)} = \ri\, g_s^{1\over 2}\,  \mu_1\, \exp \left( -\frac{A}{g_s} \right) \left\{ 1 + \sum_{n=1}^{+\infty} \mu_{n+1} g_s^n \right\}.
\ee
\noindent
Collecting results above we obtain the following contributions to $F^{(1)}$, up to two loops:
\be
\ba
A &= V_{\mathrm{h,eff}} (x_0) - \CG_0 (t), \\
\mu_1 &= -\ri\, \sqrt{\frac{ 1}{2 \pi V''_{\mathrm{h,eff}} (x_0)}}\, \exp \Big( \Phi_1 (x_0)  + \CG_1(t) \Big), \\
\mu_2 &= f_2 + \CG_2(t).
\ea
\ee

Let us now give explicit expressions for these quantities in terms of data associated to the spectral curve (\ref{scurve}). First of all, by using (\ref{gex}), (\ref{fzeroders}) and (\ref{dery}) we find
\be\label{instdiff}
A = V_{\rm h,eff}(x_0) - V_{\rm h,eff}(b) = \int_b^{x_0} \rd z \, y(z), 
\ee
\noindent
which is the instanton action (here, we use the fact that $V_{\rm h,eff}(b)=V_{\rm eff}(b)$). Notice that, as pointed out in \cite{ss}, this expression also has a geometric interpretation as the contour integral of the one--form $y(z)\, \rd z$, from the endpoint of the cut $\CC$ to the singular point $x_0$ (recall \figref{riemanninst}).

We next move to the one--loop contribution, and begin with the computation of $\Phi_1(x)$. One can find the result for $A_{0,2}(x;t)$ (which enters in the expression of $\CA_1$) simply by integrating the first formula in (\ref{annmm}) \cite{iy}
\be\label{vann}
A_{0,2}(x;t) = \log \biggl( 1 + \frac{x - (a+b)/2} {\sqrt{(x - a)(x - b)}} \biggr) - \log 2.
\ee
\noindent
Using (\ref{derom}) one further finds, 
\be\label{dervheff}
\partial_t V_{\rm h,eff}(x) = -4 \log \Bigl[ {\sqrt{x-a}} + {\sqrt{x-b}} \Bigr] +4 \log 2,
\ee
\noindent
and both these results together is all one requires to obtain
\be
\Phi_1(x) = - \log \Bigl[ (x-a)(x-b) \Bigr].
\ee
\noindent
Adding to $\Phi_1(x)$ the result for $\CG_1(t)$, which follows from (\ref{fzeroders}), it is simple to put all expressions together and obtain the contribution, $\mu_1$, of the one--loop fluctuations around the one--instanton configuration, 
\be\label{rone}
\mu_1 = -\ri\, {b-a \over 4} {\sqrt{1 \over  2 \pi M'(x_0) \Bigl[(x_0-a)(x_0-b)\Bigr]^{5\over 2}}}.
\ee
\noindent
This formula is valid for any one--cut matrix model with an unstable potential. Notice that if $x_0$ is a local maximum of $V_{\rm eff}(x)$, one will have that $M'(x_0)<0$, and hence $\mu_1$ will be real. It is also important to point out that our result (\ref{rone}) is \textit{different} from the result obtained in \cite{lvm,iy}. The reason is that, in these references, no distinction is made between correlation functions computed at $t'$ and those computed at $t$. Correspondingly, the contribution of (\ref{dervheff}) is never taken into account. While this contribution vanishes at the critical point, it is non--zero  for generic values of the parameters, making it crucial in order to obtain a generic result. In this paper we shall present substantial evidence that (\ref{rone}) is the correct result, by using the connection to the large--order behavior of perturbation theory explained in the last section.

The computation at two loops does not present any conceptual difficulty, but it is much more involved. One needs the explicit expressions
\be
\ba
A_{0,3} (x;t) &= \frac{\left( \sqrt{x-a} - \sqrt{x-b}\, \right)^3}{(a-b)^2} \left( \frac{1}{M(a) (x-a)^{\frac{3}{2}}} - \frac{1}{M(b) (x-b)^{\frac{3}{2}}} \right), \\
A_{1,1} (x;t) &= - \frac{1}{12(a-b)^2} \left( \frac{1}{M(a)} + \frac{1}{M(b)} \right) + \\
& + \frac{1}{24 (a-b)^2} \left( \frac{\left(2(x-a)+(b-a)\right) \sqrt{x-b}}{M(a) \left(x-a\right)^{\frac{3}{2}}} + \frac{\left(2(x-b)+(a-b)\right) \sqrt{x-a}}{M(b) \left(x-b\right)^{\frac{3}{2}}} \right) - \\
& - \frac{\sqrt{x-a}-\sqrt{x-b}}{8 (a-b)^2} \left( \frac{2M(a)+(a-b)M'(a)}{M^2(a) \sqrt{x-a}} - \frac{2M(b)+(b-a)M'(b)}{M^2(b) \sqrt{x-b}} \right).
\ea
\ee
\noindent
After very long but straightforward computations, one finally obtains the two--loop coefficient as
\be\label{mutwo}
\ba
\mu_2 &= \frac{1}{4(a-b) \sqrt{(x_0-a)(x_0-b)}} \left( \frac{(x_0-b) M'(a)}{M^2(a)} - \frac{(x_0-a) M'(b)}{M^2(b)} \right) - \\
&
- \frac{\sqrt{(x_0-a)(x_0-b)}}{12(a-b)^2} \left( \frac{8(x_0-a)+17(a-b)}{(x_0-a)^2 M(a)} + \frac{8(x_0-b)+17(b-a)}{(x_0-b)^2 M(b)} \right) + \\
&
+ \frac{5 \left( M''(x_0) \right)^2 - 3 M'(x_0) M^{(3)}(x_0)}{24 \left( M'(x_0) \right)^3 \sqrt{(x_0-a)(x_0-b)}} + \frac{35 \left( 2 x_0 - (a+b) \right) M''(x_0)}{48 \left( M'(x_0) \right)^2 \left( (x_0-a)(x_0-b) \right)^{\frac{3}{2}}} + \\
&
+ \frac{140 \left( 2x_0-(a+b) \right)^2 + 33 (a-b)^2}{96 M'(x_0) \left( (x_0-a)(x_0-b) \right)^{\frac{5}{2}}}.
\ea
\ee
\noindent
As one immediately realizes from the explicit expressions above, both $\mu_1$ and $\mu_2$ depend uniquely on data specified by the spectral curve. More precisely, they depend on the endpoints of the cut, $a$ and $b$, the position of the saddle--point, $x_0$, and on the moments of the function $M(p)$, evaluated at $a$, $b$ or $x_0$. It is not hard to convince oneself that the rest of the coefficients $\mu_n$ in (\ref{muex}) must also share this property. This has two important consequences. First of all, it displays the universality of the results, in the sense that two matrix models which lead to the same spectral curve will also share the same discontinuity, $\mathrm{disc}\, F$. In particular, since taking the double--scaling limit commutes with the geometric computation of the amplitudes, different models that lead to the same critical theory will also lead to the same one--instanton contribution at criticality \cite{lvm,iy}. Secondly, since the description of the B--model on mirrors of toric manifolds in \cite{mm,bkmp} only depends on the geometry of the spectral curve, we may also compute nonperturbative effects in these models by simple application of the formulae above for $F^{(1)}$: one just has to apply them to the spectral curves described in \cite{mm,bkmp}. Notice that in this paper we have restricted ourselves to the one--cut case and as such our formalism will only apply to the mirrors of local curves, worked out in \cite{mm}.

\sectiono{Application I: Quartic Matrix Model and 2d Gravity}

Before proceeding towards the realm of topological string theory, we shall test our results in the case of a rather familiar matrix model, the quartic matrix model both off and at criticality.

\subsection{The Quartic Matrix Model}

The quartic matrix model is defined by the potential 
\be
V(z) = \frac{1}{2} z^2 + \lambda z^4,
\ee
\noindent
with $\lambda$ the quartic coupling constant. The properties of this model at large $N$ were addressed long ago in \cite{bipz,biz}. The density of eigenvalues has support on the single cut $\CC = [a,b] \equiv [-2\alpha, 2 \alpha]$, where $\alpha$ is a function of $\lambda$ and the 't~Hooft parameter $t$, as
\be
\alpha^2 = \frac{1}{24 \lambda} \left( - 1 + \sqrt{1+48 \lambda t} \right).
\ee
\noindent
The spectral curve follows as
\be
y(z) = M(z) \sqrt{z^2-4\alpha^2},
\ee
\noindent
with
\be
M(z) = 1 + 8 \lambda \alpha^2 + 4 \lambda z^2.
\ee
\noindent
This function has two zeros which give two non--trivial saddle--points, namely $\pm x_0$ with 
\FIGURE[ht]{\label{vquartic}
    \centering
    \psfrag{nx}{$-x_0$}
    \psfrag{px}{$x_0$}
    \psfrag{a}{$a$}
    \psfrag{b}{$b$}
    \psfrag{C}{$\CC$}
    \epsfxsize=0.35\textwidth
    \leavevmode
    \hspace{3cm}\epsfbox{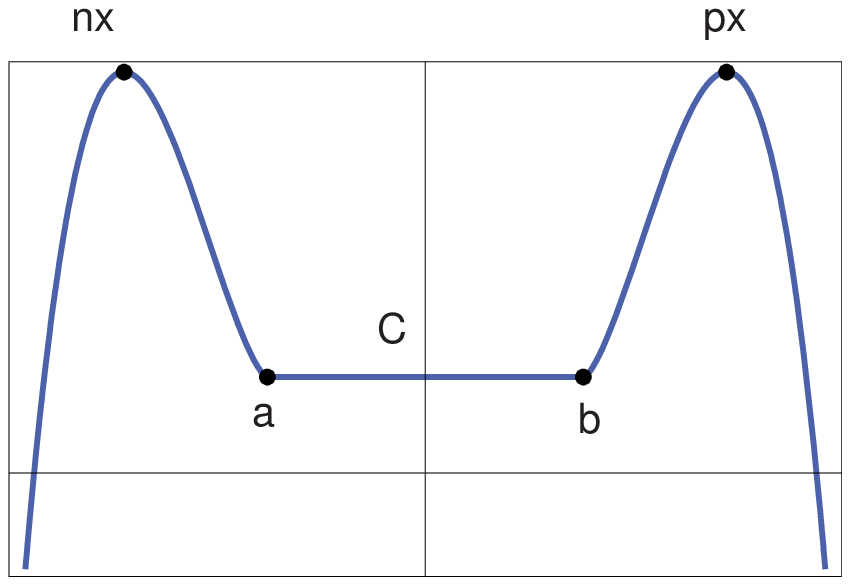}\hspace{3cm}
\caption{The effective potential $V_{\mathrm{eff}} (x)$ for the quartic matrix model. There are two saddle--points located at $x_0$ and $-x_0$.}
}
\be\label{qsaddle}
x_0^2 = -\frac{1}{4 \lambda} \left( 1+8 \lambda \alpha^2 \right).
\ee
\noindent
These two saddle--points are evident in \figref{vquartic}, where we have displayed the effective potential for the quartic matrix model. If we wish to compare the large--order prediction of our formulae with the real behavior of the perturbation theory in this model, one is required to actually compute the free energies at high genera. The set--up for such a calculation was first described in \cite{biz}, but the calculation was only carried out in that paper up to genus $g=2$. We have extended this computation to genus $g=10$ and we shall now review how to compute $F_g$ in the quartic matrix model at large $g$.

The calculation of the $1/N$ expansion of the free energy in the quartic matrix model was set up in \cite{biz} using the method of orthogonal polynomials. 
A review of such method and subsequent calculation would lead us too far apart from the main line of this work, so that in the following we restrict ourselves to presenting an algorithmic prescription to compute $F_g$ which summarizes the results of \cite{biz}. The interested reader should consult the original reference \cite{biz} for full details. Also, for simplicity, we set $t=1$ in the following and will follow the exact same conventions as in \cite{biz}. In particular, in this section our convention for the free energy, following \cite{biz}, is that $F=-\log Z$. There are several components that make up the calculation of $F_g$. It starts with the so--called pre--string equation
\be\label{prestring}
R_n \bigl\{ 1 + 4 \lambda \left( R_n + R_{n-1} + R_{n+1} \right) \bigr\} = n g_s,
\ee
\noindent
for the coefficients $R_n$ which determine the partition function in the orthogonal polynomial formalism. One then considers a continuous version of these coefficients, corresponding to a family of polynomials, $r_{2s} (x;\lambda)$, which, in light of the pre--string equation (\ref{prestring}), satisfy a simple algebraic, recursive relation. For $s=0$
\be
r_0 (x;\lambda) = \frac{1}{24 \lambda} \left( - 1 + \sqrt{1 + 48 \lambda x} \right),
\ee
\noindent
while for $s>0$ the pre--string equation yields the recursive expression
\be
r_{2s} (x;\lambda) + 4\lambda \sum_{m+n=s} r_{2m} (x;\lambda) \left( r_{2n} (x;\lambda) + 2 \sum_{k+p=n} \frac{r_{2k}^{(2p)} (x;\lambda)}{(2p)!} \right) = 0.
\ee
\noindent
In this way it is rather simple to compute the polynomials $r_{2s}(x;\lambda)$ to very high $s$. These polynomials are crucial in other to find $F_g$. Indeed, the general formula for the total free energy is \cite{biz}
\be\label{febiz}
\ba
g_s^2 F(\lambda) &= - \int_0^1 \rmd x\, \left( 1-x \right) \log \Xi(x;\lambda) + \CH(\lambda) - \\
&- \sum_{p=1}^{+\infty} g_s^{2p}\, \, \frac{B_{2p}}{(2p)!}\, \frac{\rmd^{2p-1}}{\rmd x^{2p-1}} \Big( \left( 1-x \right) \log \Xi(x;\lambda) \Big) \bigg|_{x=0}^{x=1},
\ea
\ee
\noindent
where the function $\Xi(x;\lambda)$ is precisely built using the $r_{2s}(x;\lambda)$ polynomials as
\be
\Xi(x;\lambda) = \sum_{s=0}^{+\infty} g_s^{2s}\, \frac{r_{2s}(x;\lambda)}{x}.
\ee
\noindent
In the expression above, $B_{2p}$ are Bernoulli numbers and $\CH(\lambda)$ is the function
\be
\CH(\lambda) = -\frac{1}{2}\, g_s \left[ \log \int_{-\infty}^{+\infty} \rmd \mu\, \rme^{- \frac{1}{2}\mu^2 - g_s \lambda \mu^4} - \log \int_{-\infty}^{+\infty} \rmd \mu\, \rme^{- \frac{1}{2}\mu^2 + g_s \lambda \mu^4} \right].
\ee
\noindent
An expansion of (\ref{febiz}) in powers of $g_s$ then yields explicit expressions for $F_g$. Moreover, this is an algorithmic prescription of calculation, which may be simply implemented with a symbolic computation program. This calculation was carried out analytically up to $g=2$ in \cite{biz} and we have implemented it in a computer program, obtaining in this way 
explicit results up to $g=10$. A partial list of our $F_g$ can be found in the appendix. Here, let us just recall that \cite{biz} conjectured that, for genus $g \ge 2$, the general structure should be of the form
\be
F_g (\alpha^2) = \frac{\left( 1-\alpha^2 \right)^{2g-1}}{\left( 2-\alpha^2 \right)^{5(g-1)}}\, \CP_g (\alpha^2),
\ee
\noindent
with $\CP_g (\alpha^2)$ a polynomial in $\alpha^2$ such that
\be
\CP_g (\alpha^2=1) = \frac{1}{2 \cdot 6^{2g-1}}\, \frac{(4g-3)!}{g! (g-1)!}.
\ee
\noindent
We have checked this conjecture up to genus $g=10$ and further found that the polynomial $\CP_g (\alpha^2)$ is of order $3g-4$ in $\alpha^2$.

\subsection{Instanton Effects and Large--Order Behavior}

Let us now present explicit formulae for the terms contributing to the one--instanton sector, in the quartic matrix model. As we did before, in the following we will set $t=1$ for simplicity. The first thing to notice is that, since the potential is symmetric, there are {\it two} instanton solutions, corresponding to eigenvalue tunneling from $\CC$ to the two saddles $\pm x_0$ (see \figref{vquartic}). Both instantons have the same action, which is computed via direct integration of the spectral curve
\be
A= -\frac{\sqrt{3}\, \alpha^2}{4 \left( 1-\alpha^2 \right)}\, \sqrt{4-\alpha^4} - 2 \log \left[ \sqrt{3} \sqrt{-2+\alpha^2} + \sqrt{-2-\alpha^2} \right]  + \log 4 \left( 1-\alpha^2 \right),
\ee
\noindent
and therefore contribute equally to the large--order behavior. The one--loop contribution $\mu_1$ can be easily obtained from the general formula we derived before, but it has an extra factor of $2$ in order to account for the two instantons. It reads,
\be
\mu_1 = - \frac{1}{3^{\frac{3}{4}} \sqrt{\pi}}\, \frac{1-\alpha^2}{\left( 2-\alpha^2 \right)^{\frac{5}{4}} \left( 2+\alpha^2 \right)^{\frac{1}{4}}}.
\ee
\noindent
After some tedious but straightforward analysis, one likewise obtains for the two--loop contribution $\mu_2$
\be
\mu_2 = \frac{1}{4\sqrt{3}}\, \frac{1}{\left( 2-\alpha^2 \right)^{\frac{5}{2}} \left( 2+\alpha^2 \right)^{\frac{3}{2}}}\, \Big( 40-12\alpha^2-21\alpha^4-10\alpha^6 \Big).
\ee
\noindent
Using these formulae, we see that $b=-5/2$ in (\ref{curlfa}), and the asymptotics of $F_g(\lambda)$ is then given by
\be
\label{fgmmas}
F_g(\lambda) \sim \frac{\mu_1}{\pi}\ A^{-2g+5/2}\ \Gamma\left( 2g - {5\over2} \right) \left[ 1 + \frac{\mu_2 A}{2g} + \CO \left( \frac{1}{g^2} \right) \right]. 
\ee

The goal is now to compare this ``theoretical'' large--order prediction with the actual, ``experimental'' behavior of the $1/N$ expansion, using the results we have obtained for the free energies $F_g$ up to $g=10$, in the quartic matrix model. We first focus on the range of values of $\lambda$ where the instanton action, as well as the $F_g$, are real. This is precisely the interval between $\lambda=0$ and the critical point $\lambda=-{1\over 48}$ (we will come back to this critical point in the next subsection). In \figref{fig:qainst}--\figref{fig:qc1} we have displayed the asymptotic values of the instanton action as well as the one and two--loop results for the quartic potential. This is done at specific values of the coupling. The graphs include results extracted from the original sequence $F_g$ (the uppermost sequence of data, colored in red), and its Richardson transforms (colored in orange, blue and green), alongside with the prediction from instanton calculus. In \figref{fig:qcw} we have plotted the asymptotic values of $\mu_1$ and $\mu_2$, obtained as a function of $\lambda$ from the third Richardson transform, divided by the corresponding prediction from instanton calculus. It is rather clear that this quotient is very close to $1$, with a small error of roughly $0.1\%$ over most of moduli space. The larger error found at $\lambda\approx 0$ is due to numerical difficulties related to the divergence of the instanton action in this region. Indeed, at very small $\lambda$, the Richardson transformations converge too slowly to fall on a horizontal line at low genus---in this case, we would need higher--genus data to obtain better agreement with the predictions. In any case, the complete set of displayed numerical results strongly supports our analytical predictions.
\FIGURE[ht]{\label{fig:qainst}
    \centering
    \epsfxsize=0.5\textwidth
    \leavevmode
    \mbox{\hspace{-1cm}\epsfbox{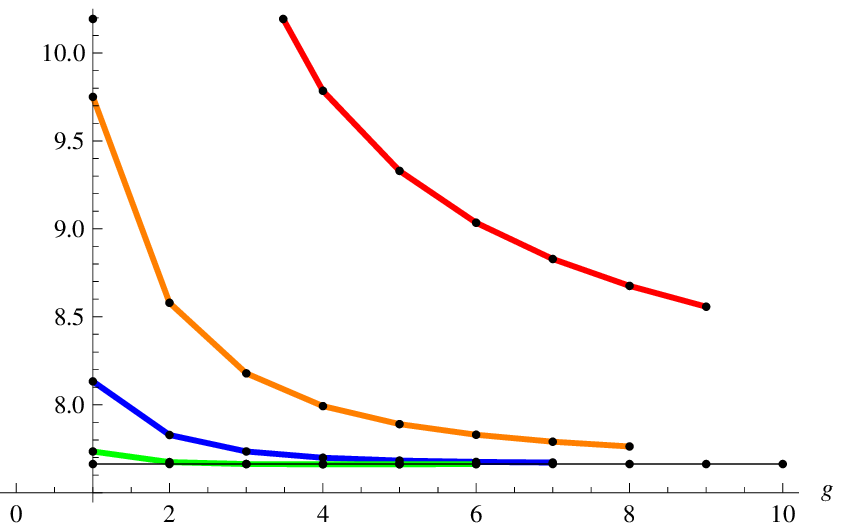}\quad
    \epsfbox{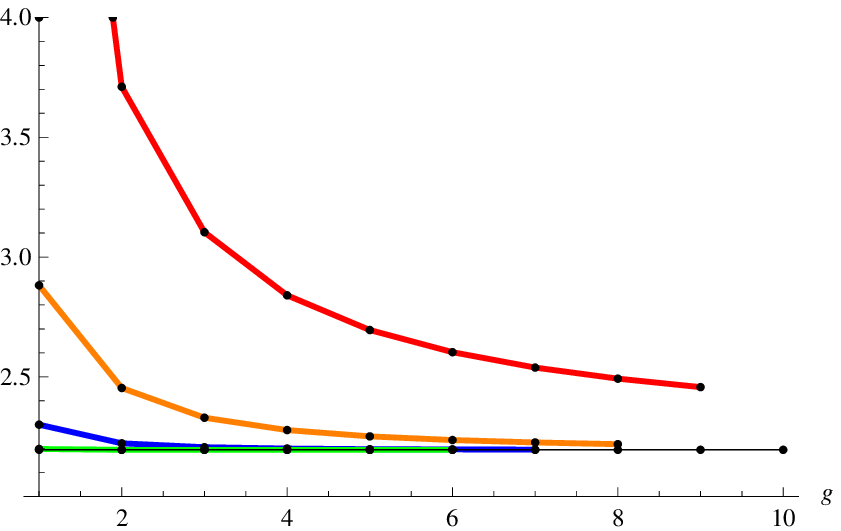}\hspace{-1cm}}
\caption{The sequence $\sqrt{1/Q_g}$ with $Q_g$ as defined in \eqref{richainst} and the corresponding Richardson transforms for the quartic matrix model, at fixed values $\lambda=-0.005$ (left) and $\lambda=-0.01$ (right). The prediction for the leading asymptotics is given by the instanton action $A(\lambda)$, shown as a straight line. The error for $g=10$ is $0.01\%$ at $\lambda=-0.005$, respectively $0.0047\%$ at $\lambda=-0.01$.}
}
\FIGURE[ht]{\label{fig:qc0}
    \centering
    \epsfxsize=0.5\textwidth
    \leavevmode
    \mbox{\hspace{-1cm}\epsfbox{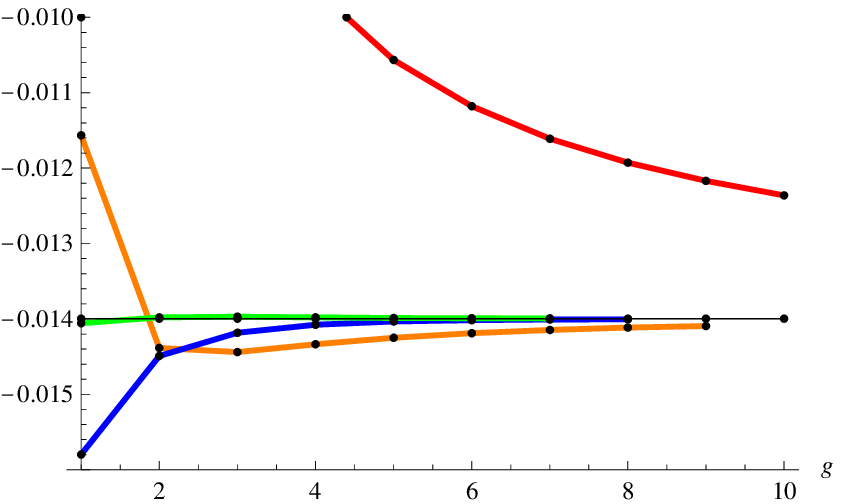}\quad
    \epsfbox{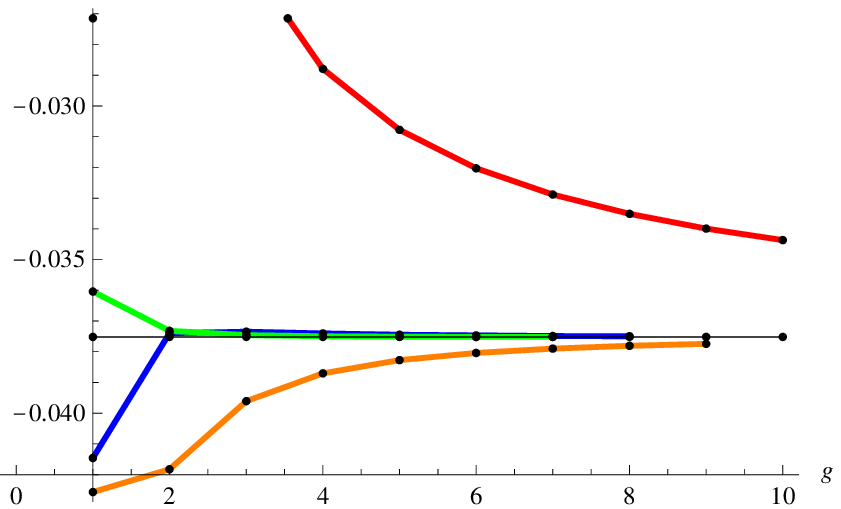}\hspace{-1cm}}
\caption{The sequence ${\pi F_g A^{2g-{5\over 2}}/\Gamma(2g-{5\over 2})}$ and its Richardson transforms for the quartic matrix model, at fixed values $\lambda=-0.005$ (left) and $\lambda=-0.01$ (right). The prediction for the asymptotic value is the one--loop result $\mu_1$ (straight line). The error is $0.003\%$ at $\lambda=-0.005$, respectively $0.002\%$ at $\lambda=-0.01$.}
}
\FIGURE[ht]{\label{fig:qc1}
    \centering
    \epsfxsize=0.5\textwidth
    \leavevmode
    \mbox{\hspace{-.5cm}\epsfbox{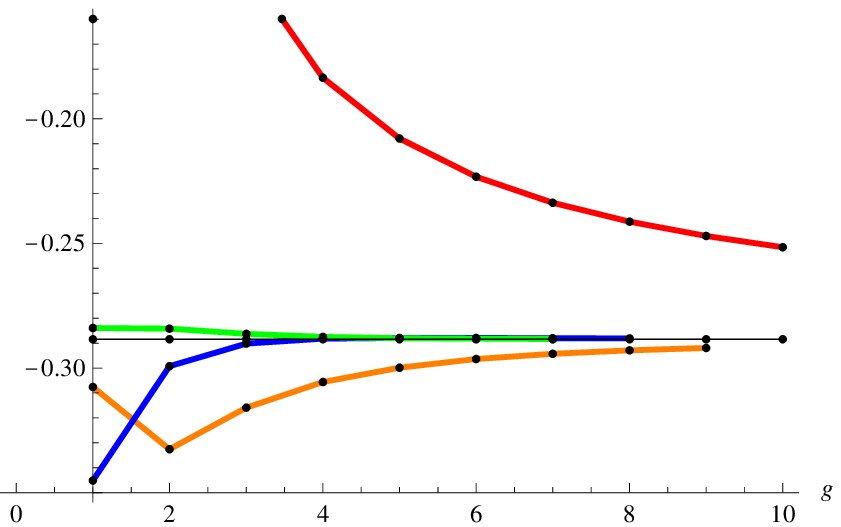}\quad
    \epsfbox{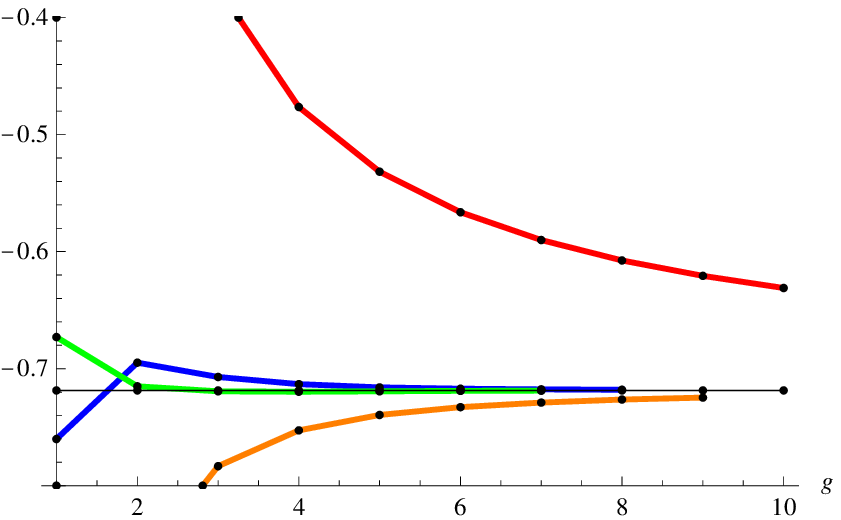}\hspace{-.5cm}}
\caption{The sequence \eqref{richc1} and its Richardson transforms for the quartic matrix model, at $\lambda=-0.005$ (left) and $\lambda=-0.01$ (right). The prediction for the leading asymptotics is given by the two--loop result $\mu_2$. The error is $0.05\%$ at $\lambda=-0.005$, respectively $0.016\%$ at $\lambda=-0.01$.}
}
\FIGURE[ht]{\label{fig:qcw}
    \centering
    \epsfxsize=0.5\textwidth
    \leavevmode
    \mbox{\hspace{-.5cm}\epsfbox{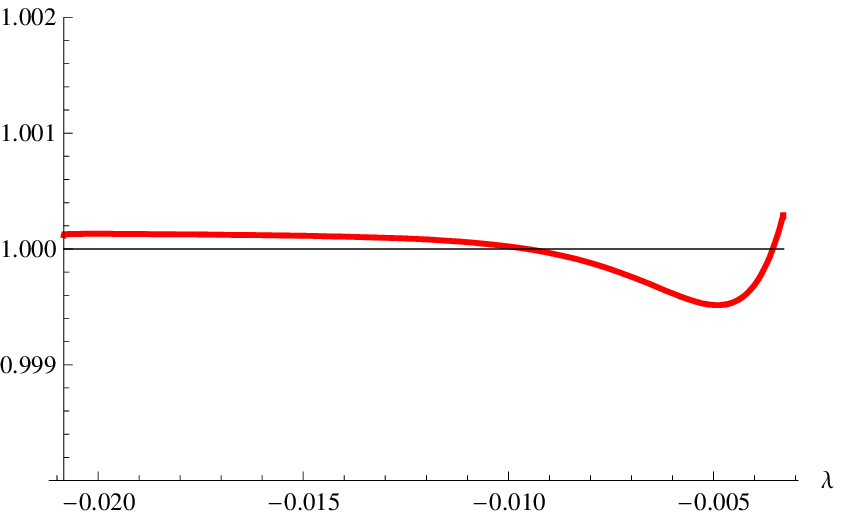}\quad
    \epsfbox{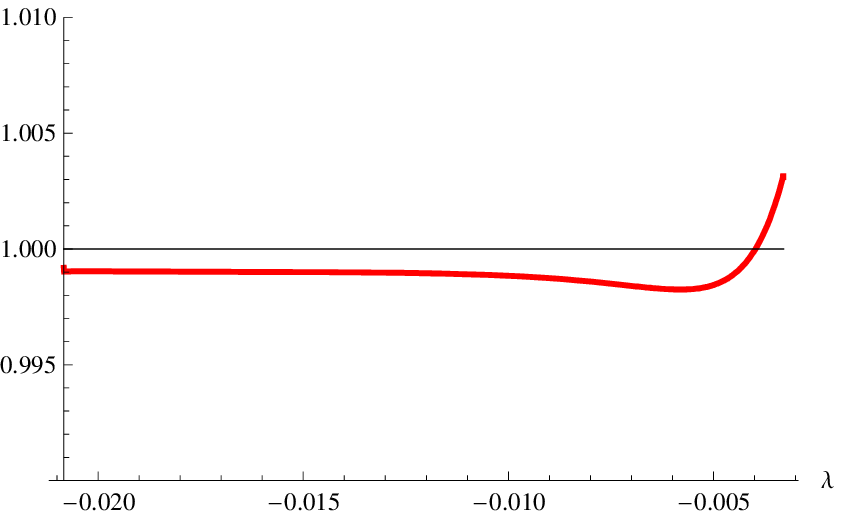}\hspace{-.5cm}}
\caption{The left figure shows the asymptotic value of ${\pi F_g A^{2g-{5\over 2}}/\Gamma(2g-{5\over 2})}$ for the quartic matrix model, as extracted as a function of $\lambda$ by the third Richardson transform, divided by the analytic prediction $\mu_1$. The figure on the right shows the analogous quotient for $\mu_2$. For $\lambda<-0.004$ the error is always less than $0.06\%$.}
}

Let us now consider the range of moduli space where $\lambda>0$. In this region the amplitudes $F_g$ are still real, as are the endpoints of the cut $\pm 2\alpha$. However, the saddle--points $x_0$ given in \eqref{qsaddle} now become purely imaginary and conjugate to each other. This implies that there are now \textit{four} instanton solutions, corresponding to eigenvalues tunneling from both endpoints of the cut to both of the saddle--points, as depicted in \figref{fig:q4inst}. The corresponding instanton actions are complex conjugate by a constant shift of $\pm \ri\pi$. We therefore expect the leading asymptotics to be of the form \eqref{acos}, implying that
\be\label{quotcos}
{\pi F_g|A|^{2g-{5\over 2}}\over |\mu_1|\Gamma(2g-{5\over 2})} = 2\cos\left((2g-{5\over 2})\theta_A+\theta_{\mu_1}\right)\left(1+\CO\left({1\over g}\right)\right),
\ee
\noindent
where $\theta_A$ and $\theta_{\mu_1}$ have been defined in \eqref{thetas}. This is indeed the case, as one can see from \figref{fig:qpos} showing the quotient in the left hand side of (\ref{quotcos}) together with the prediction for $2\cos\left((2g-{5\over 2})\theta_A+\theta_{\mu_1}\right)$, at two positive values $\lambda=0.004$ and $\lambda=3$.
\FIGURE[ht]{\label{fig:q4inst}
    \psfrag{x0}{$x_0$}\psfrag{a}{$a$}\psfrag{b}{$b$}\psfrag{bx0}{$x_0^*$}
    \centering
    \epsfxsize=0.3\textwidth
    \leavevmode
    \hspace{3cm}\epsfbox{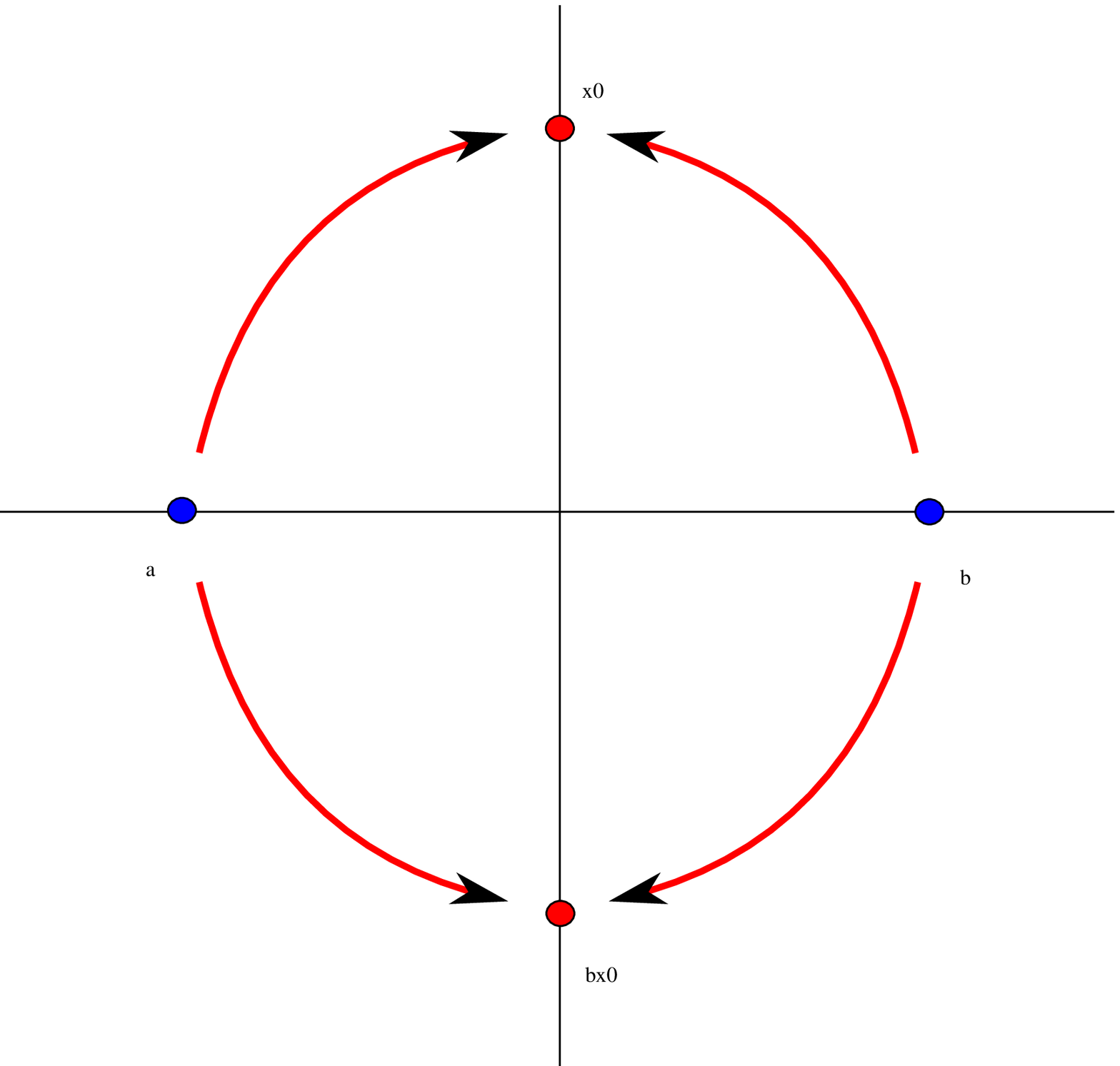}\hspace{3cm}
\caption{This figure shows the instanton effects for the quartic matrix model at \textit{positive} coupling $\lambda$. The endpoints of the cut, $a$ and $b$, are real while the two saddle--points of the effective potential are purely imaginary and complex conjugate to each other. There are two pairs of complex conjugate instantons, corresponding to eigenvalues tunneling from either end of the cut to the saddles $x_0$ and $x_0^*$.}
}
\FIGURE[ht]{\label{fig:qpos}
    \centering
    \epsfxsize=0.5\textwidth
    \leavevmode
    \mbox{\hspace{-.5cm}\epsfbox{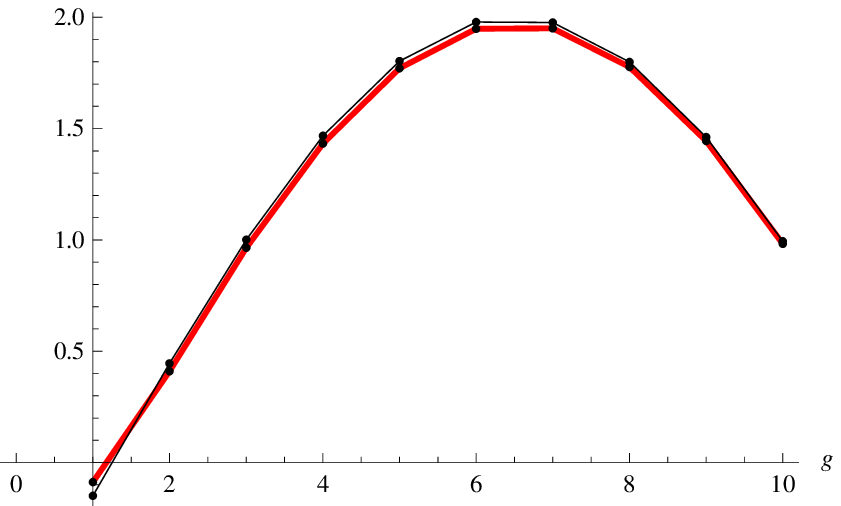}\quad
    \epsfbox{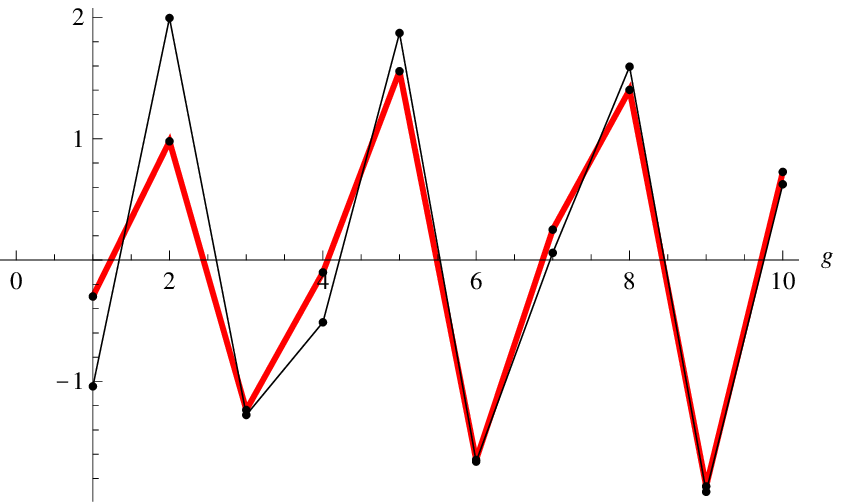}\hspace{-.5cm}}
\caption{The sequence $\pi F_g|A|^{2g-{5\over 2}}/\left(|\mu_1|\Gamma(2g-{5\over 2})\right)$ for the quartic matrix model, together with the prediction for the leading asymptotics $2\cos\left((2g-{5\over 2})\theta_A(\lambda)+\theta_{\mu_1}(\lambda)\right)$ (thin black line), at $\lambda=0.004$ (left), respectively $\lambda=3$ (right). At the highest depicted values of $g$ the error is of the order of $2\%$ ($\lambda=0.004$), respectively $5\%$ ($\lambda=3$).}
}

\subsection{2d Gravity and the Painlev\'e I Equation}

A rather well--known result (see \cite{dfgz} for an excellent review) is that the quartic matrix model has a critical point at
\be
\lambda_c = - \frac{1}{48}.
\ee
\noindent
At this critical value of $\lambda$, the saddles $\pm x_0$ collide with the two endpoints of the cut $\pm 2 \alpha$. One may further use the matrix model near this point in order to define two--dimensional gravity by means of a double--scaling limit. In this specific limit, one takes
\be
\lambda \rightarrow \lambda_c, \qquad g_s \rightarrow 0,
\ee
\noindent
in such a way that the variable 
\be
z = - \frac{1}{\lambda_c} \left( \lambda-\lambda_c \right) g_s^{-4/5}
\ee
\noindent
is kept fixed. In this limit it follows that the total, perturbative free energy of the matrix model becomes the free energy of pure 2d gravity
\be
F(g_s, \lambda) \rightarrow F_{\rm ds}(z).
\ee
\noindent
Furthermore, in this limit, the pre--string equation of the quartic matrix model (\ref{prestring}) precisely becomes the Painlev\'e I equation
\be\label{qpone}
u^2 - {1\over 3} u'' = z,
\ee
\noindent
governing the specific heat of the model
\be\label{sheat}
u(z) = - F''_{\rm ds}(z).
\ee
\noindent
These results may be used to obtain the perturbative expansion of $F_{\rm ds}(z)$, at any given order. It turns out that the free energy obtained in this way is actually doubled, since it gets contributions from the two collisions at $\pm x_0$. This is of course due to the symmetry of the potential, which we have discussed before. In order to remove the doubling it is enough to change the normalization of the quantities appearing above, by
\be\label{change}
z \rightarrow 2^{2\over 5} z, \qquad u \rightarrow 2^{{1\over 5}} u, \qquad F_{\rm ds} \rightarrow 2F_{\rm ds}.
\ee
\noindent
Proceeding in this way one is led to the Painlev\'e I equation with the normalization 
\be\label{pone}
u^2 - {1\over 6} u'' = z,
\ee
\noindent
while the double--scaled free energy still satisfies (\ref{sheat}). The perturbative expansion of the specific heat has the form
\be
u(z) = z^{1\over 2} \sum_{g=0}^{+\infty} u_g\, z^{-5g/2},
\ee
\noindent
so that the Painlev\'e I equation becomes equivalent to the following difference equation for the coefficients $u_g$
\be\label{diffeq}
u_{g} = \frac{25(g-1)^2-1}{48} u_{g-1} - {1\over 2} \sum_{\ell=1}^{g-1} u_\ell u_{g-\ell}, \qquad u_0 = 1.
\ee
\noindent
The coefficients $a_g$, which appear in the perturbative expansion of the double--scaled free energy as
\be\label{fds}
F_{\rm ds}(z)=-{4\over 15}z^{5/2}-{1\over 48} \log\, z +\sum_{g\ge 2} a_g\, z^{-5(g-1)/2},
\ee
can then be obtained from $u_g$ through the simple relation
\be
a_g=-{4\over (5g-5)(5g-3)}u_g.
\ee
\noindent
As a result one finds, for example,
\be\label{cubicds}
F_{\rm ds}(z) = - {4\over 15} z^{5\over 2} - {1\over 48} \log z + {7\over 5760} z^{-{5\over 2}} + {245\over 331776} z^{-5} + \cdots.
\ee
\noindent
We are now in a position where we may obtain a prediction for the asymptotics of the coefficients of this series, $a_g$, by simply evaluating the expressions we obtained for the quartic matrix model near the critical point, and taking into account the change of normalization in (\ref{change}). In this way we find
\be
\label{2dinst}
{A \over g_s}= \frac{8\sqrt{3}}{5}\, z^{\frac{5}{4}}.
\ee
\noindent
Moreover, for $\mu_1$ we obtain 
\be
\sqrt{g_s}\, \mu_1 = \frac{1}{8 \cdot 3^{\frac{3}{4}} \sqrt{\pi}}\, z^{-\frac{5}{8}},
\ee
\noindent
while at two loops we get the result
\be
g_s\, \mu_2 = - \frac{37}{64 \sqrt{3} }\, z^{-\frac{5}{4}}.
\ee
\noindent
Altogether, this means that the one--instanton contribution to the double--scaled free energy, up to two--loop order, is
\be\label{dsoneins}
F^{(1)}_{\rm ds} = \frac{\rmi}{8 \cdot 3^{\frac{3}{4}} \sqrt{\pi} }\, z^{-\frac{5}{8}}\, \exp \left( - \frac{8\sqrt{3}}{5 }\, z^{\frac{5}{4}} \right) 
\left\{ 1 - \frac{37}{64 \sqrt{3}}\, z^{-\frac{5}{4}} + \cdots \right\}.
\ee
\noindent
The result for the one--loop coefficient, $\mu_1$, was first obtained by David in \cite{davidtwo} and later re--derived in \cite{lvm}. We have obtained $\mu_2$ directly from an instanton computation in the matrix model, but we may also verify our result by computing the one--loop instanton expansion directly from the Painlev\'e I equation. This expansion has been studied in detail in \cite{jk}, where it has been used to analyze the asymptotics of the perturbative answer. The calculation of this expansion goes as follows. As noticed in \cite{eyzj,dfgz}, the discontinuity of the double--scaled free energy (\ref{dsoneins}) can be computed by linearizing the string equation (\ref{pone}) around the perturbative, asymptotic solution. If we denote
\be
\epsilon(z) ={\rm disc}\, u(z), \qquad F^{(1)}_{\rm ds}=-\epsilon''(z), 
\ee
one finds the {\it linear} and {\it homogeneous} differential equation 
\be
\label{epsde}
\epsilon''(z) - 12 u_0(z) \epsilon (z) =0, 
\ee
where
\be
u_0(z)=z^{1\over 2} \biggl( 1-{1\over 48} z^{-{5 \over 2}} -{49 \over 4608} z^{-5} -{1225 \over 55 296} z^{-{15\over 2}}+\cdots\biggr). 
\ee
It is easy to solve (\ref{epsde}) at $z\rightarrow \infty$ after ``peeling off" the exponential piece, 
\be
\epsilon(z) =  c \, z^{-\frac{1}{8}}\, \exp \left( - \frac{8\sqrt{3}}{5 }\, z^{\frac{5}{4}} \right) \biggl( 1+ \sum_{k=1}^{\infty} 
\epsilon_k z^{-{5k \over 4}}\biggr). 
\ee
The overall coefficient $c$ cannot be deduced from the differential equation (\ref{epsde}) due to its homogeneity, but the $\epsilon_k$ can be easily found in terms of the coefficients of the asymptotic expansion of $u_0$. One finds, for the very first terms, 
\be
\epsilon (z) =c \, z^{-\frac{1}{8}}\, \exp \left( - \frac{8\sqrt{3}}{5 }\, z^{\frac{5}{4}} \right) \biggl( 1 - {5 \over 64 {\sqrt {3}}} z^{-{5\over 4}} 
+ {75 \over 8192} z^{-{5\over 2}} - {341329 \over 23592960 {\sqrt{3}}} z^{-{15\over 4}} +\cdots\biggr). 
\ee
Of course, the coefficient $c$ may still be fixed with the explicit result for the one--loop coefficient $\mu_1$. Assembling all together, one finds the {\it full} perturbative expansion of the free energy around the one--instanton configuration, 
\be
F^{(1)}_{\rm ds} = \frac{1}{8 \cdot 3^{\frac{3}{4}} \sqrt{\pi} }\, z^{-\frac{5}{8}}\, \exp \left( - \frac{8\sqrt{3}}{5 }\, z^{\frac{5}{4}} \right) 
\left\{ 1 - \frac{37}{64 \sqrt{3}}\, z^{-\frac{5}{4}} + {6433 \over 24576} z^{-{5\over 2}} -{12741169 \over 23592960 {\sqrt{3}}} z^{-{15\over 4}} +\cdots \right\}.
\ee

We can now use (\ref{dsoneins}), together with (\ref{lostring}), in order to obtain a prediction concerning the large--order behavior of the perturbative coefficients of the double--scaled free energy, as
\be\label{pred}
a_g \sim {16 {\sqrt {30}} \over 125 \, \pi^{3\over 2}} \left( {25\over 192} \right)^g \Gamma\biggl(2g-{5\over 2}\biggr) \left[ 1 - \frac{37}{80 g} -\frac{3927}{12800 g^2} - \frac{3618769}{15360000 g^3} + \cdots \right]. 
\ee
\noindent
Since one can compute these coefficients up to very large order, by using the Painlev\'e I equation, we can now perform truly precise tests of some of our proposals. In \figref{fig:painleve} we show numerical checks for both the one and two--loop predictions, up to genus 400. Indeed both leading and subleading asymptotics of the coefficients $a_g$ clearly agree, to a very high degree of precision, with our prediction \eqref{pred}. This also leads us to an important point. It is sometimes stated in the literature, \textit{e.g.}, \cite{lvm, iy}, that the one--instanton amplitude (\ref{dsoneins}) cannot be deduced from the Painlev\'e I equation. The reason for this assertion is simply that the linearized equation (\ref{epsde}) for $\epsilon$ does not allow the calculation of $\mu_1$. But it is clear, in view of the connection between large--order behavior and instanton effects, that there is a more subtle relation between the one--instanton amplitude and the perturbative result. In fact, one could instead have derived this amplitude from the asymptotics of the coefficients $a_g$, themselves derived from Painlev\'e I. It is easy to see that from the difference equation (\ref{diffeq}) one may obtain
\be
\label{leadas}
a_g \sim  \biggl( {25\over 192}\biggr)^g (2g)!,
\ee
\noindent
a result which at leading order precisely agrees with (\ref{pred}). A careful study of the difference equation (\ref{diffeq}) beyond 
(\ref{leadas}) \cite{jk} confirms indeed the result (\ref{pred}) for the asymptotics of $a_g$, and in particular makes possible to extract the one--instanton amplitude directly from large order.
\FIGURE[ht]{\label{fig:painleve}
    \centering
    \epsfxsize=0.5\textwidth
    \leavevmode
    \mbox{\hspace{-.5cm}\epsfbox{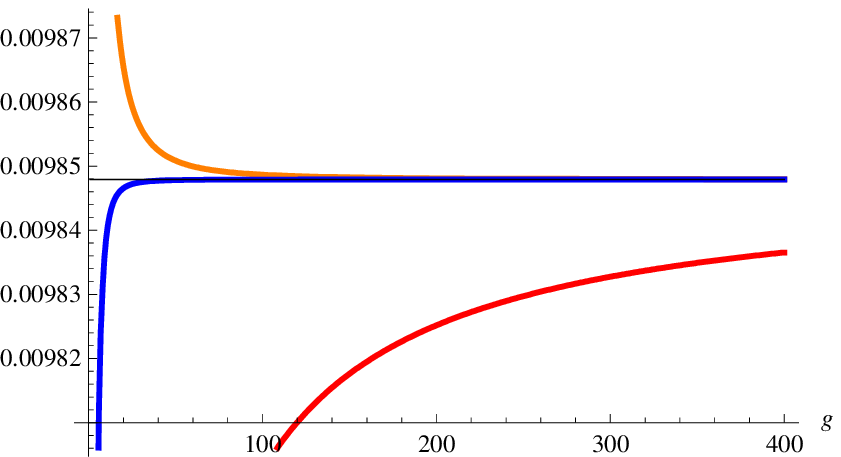}\quad
    \epsfbox{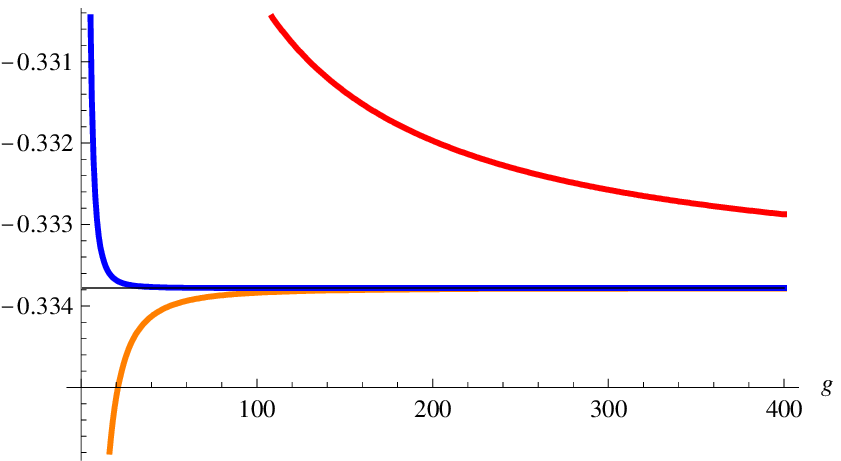}\hspace{-.5cm}}
\caption{The left figure shows the sequence $a_g A^{(2g-5/2)}/\Gamma(2g-5/2)$ for 2d gravity alongside with its Richardson transforms, up to $g=400$, 
clearly converging to the one--loop prediction $1/(8\cdot 3^{3/4}\pi^{3/2}) =0.009847$. The right figure shows the modified sequence \eqref{richc1} and its Richardson transforms for $a_g$, converging towards the two--loop result $-37/(64\sqrt{3})=-0.33378$, again up to $g=400$. The error at this genus is of order $10^{-9}\%$.}
}

\sectiono{Application II: Topological Strings on Local Curves}

We shall now proceed into the realm of topological string theory, beginning with the case of topological strings on local curves.

\subsection{Topological Strings and Matrix Models}

Topological strings are defined as twisted $\CN=2$ sigma--models coupled to 2d gravity, and they provide a vast generalization of noncritical/minimal strings (where one considers conformal matter with $c\le 1$ coupled to 2d gravity). As it is well known, topological strings come in two types, called the A--model and the B--model, which are related by mirror symmetry. The A--model provides a physical formulation of Gromov--Witten theory, while the B--model is deeply related to the theory of deformation of complex structures. In the last years it has become apparent that topological strings on CY threefolds share many of the remarkable properties of noncritical string theories, like integrability, but their geometrical structures are much richer.

It is natural to ask if topological strings on CY threefolds have a matrix model description, at least in some cases. In \cite{dv}, Dijkgraaf and Vafa showed that the B--model, on certain CY geometries, is indeed equivalent to a matrix model (see \cite{mmhouches} for a review of this development). This class of target geometries is of the form
\be\label{scts}
uv=H(X,Y),
\ee
\noindent
where $H(X,Y)$ is a polynomial in $X, Y \in \BC$, and $u, v \in \BC$. The nontrivial information about this geometry turns out to be encoded in the Riemann surface $\Sigma$ described by $H(X,Y)=0$. An important insight of the analysis of Dijkgraaf and Vafa is that the spectral curve of the matrix model is precisely the Riemann surface $\Sigma$, providing in this way a beautiful example in which the master field of the $1/N$ expansion generates the target geometry of a string theory.

More recently, the correspondence between matrix models and topological strings was extended to toric CY threefolds \cite{mm,bkmp}. This class of examples is very interesting since (in contrast to the geometries considered in \cite{dv}) they have mirror geometries. These geometries are CY threefolds described by an equation of the form (\ref{scts}), but where the variables $X,Y$ now belong to $\BC^*$. The proposal of \cite{mm,bkmp} is to regard these geometries as spectral curves of a matrix model. The open and closed string amplitudes of the B--model are then computed by the $1/N$ expansion associated to the spectral curve. Notice that one does not need to specify the matrix integral in order to compute these amplitudes; using the results of \cite{ackm,eo} it is enough to specify the spectral curve in order to compute the $1/N$ expansion. 

In the previous section we have computed one--instanton effects in one--cut matrix models in terms of data associated to the spectral curve. We can then use the correspondence of \cite{mm,bkmp} to apply our results to topological string theories described by this class of matrix models. The restriction to the one--cut case still leaves a rather general class of CY backgrounds to explore, the so--called local curves. A special limit of the theory of local curves gives the theory of simple Hurwitz numbers studied for example in \cite{ksw}, which will be addressed in the next section.

\subsection{Topological Strings on Local Curves}

Local curves are toric CY manifolds of the form
\be
X_p=\CO(p-2) \oplus \CO(-p) \rightarrow \BP^1, \quad p \in \BZ.
\ee
\noindent
Topological string theory on $X_p$ has received a lot of recent attention (see, \textit{e.g.}, \cite{aosv} and references therein). As explained in \cite{bp}, the A--model on $X_p$ has to be defined equivariantly, and the most natural choice (the equivariant CY case) corresponds to the antidiagonal action on the bundle (we refer the reader to \cite{bp} for further details). Of more interest to us in the present work is that the free energies at genus $g$ on this geometry, $F^{X_p}_g(t)$, depend on a single complexified K\"ahler parameter $t$, associated to the complexified area of $\BP^1$. They can be computed in both the A and B--models.

In the A--model, the total partition function is given by
\be\label{totalfgen}
Z_{X_p} = \exp \left( F_{X_p}(g_s,t) \right), \qquad F_{X_p}(g_s,t) = \sum_{g=0}^{+\infty} g_s^{2g-2} F^{X_p}_g(t),
\ee
\noindent
and near $t\rightarrow \infty$, $F^{X_p}_g(t)$ has the expansion 
\be
F^{X_p}_g(t)=\sum_{d=1}^{\infty} N_{g,d}\, \re^{-dt}, 
\ee
where $N_{g,d}$ are the Gromov--Witten invariants of the CY manifold $X_p$ at genus $g$ and degree $d$. 

The total partition function $Z_{X_p}$ can be computed as a sum over partitions, by making use of the topological vertex formalism as described in \cite{topvertex}. In order to write the explicit resulting formula, we first have to introduce some notation. To begin with, define the q--number $[n]$ as
\be\label{qnumb}
[n]=q^{n/2} -q^{-{n/2}}, \qquad q=\re^{g_s}.
\ee
\noindent
A representation, $R$, of $U(\infty)$ is encoded by a Young tableau, labeled by the lengths of its rows $\{l_i\}$. The quantity
\be
\ell(R) =\sum_i l_i
\ee
\noindent
is the total number of boxes in the tableau. Another important quantity associated to a given tableau is
\be\label{kappar}
\kappa_R=\sum_i l_i(l_i-2i+1).
\ee
\noindent
We finally introduce the quantity
\be
\label{wr}
W_R =q^{-\kappa_R/4} \prod_{\tableau{1} \in R} {1\over [{\rm hook}(\tableau{1})]},
\ee
\noindent
with ${\rm hook}(\tableau{1})$ the hook--length. With all this notation at hand, we may finally write the explicit expression for the topological string partition function on $X_p$, which is given by
\be\label{totalz}
Z_{X_p} = \sum_R W_R W_{R^t} q^{(p-1) \kappa_R/2}Q^{\ell(R)}, \quad Q=(-1)^p \re^{-t},
\ee
\noindent
where $R^t$ denotes the transposed Young tableau (\textit{i.e.}, the tableau where we have exchanged the rows with the columns). 

Although (\ref{totalz}) gives an all--genus expression, it is effectively an expansion in powers of $Q$. In order to obtain an expression for each $F_g^{X_p}(t)$ to all orders in $Q$, one usually appeals to mirror symmetry and the B--model. However, standard techniques of mirror symmetry do not work well when applied to local curves. The Riemann surface encoding the mirror geometry for local curves was proposed in \cite{mm} based on the direct analysis of the sum over partitions presented in \cite{cgmps}, and later on some aspects of this mirror construction where confirmed from a more mathematical point of view \cite{fj}. The B--model geometry is encoded in the spectral curve
\be\label{scapart}
y(\lambda) = {2 \over \lambda} \biggl( \tanh^{-1} \biggl[ {{\sqrt{(\lambda-a)(\lambda-b)}}\over \lambda - {a+b\over 2}} \biggr] - p \tanh^{-1} \biggl[ {{\sqrt{(\lambda-a)(\lambda-b)}}\over \lambda + {\sqrt{ab}}} \biggr] \biggr), 
\ee
\noindent
which has genus zero. Although this curve is not algebraic, it is easy to see that when written in terms of the variables $X=\lambda, Y=\re^y$ one obtains an algebraic equation for the $\BC^*$ variables $X,Y$, which leads to the mirror CY threefold (\ref{scts}). The advantage of writing the curve in the nonalgebraic form (\ref{scapart}) is that, as explained in \cite{mm,bkmp}, one can apply {\it verbatim} the standard matrix model technology that we use in this paper. The curve (\ref{scapart}) may also be written in the form (\ref{scurve}), with a moment function $M(\lambda)$ which has various nontrivial zeroes where the spectral curve is singular. The endpoints of the cut, $a$ and $b$, are given by
\be\label{endpoints}
a = (1-\zeta)^{-p} (1-\zeta^{1\over 2})^2, \quad b = (1-\zeta)^{-p} (1+\zeta^{1\over 2})^2,
\ee
\noindent
where $\zeta$ is related to $Q$ by the mirror map \cite{cgmps,fj}
\be\label{mirror}
Q = (1-\zeta)^{-p(p-2)} \zeta.
\ee
\noindent
It was further conjectured in \cite{mm} that the free energies $F^{X_p}_g(t)$ can be obtained as the standard genus $g$ free energies of a matrix model with spectral curve (\ref{scapart}). And it was conjectured in \cite{cgmps} that, for $g \ge 2$, these free energies may be written as
\be\label{qansatz}
F^{X_p}_g(t) = {\CP_g (\zeta,p) \over (\zeta-\zeta_c)^{5(g-1)}}, \quad \CP_g(\zeta,p) = \sum_{i=1}^{5(g-1)} a_{g,i}(p)\, \zeta^i,
\ee
\noindent
where
\be\label{critical}
\zeta_c={1\over (p-1)^2}
\ee
is a critical point of the model. In fact, at this point, a zero $x_0$ of $M(\lambda)$ collides with the endpoint of the cut $b$, and we are left with a critical theory in the universality class of pure 2d gravity \cite{cgmps}. If one further takes the double--scaling limit,
\be
\zeta \rightarrow \zeta_c, \quad g_s \rightarrow 0, \quad z\, \, {\rm fixed},
\ee
\noindent
where 
\be\label{scaled}
z^{5/2} = g_s^{-2} {(p-1)^8 \over 4 (1-\zeta_c)^3}  (\zeta_c-\zeta)^{5},
\ee
\noindent
then the total free energy (\ref{totalfgen}) becomes the free energy of pure 2d gravity.

\subsection{Instanton Effects and Large--Order Behavior}

In \cite{mm} the matrix model description, based on the spectral curve (\ref{scapart}), was used to study nonperturbative effects in this topological string theory. The spectral curve (\ref{scapart}) has a nontrivial saddle $x_0$, which is the solution to
\be
M(x_0)=0.
\ee
\noindent
For the cases $p=3$ and $p=4$ the relevant solutions have been determined in \cite{mm}; they are given by
\be
x_0 = {4ab\over (\sqrt{a}-\sqrt{b})^2}, \qquad p=3,
\ee
\noindent
and
\be
x_0 = {2\sqrt{a}b \over \sqrt{a}-\sqrt{b}}, \qquad p=4.
\ee
\noindent
In \cite{mm} it was argued that this saddle controls the large--order behavior of $F^{X_p}_g(t)$, at any value of $t$. We shall now show that this is indeed the case, and that the one and two--loop results $\mu_{1,2}$ computed in terms of the spectral curve (\ref{scapart}) control the subleading large $g$ asymptotics. The instanton action for an eigenvalue tunneling from $b$ to $x_0$ has already been computed in \cite{mm}; it is given by the rather formidable expression
\be\label{instp}
A(Q) = F(x_0) - F(a),
\ee
\noindent
where
\be\label{bigf}
\ba
F(x) &= -\log \,(f_1(x)) \biggl( \log \, (f_1(x)) - 2\log \Bigl( 1 + {2 f_1(x) \over ({\sqrt{a}} - {\sqrt{b}})^2} \Bigr) + \log \Bigl( 1 + {2 f_1(x) \over ({\sqrt{a}} + {\sqrt{b}})^2} \Bigr) \biggr) - \\
&- 2 {\rm Li}_2 \Bigl( - {2 f_1(x) \over ({\sqrt{a}} - {\sqrt{b}})^2} \Bigr) -
2 {\rm Li}_2 \Bigl( - {2 f_1(x) \over ({\sqrt{a}} + {\sqrt{b}})^2} \Bigr) - \log {(a-b)^2 \over 4} \log \, x - \\
&- p \log \,(f_2(x)) \biggl( \log \, (f_2(x)) + 2\log \Bigl( 1 - {f_2(x) \over 2 {\sqrt{ab}}} \Bigr) - \log \Bigl( 1 - {2 f_2(x) \over ({\sqrt{a}} + {\sqrt{b}})^2} \Bigr) \biggr) - \\
&- 2 p {\rm Li}_2 \Bigl( - {f_2(x) \over 2 {\sqrt{ab}}} \Bigr) + 2 p
{\rm Li}_2 \Bigl( {2 f_2(x) \over ({\sqrt{a}} + {\sqrt{b}})^2} \Bigr) + {p \over 2} (\log x)^2 + p \log ({\sqrt{a}} + {\sqrt{b}})^2 \log\, x,
\ea
\ee
\noindent
and 
\be
\ba
f_1(x) &= {\sqrt{(x-a)(x-b)}} + x - {a+b\over 2}, \\
f_2(x) &= {\sqrt{(x-a)(x-b)}} + x + {\sqrt{ab}}.
\ea
\ee
\noindent
In these expressions $a$ and $b$ are the endpoints of the cut as usual, given in \eqref{endpoints}.

The one and two--loop coefficients are again given by the general expressions \eqref{rone} and \eqref{mutwo}, as in the previous section. We shall now compare the analytic results to the large--order behavior of the perturbation series for the case of the local curve $X_3$ ($p=3$). The leading and subleading asymptotic behavior of $F_g$ should be given by the same structure found for the quartic matrix model \eqref{fgmmas}. As explained in section \ref{sec:num}, we can independently test the predictions for the instanton action, as well as the one and the two--loop results, by  applying Richardson transformations to the modified sequences \eqref{richainst}--\eqref{richc1}. Notice that all these quantities depend on the B--model modulus $\zeta$. For simplicity, we shall restrict our analysis to the range $0<\zeta<\zeta_c={1\over 4}$, where the endpoints of the cut, as well as the instanton action, are real. \figref{fig:lcainst} shows the inverse square root of the sequence $Q_g$ in (\ref{richainst}) and its first three Richardson transforms, at two specific values of $\zeta$. The straight line is the prediction for the instanton action, $A$. As is evident from the plot, and even though we only use data up to genus $g=8$, the third Richardson transform already falls on the straight line. The mismatch between numerical extrapolation and the prediction is of order $0.02\%$. Analogously, we may check the one and two--loop results. In \figref{fig:lcc0} and \figref{fig:lcc1} we plot the modified sequences, \eqref{richc0} and \eqref{richc1}, together with the corresponding Richardson transforms, again at two fixed values of the K\"ahler modulus. As explained in section \ref{sec:num}, the predictions for their leading asymptotics are $\mu_1$, respectively $\mu_2$, which are shown in the figures as straight lines. Again, this is confirmed by the Richardson transforms, clearly converging to the prediction from instanton calculus. The error in here is of order $1\%$.

Similar graphs can be produced at any other point in moduli space. \figref{fig:lcainstw} shows the asymptotic value of the instanton action, as approximated by the third Richardson transform, divided by the corresponding analytical prediction, and plotted as a function of the modulus over $0<\zeta<1/4$. This quotient is indeed very close to one, as it should be from our discussion. Similarly, in \figref{fig:lccw} we plot the asymptotic results for $\mu_1$ and $\mu_2$, divided by the corresponding analytic predictions, as functions of $\zeta$. Notice that while the agreement is excellent over most of moduli space, as one approaches $\zeta \sim 0$ the deviation from the predicted value increases. This is again due to the divergence of the instanton action at this particular point of moduli space. Indeed in this region, the Richardson transforms converge too slowly to fall on one line, at low genus $g<10$. In order to obtain full agreement one would need higher--genus data, which is out of our scope in this paper.

We have performed similar checks of our predictions for the local curve $X_4$, also obtaining agreement to very high precision, and further strengthening our analytical results.
\FIGURE[ht]{\label{fig:lcainst}
    \centering
    \epsfxsize=0.5\textwidth
    \leavevmode
    \mbox{\hspace{-.5cm}\epsfbox{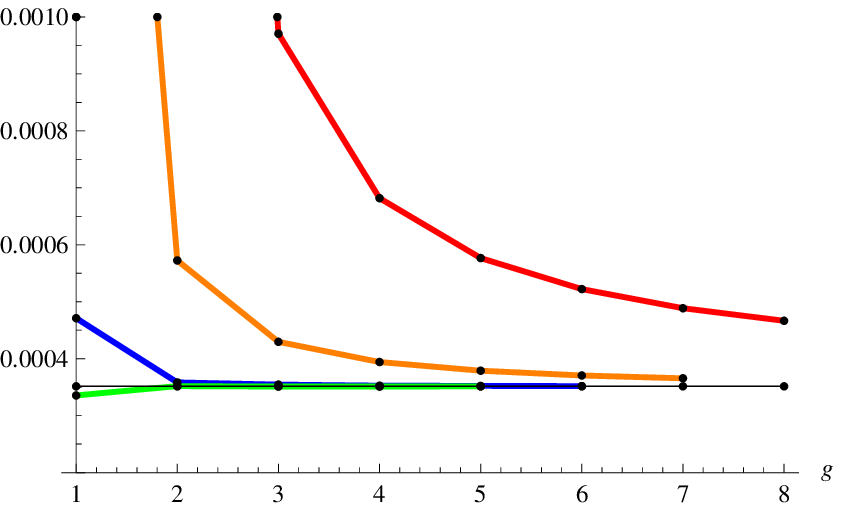}\quad
    \epsfbox{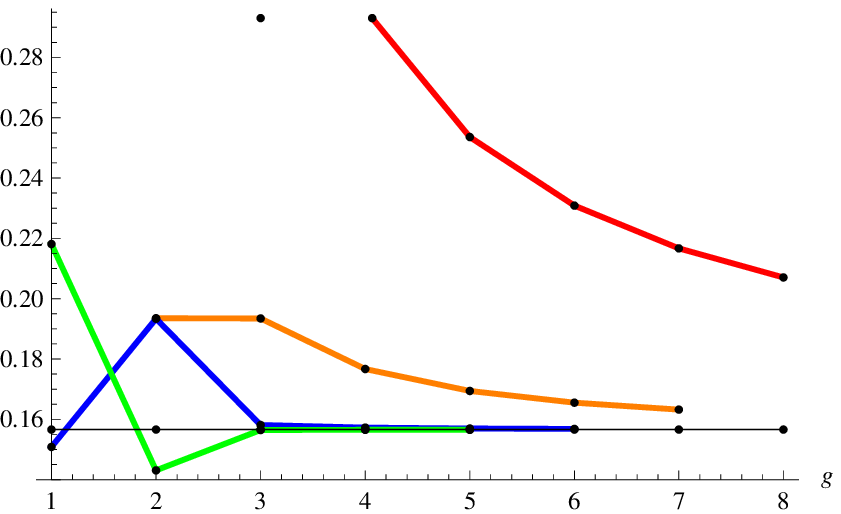}\hspace{-.5cm}}
\caption{The sequence $\sqrt{1/Q_g}$ and the corresponding Richardson transforms for the local curve $X_3$, at fixed values $\zeta=0.24$ (left) and $\zeta=0.15$ (right). The leading asymptotics are predicted to be given by the instanton action $A(\zeta)$, shown as a straight line. The error for the available degree $g=8$ is $0.014\%$ at $\zeta=0.24$, respectively $0.025\%$ at $\zeta=0.15$.}
}
\FIGURE[ht]{\label{fig:lcc0}
    \centering
    \epsfxsize=0.5\textwidth
    \leavevmode
    \mbox{\hspace{-.5cm}\epsfbox{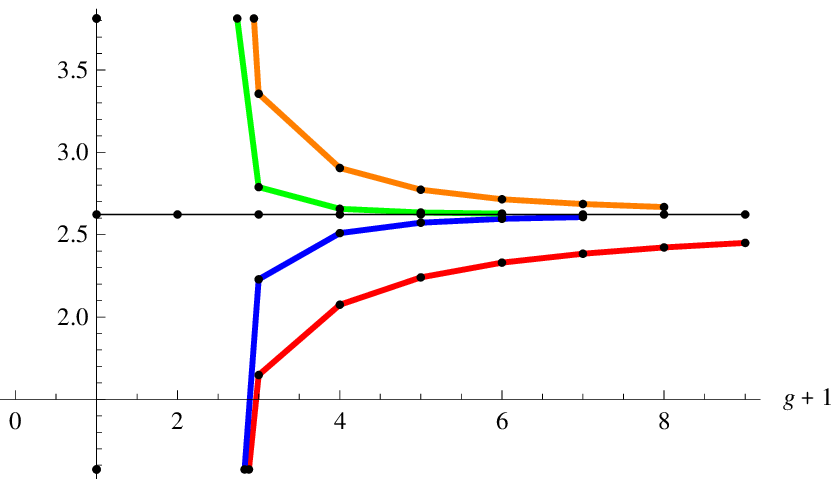}
    \epsfbox{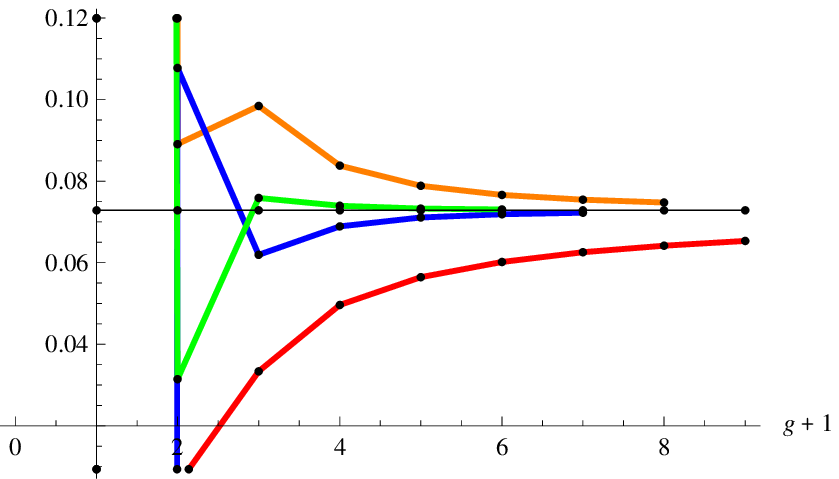}\hspace{-.5cm}}
\caption{The sequence ${\pi F_g A^{2g-{5\over 2}}/\Gamma(2g-{5\over 2})}$ and its Richardson transforms for the local curve $X_3$, at fixed values $\zeta=0.24$ (left) and $\zeta=0.15$ (right). The prediction for the asymptotic value is the one--loop result $\mu_1$, shown as a straight line. The error is $0.49\%$ at $\zeta=0.24$, respectively $0.58\%$ at $\zeta=0.15$.}
}
\FIGURE[ht]{\label{fig:lcc1}
    \centering
    \epsfxsize=0.5\textwidth
    \leavevmode
    \mbox{\hspace{-.5cm}\epsfbox{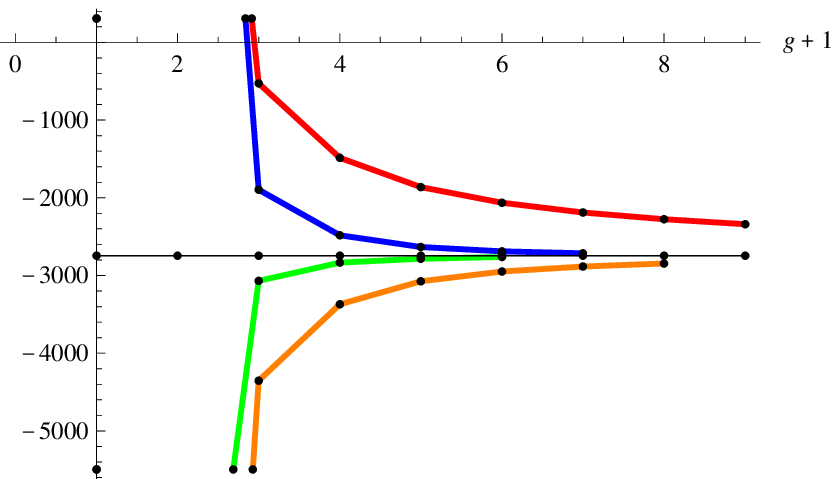}
    \epsfbox{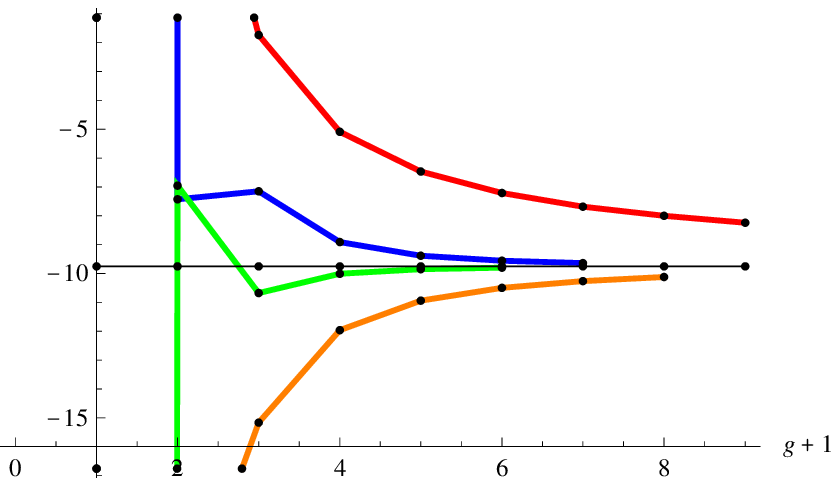}\hspace{-.5cm}}
\caption{The sequence \eqref{richc1} for the local curve $X_3$ and its Richardson transforms, at $\zeta=0.24$ (left) and $\zeta=0.15$ (right), with leading asymptotics predicted to be given by the two--loop result $\mu_2$. The error is $1.38\%$ at $\zeta=0.24$, respectively $1.04\%$ at $\zeta=0.15$.}
}
\FIGURE[ht]{\label{fig:lcainstw}
    \centering
    \epsfxsize=0.5\textwidth
    \leavevmode
    \hspace{3cm}\epsfbox{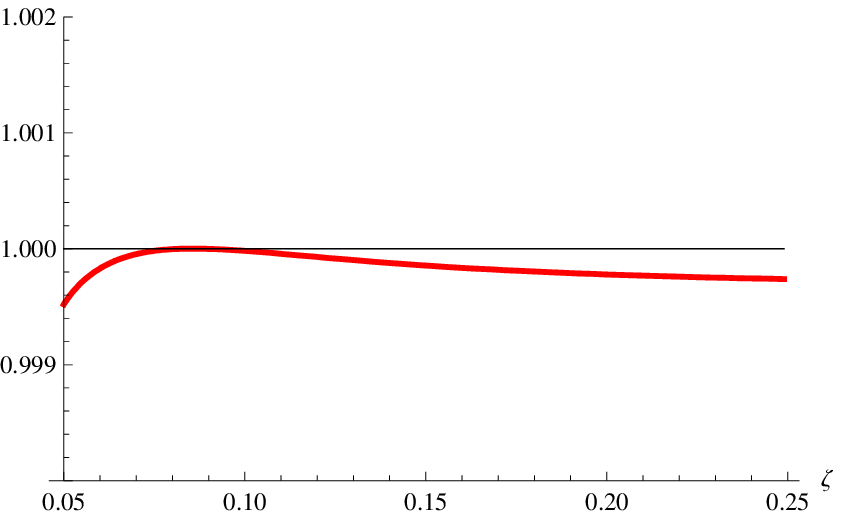}\hspace{3cm}
\caption{The asymptotic value of ${\sqrt{1/Q_g}}$ for the local curve $X_3$ as extracted from the third Richardson transform as a function of $\zeta$, divided by the analytic prediction for the instanton action. For $\zeta>0.05$, the error is always less than $0.03\%$.}}
\FIGURE[ht]{\label{fig:lccw}
    \centering
    \psfrag{g}{$g<0$}
    \psfrag{x}{$g>0$}
    \epsfxsize=0.5\textwidth
    \leavevmode
    \mbox{\hspace{-.5cm}\epsfbox{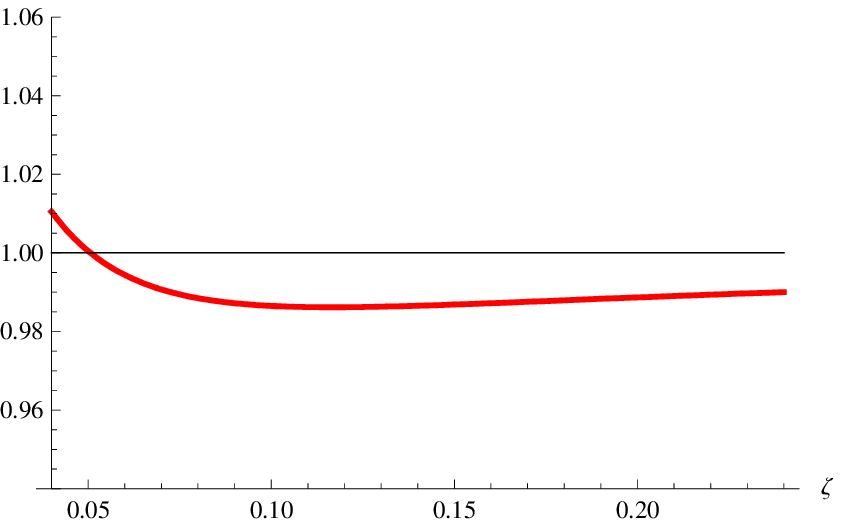}\quad
    \epsfbox{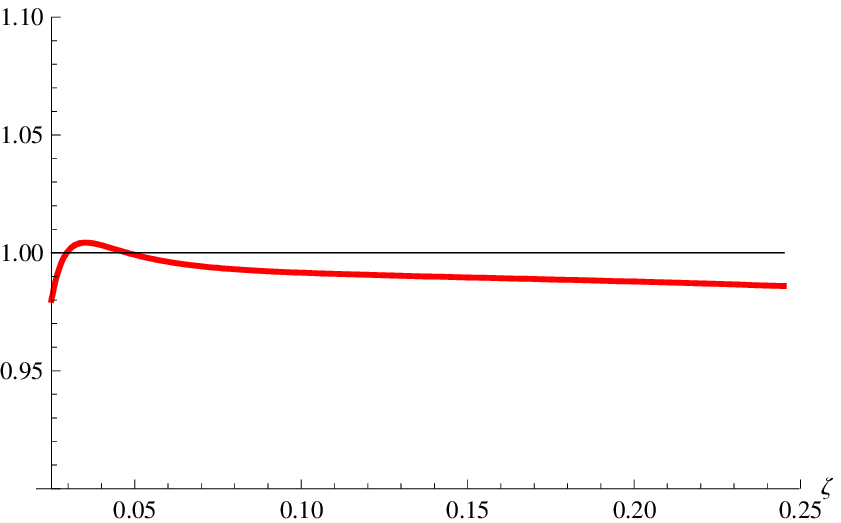}\hspace{-.5cm}}
\caption{The left figure shows $\mu_1$ for the local curve as extracted from the perturbative series using the third Richardson transform of the sequence \eqref{richc0}, divided by the corresponding analytical prediction, and plotted over the range $0<\zeta<1/4$. Similarly, the second figure shows the asymptotic result for $\mu_2$ as obtained from the perturbative series using \eqref{richc1}, again divided by the corresponding analytic prediction. The typical error is about $1.5\%$.}}

\subsection{Spacetime Interpretation of the Instanton Effects}

As we have seen in (\ref{instdiff}), the instanton action can be computed as a contour integral from the endpoint of the cut to the saddle $x_0$. This contour integral measures the potential difference between the cut $\CC$ and $x_0$. When the spectral curve corresponds to a double--scaled matrix model, this instanton action should correspond to the disk amplitude for a D--instanton in noncritical string theory. These D--instanton configurations have been identified in terms of ZZ branes, and it has also been verified that indeed the matrix model computation agrees with the disk amplitude for a ZZ brane \cite{akk}. Equivalently, the ZZ disk amplitude can be calculated as the difference between the disk amplitudes for two FZZT branes located, respectively, at the branch cut of the curve and at the pinched point of the curve. It turns out that, for topological string theory on local curves, there is a similar interpretation of the instanton action in terms of D--branes, as well as a spacetime interpretation in terms of domain walls. 

The natural branes for the A--model on a toric CY manifold are the Harvey--Lawson branes, first studied in this context in \cite{av}. The mirrors of these branes are just points in the spectral curve of the B--model. Two branes located at points $z_0$ and $z_1$ define an interpolating domain wall in the underlying type II theory. The tension of this domain wall is given by the difference of D--brane superpotentials \cite{av}
\be\label{diffsup}
W(z_1)-W(z_0)=\int_{z_0}^{z_1} \rd z\, y(z).
\ee
\noindent
When $z_0$ and $z_1$ correspond, respectively, to the endpoint of the cut and the saddle $x_0$, (\ref{diffsup}) is exactly the instanton action computed in (\ref{instdiff}). The connection between instanton actions in the matrix model and tensions of domain walls was already made in \cite{dv} for the backgrounds considered therein. At the same time, (\ref{diffsup}) can be regarded as the difference between two disk amplitudes for D--branes located at $z_1$ and $z_0$. We then see that the role of FZZT branes in noncritical string theory is played by the Harvey--Lawson branes in topological string theory on local CY threefolds. Indeed, it can be easily seen \cite{mm} that, in the case of local curves, the toric branes become FZZT branes near the critical point describing 2d gravity. On the other hand, the saddle $x_0$ that we have been considering (and which leads to an extremum of the superpotential) gives a topological string analogue of the ZZ brane. 

A more invariant way of writing (\ref{diffsup}), by taking into account the full six--dimensional geometry of the CY, is
\be\label{generalA}
A = \int_{\Gamma} \Omega, \qquad \Gamma=[C_1 -C_0], 
\ee
\noindent
where $\Omega$ is the holomorphic $(3,0)$ form on the CY, and $\Gamma$ is a three--cycle interpolating between the two--cycles $C_{0,1}$ associated to $z_{0,1}$ in the full geometry. This is indeed the general form for disk amplitudes of B--branes presented in \cite{mw}. 

It is interesting to notice that usually the nonperturbative effects due to B--branes considered in the literature involve the hypermultiplet moduli, since a B--brane supported on a curve will couple to the K\"ahler form, and not to $\Omega$ \cite{nov,n}. This type of D--instanton effects (which in some cases can be computed exactly \cite{robles}) cannot however be related to the large--order behavior of the topological string amplitudes, which depend on the vector multiplet moduli. On the other hand, domain walls interpolating between two B--branes {\it can} couple to $\Omega$ and therefore have the right structure to control the large--order behavior of topological string perturbation theory. In this paper we have checked this for a restricted class of toric geometries, but we expect this fact to be true in the more general case, for an appropriate choice of the domain wall.

\sectiono{Application III: Hurwitz Theory}

We finally proceed to our last example, Hurwitz theory.

\subsection{Hurwitz Theory}

Hurwitz theory studies branched covers of Riemann surfaces. Here, we shall restrict ourselves to the coverings of a sphere $\BP^1$ (the ``target") by surfaces of genus $g$ (the ``worldsheets''). The covering maps will be restricted to have only simple branch points. The number of disconnected coverings of degree $d$ with these topological characteristics is counted by the so--called simple Hurwitz number, which we denote by $H_{g,d}^{\BP^1} (1^d)$. It can be computed, in classical Hurwitz theory, in terms of representation theory of the symmetric group:
\be\label{hurwitznumbers}
H_{g,d}^{\BP^1} (1^d) =\sum_{\ell(R)=d} \biggl( {d_R \over \ell(R)!}\biggr)^2  (\kappa_R/2)^{2g-2+ 2d}.
\ee
\noindent
Here the sum is over Young tableaux $R$, with a fixed number of boxes $\ell(R)$ equal to the degree $d$, and $d_R$ is the dimension of $R$ regarded as a representation of the symmetric group $S_d$. The quantity $\kappa_R$ was defined in (\ref{kappar}).

We can now define the total partition function of Hurwitz theory as a generating functional for simple Hurwitz numbers,
\be
Z^{\rm H}(t_H,g_H)=\sum_{g\ge 0} g_H^{2g-2} \sum_{d\ge 0} {H_{g,d}^{\BP^1} (1^d)  \over (2g-2+2d)!}Q^d,
\ee
\noindent
where $Q=\re^{-t_H}$ and $g_H$ can be regarded as formal parameters keeping track of the degree and the genus, respectively. This partition function can be written as
\be\label{zksw} 
Z^{\rm H}(t_H, g_H)=\sum_R\biggl( {d_R \over |\ell(R)|!}\biggr)^2 g_H^{-2\ell(R)} \re^{g_H \kappa_R/2} Q^{\ell(R)}. 
\ee
\noindent
The free energy $\log\, Z^{\rm H}$ describes {\it connected}, simple Hurwitz numbers $H_{g,d}^{\BP^1} (1^d)^{\bullet}$,
\be\label{freehurwitz}
F^{\rm H} = \log Z^{\rm H} = \sum_{g\ge 0} g_H^{2g-2} \sum_{d\ge 0}  {H_{g,d}^{\BP^1} (1^d)^{\bullet} \over (2g-2+2d)!}Q^d,
\ee
\noindent
and it has the genus expansion
\be\label{genushurwitz}
F^{\rm H}(g_H,t_H) = \sum_{g=0}^{\infty} g_H^{2g-2} F_g^{\rm H}(Q_H).
\ee
\noindent
This theory is in fact a topological string theory in disguise. It can be realized as a special limit of the type--A theory on local curves $X_p$ with K\"ahler parameter $t$ that we studied in the previous section \cite{cgmps}, namely the limit
\be
p\rightarrow \infty,\qquad t\rightarrow\infty,\qquad g_s\rightarrow 0,
\ee
\noindent
while the new parameters $g_H$ and $t_H$, which are defined by
\be
g_H=p g_s,\qquad \re^{-t_H}=(-1)^p p^2\re^{-t},
\ee
\noindent
are kept fixed. As in the case of the theory on local curves, there is a B--model mirror to this theory. Its natural coordinate $\chi$ is related to the A--model coordinate $Q=\re^{-t_H}$ by the mirror map
\be\label{chiQ}
\chi\re^{-\chi} =Q,
\ee
\noindent
which can indeed be understood as an appropriate limit of (\ref{mirror}) for $p\rightarrow \infty$ \cite{cgmps}. The inverse mirror map is provided by Lambert's $W$ function \cite{lambert},
\be\label{lambert}
\chi=-W(-Q) = \sum_{k=1}^{\infty} {k^{k-1}\over k!} Q^k,
\ee
\noindent
which has convergence radius $Q_c=\re^{-1}$ or $\chi=1$. The large--radius region corresponds to $Q\rightarrow 0$ (and also to $\chi\rightarrow 0$). The spectral curve characterizing the B--model is of the form
\be\label{hurwitzcurve}
y(h)=2 \tanh^{-1} \biggl[ 2{ \sqrt{(a-h)(b-h)}  \over 2h-(a+b)}\biggr]-{ \sqrt{(a-h)(b-h)}},
\ee
\noindent
where the endpoints of the cut are given by
\be
b=\bigl(1 + \chi^{1\over 2}\bigr)^2, \quad a=\bigl(1 - \chi^{1\over 2}\bigr)^2.
\ee
\noindent
The above spectral curve can also be read from the saddle--point description of the sum over partitions (\ref{zksw}) given in \cite{ct,ksw}. 

Hurwitz theory has been extensively studied in the mathematical literature, and these studies have unveiled interesting properties. As shown in \cite{gjv}, the higher--genus free energies $F_g^{\rm H}(Q)$, when expressed in terms of the mirror coordinate $\chi$, have a very simple structure, namely 
\be\label{hansatz}
\ba
F^{\rm H}_0 (\chi) &= \frac{\chi^3}{6} - \frac{3\chi^2}{4} + \chi, \\
F^{\rm H}_1(\chi)&=-{1\over 24} \Bigl( \log (1-\chi) + \chi\Bigr),\\
F^{\rm H}_g(\chi) &={P_g(\chi) \over (1-\chi)^{5(g-1)}}, \quad P_g(\chi) =\sum_{i=2}^{3g-3} c_{g,i} \, \chi^i,\quad g\ge 2.
\ea
\ee
\noindent
Moreover, the polynomials $P_g(\chi)$ have the property
\be
P_g(1) = 4^{g-1} a_g, \quad g\ge 2,
\ee
\noindent
where $a_g$ is the genus $g$ free energy of 2d gravity appearing in (\ref{fds}). Therefore, in the double--scaling limit
\be\label{hurwitzds}
\chi\rightarrow 1, \qquad g_H \rightarrow 0, \qquad g_H^{-2} (1-\chi)^5=4 \kappa^{5\over 2},
\ee
\noindent
the total free energy of Hurwitz theory becomes (\ref{fds}) 
\be
F^{\rm H}(g_H, t_H) \rightarrow F_{\rm ds}(\kappa), 
\ee
\noindent
and one recovers 2d gravity at the critical point. This was first pointed out at genus zero in \cite{ksw} and then established at all genera in \cite{cgmps}, using the results of \cite{gjv}. 

Another interesting result concerning Hurwitz theory was obtained in \cite{toda}, where the total free energy was shown to satisfy the Toda equation, 
\be\label{todaF}
\exp\left(F^{\rm H}(g_H,t_H+g_H)+F^{\rm H}(g_H,t_H-g_H)-2F^{\rm H}(g_H,t_H)\right)=g_H^2\re^t\partial_{t_H}^2F^{\rm H}(g_H,t_H).
\ee
\noindent
This equation is the analogue for this model of the pre--string equation (\ref{prestring}) for the quartic matrix model. One can directly derive from (\ref{todaF}) that the double--scaled specific heat satisfies the Painlev\'e I equation (\ref{pone}), providing in this way yet another derivation of the result in \cite{cgmps}. We have used the Toda equation to compute Hurwitz amplitudes up to genus $16$, and some of these results are presented in appendix \ref{sec:aphurw}.

\subsection{Instanton Effects and Large--Order Behavior}

Let us now turn to the computation of the one--instanton quantities. From the curve (\ref{hurwitzcurve}) we find the moment function, 
\be\label{hurwitzmoment}
M(h)={2 \over { \sqrt{(a-h)(b-h)}}} \tanh^{-1} \biggl[ 2{ \sqrt{(a-h)(b-h)}  \over 2h-(a+b)}\biggr]-1, 
\ee
\noindent
where the nontrivial saddle--point is defined by
\be
M(h_0)=0, 
\ee
\noindent
or
\be\label{hsaddle}
{2\over { \sqrt{(a-h_0)(b-h_0)}}} \tanh^{-1}\biggl[2{ \sqrt{(a-h_0)(b-h_0)}  \over 2h_0-(a+b)}\biggr]=1.
\ee
\noindent
This equation can be written in a simpler way by defining $w$ as
\be
h_0=4\sqrt{\chi}\cosh^2\Bigl({w\over 2}\Bigr)+(1-\sqrt{\chi})^2.
\ee
\noindent
In terms of these variables, equation (\ref{hsaddle}) simply reads
\be
{w\over \sinh(w)}=\sqrt{\chi}.
\ee
\noindent
Even though we cannot solve analytically for $h_0(\chi)$, we can solve \eqref{hsaddle} to find $h_0(\chi)$ near $\chi=0,1$ as a power series. Near the critical point $\chi=1$, it is easy to see that $h_0(\chi)$ has a Taylor series 
expansions in powers of $\xi=1-\chi$
\be
h_0(\chi) = 4+ \xi+{4\over 5}\xi^2 +\cdots. 
\ee
\noindent
Near $\chi=0$, the power series solution is more complicated. At leading order it is easy to find that 
\be\label{wz}
w\sim-{1\over 2}\log(\chi) + \log(-\log(\chi)), 
\ee
\noindent
which yields
\be
h_0(\chi)\sim -\log \, \chi + 2 \log (-\log \, \chi), \qquad \chi \rightarrow 0. 
\ee
\noindent
The corrections to the leading asymptotics (\ref{wz}) can be obtained following a method exposed, for example, in \cite{deBruijn}. The full solution can be written as
\be
w = \log(-\log({\sqrt{\chi}\over 2}))-{1\over 2}\log(\chi)+v,
\ee
\noindent
where $v$ is a power series
\be
v=\sum_{j,k,m}c_{jkm}\mu^j\sigma^k\tau^m
\ee
\noindent
in the variables
\be
\sigma={1\over \log(\chi)}, \qquad \tau={\log(-\log(\chi))\over \log(\chi)}, \qquad \mu=\left({\chi\over \log(\chi)}\right)^2.
\ee
\noindent
The coefficients $c_{jkm}$ can be explicitly written as
\be
\ba
c_{jkm}& = - \oint \frac{\rd z}{2\pi\ri}\, {\re^{(j-1)z}z^{k+1}(-1)^m\over (\re^{-z}-1)^{j+k+m+1}}\, {(j+k+m)!\over j!k!m!} +\\
& + \oint \frac{\rd z}{2\pi\ri}\, {\re^{jz}z^k(-1)^m\over (\re^{-z}-1)^{j+k+m}}\, {(j+k+m-1)!\over j!(k-1)!m!}.
\ea
\ee

The instanton action
\be
A(\chi)=\int_b^{h_0(\chi)} \rd h\, y(h)
\ee
can now be computed explicitly as a function of $h_0$ as
\be
A(\chi)=(b-a)\left(\gamma \cosh^{-1}(\gamma)-\sqrt{\gamma^2-1}\right)-{(a-b)^2\over 8}\left(\gamma\sqrt{\gamma^2-1}-\cosh^{-1}(\gamma)\right),
\ee
\noindent
where
\be
\gamma={1\over b-a} (2h_0(\chi)-a-b).
\ee
\noindent
Using the above results for the behavior of $h_0$ near $\chi=0,1$, we can also find the behavior of the instanton action near these points. At the critical point, one finds 
\be
{1\over g_H} A(\chi) \rightarrow {8 {\sqrt 3} \over 5} \kappa^{5/4} +\cdots, \quad \chi\rightarrow 1,
\ee
where $\kappa$ is the double--scaled variable introduced in (\ref{hurwitzds}). Of course, this is the expected universal, double--scaled result of (\ref{2dinst}). Near $\chi=0$, we find
\be
A(\chi) \sim {1\over 2} (\log\, \chi)^2, \quad  \chi\rightarrow 0.
\ee

We may now compare our predictions with the numerical asymptotics of $F_g$. The moduli space of $\chi$ can be divided into the six segments shown in \figref{fig:hurwitzregions}, and we have tested our predictions in each of them. They have the following characteristics:
\begin{itemize}
\item The simplest case to study is the real interval $0<\chi<1$. Here, there is one single instanton with real action, corresponding to an eigenvalue tunneling from $b$ to the saddle $x_0$ on the right of the cut, and all $F_g$ are also real.
\item As $\chi$ moves to the right of $[0,1]$, beyond the critical point at $\chi=1$, the only solutions to $M(h)=0$ are located inside the cut and the instanton action becomes purely imaginary while $\mu_{1,2}$ remain real. The $F_g$ oscillate in sign.
\item It turns out that there is no systematic difference between the regions I--VI away from the real axis. The instanton action as well as $\mu_{1,2}$ and of course $F_g$ are generically complex, in spite of which our predictions continue to hold.
\item When $\chi$ lies on the negative real line, the endpoints of the cut move away from the real axis and become complex conjugate. There are now two saddle--point solutions, $x_0$ and $x^*_0$, complex conjugate to each other, and accordingly \emph{two} instanton solutions with conjugate actions, one corresponding to an eigenvalue tunneling from $b$ to $x_0$ and another from $a=b^*$ to $x^*_0$, as shown in \figref{fig:hurwinst}. Therefore the $F_g$ are real, with asymptotics of the form \eqref{acos} involving a cosine. Notice that this is very similar to a mechanism for the local curve, first observed in \cite{mm}.
\end{itemize}
\noindent
As before, the one and two--loop coefficients are given by \eqref{rone} and \eqref{mutwo}, evaluated for the moment function \eqref{hurwitzmoment}. The saddle--point solution has to be evaluated numerically. The instanton action, as well as $\mu_1$ and $\mu_2$, are well--defined over the whole complex plane of the modulus $\chi$. \figref{fig:hurw1} and \figref{fig:hurw2} show the inverse square root of the sequence $Q_g$ in \eqref{richainst} and of the corresponding Richardson transforms, alongside with the prediction of instanton calculus for the instanton action, at values of the modulus $\chi=0.5,\ \chi=1.5+\ri,$ and $\chi=-1-0.5\ri$. In \figref{fig:hurw4} we compare the sequence $\pi F_g |A|^{2g -5/2}/({\Gamma(2g-{5\over 2})|\mu_1|})$ for Hurwitz theory, together with the prediction $2 \cos\left(\left(2g-{5\over 2}\right)\theta_A+\theta_{\mu_1}\right)$, at $\chi=-0.5$ and $\chi=-3$. \figref{fig:hurwc} and \figref{fig:hurwcw} show the modified sequences, \eqref{richc0} and \eqref{richc1}, with leading asymptotics given by the one and two--loop fluctuations around the one--instanton configuration, together with the analytic prediction. Indeed, the agreement is again quite spectacular.
\FIGURE[ht]{\label{fig:hurwitzregions}
    \centering
    \epsfxsize=0.45\textwidth
    \epsfysize=0.3\textwidth
    \leavevmode
    \hspace{2cm}\epsfbox{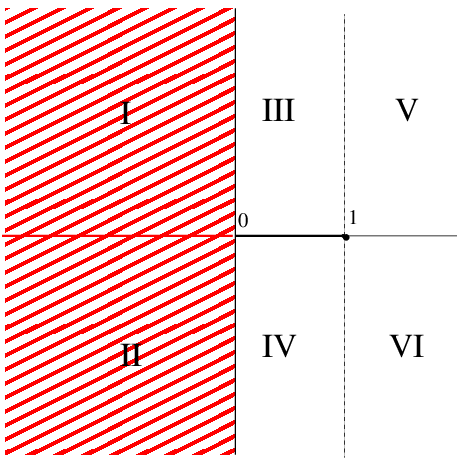}\hspace{2cm}
\caption{The moduli space of Hurwitz theory.}
}
\FIGURE[ht]{\label{fig:hurwinst}
    \centering
    \epsfxsize=0.4\textwidth
	\epsfysize=0.3\textwidth
    \leavevmode
    \hspace{3cm}\epsfbox{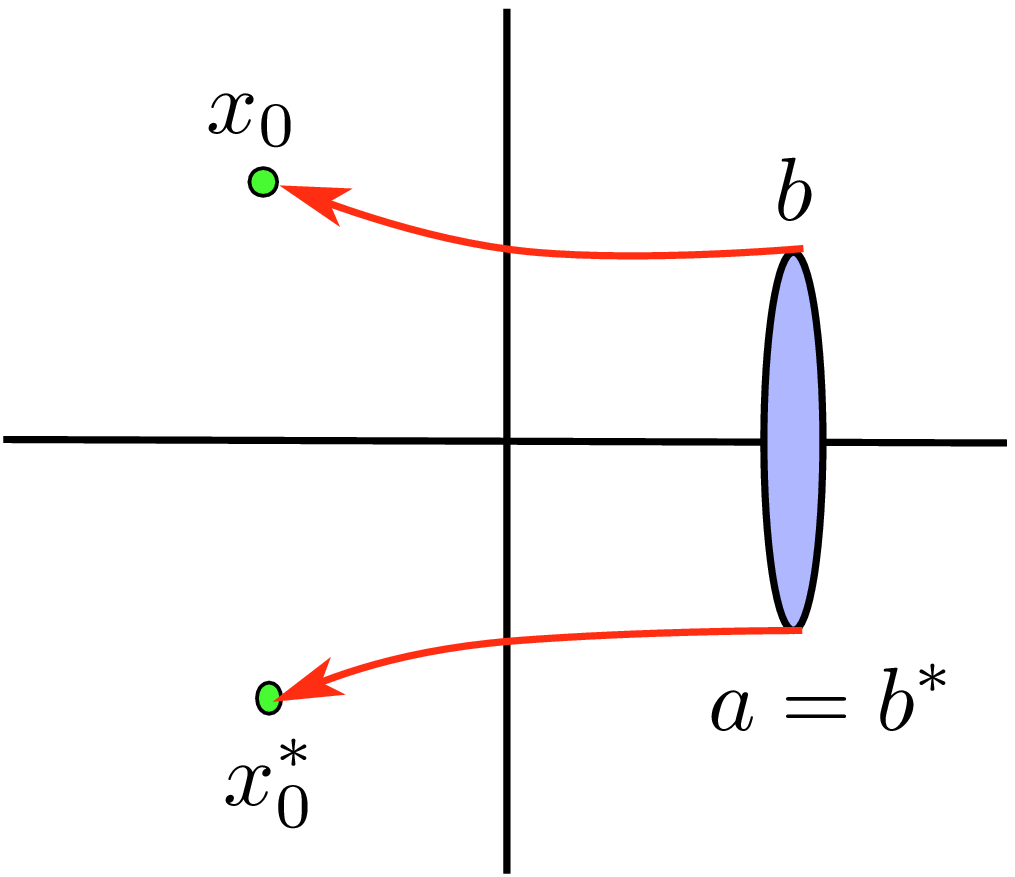}\hspace{3cm}
\caption{At negative parameter $\chi$, the endpoints of the cut become complex conjugate to each other and a second instanton solution appears, going from $a=b^*$ to $x_0^*$.}
}
\FIGURE[ht]{\label{fig:hurw1}
    \centering
    \epsfxsize=0.5\textwidth
    \leavevmode
    \mbox{\hspace{-.5cm}\epsfbox{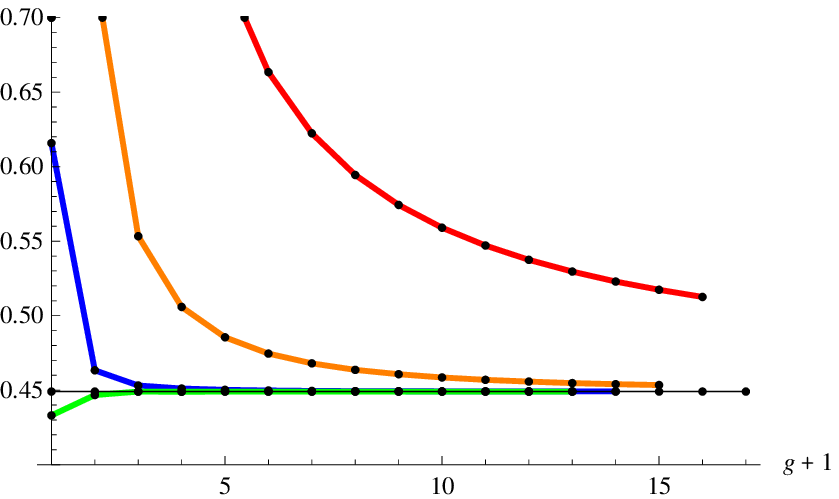}\quad\epsfbox{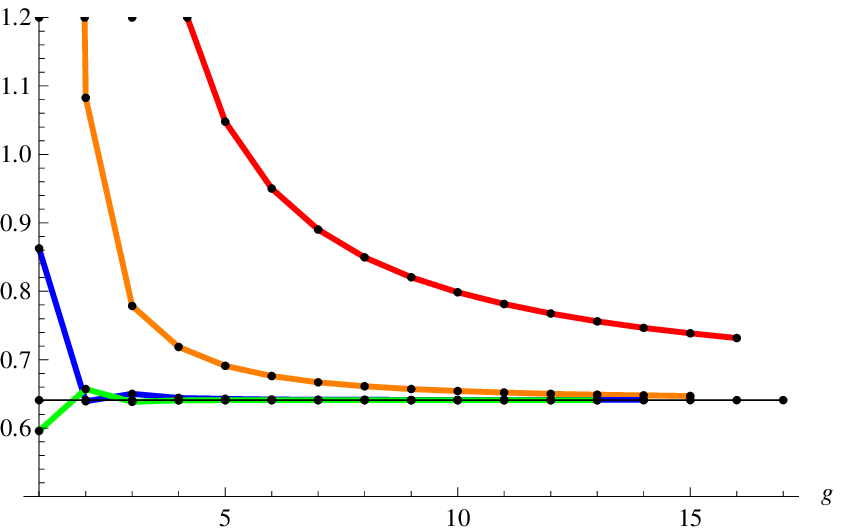}\hspace{-.5cm}}
\caption{The left figure shows the sequence $({\rm Re}(Q_g^{-1}))^{1/2}$ for Hurwitz theory, together with its Richardson transforms. The straight line shows the corresponding prediction $({\rm Re}(A^2))^{1/2}$, at $\chi=0.5$. On the right, the same for $\chi=1.5+\ri$. The available degree is $g=16$, the error is $4 \times 10^{-6}\%$ at $\chi=0.5$, and $7 \times 10^{-6}\%$ at $\chi=1.5+\ri$.}
}
\FIGURE[ht]{\label{fig:hurw2}
    \centering
    \epsfxsize=0.5\textwidth
    \leavevmode
    \mbox{\hspace{-.5cm}\epsfbox{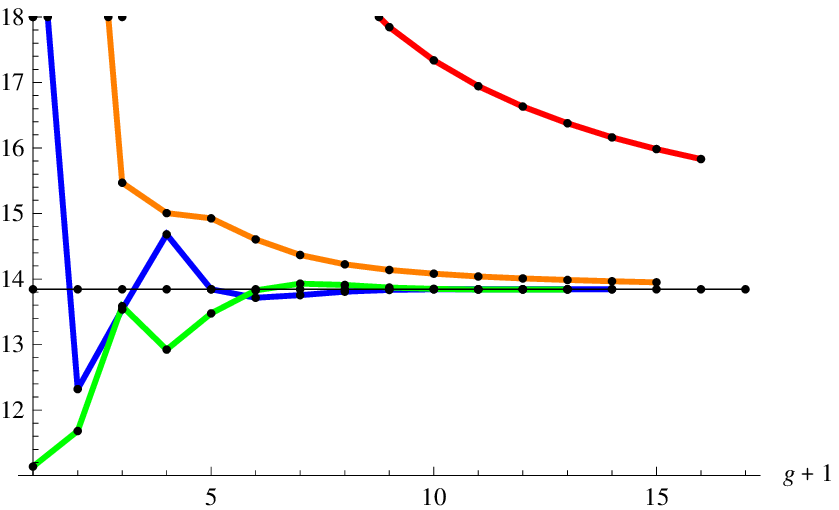}\quad\epsfbox{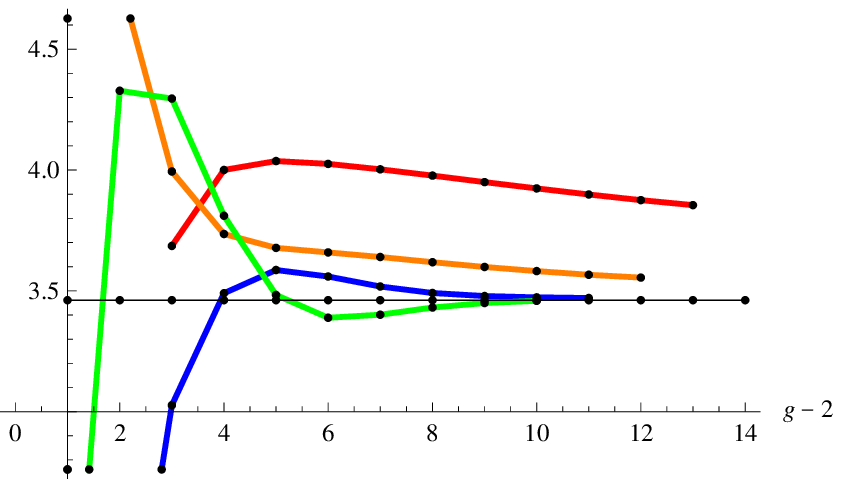}\hspace{-.5cm}}
\caption{On the left, the sequences $({\rm Re}(Q_g^{-1}))^{1/2}$ with its Richardson transforms and the prediction $({\rm Re}(A^2))^{1/ 2}$, at $\chi=-1-0.5\ri$ (straight line). On the right, we show the same for the imaginary parts. The errors at $g=16$ are $0.01\%$ and $0.08\%$, respectively.}
}
\FIGURE[ht]{\label{fig:hurw4}
    \centering
    \epsfxsize=0.5\textwidth
    \leavevmode
    \mbox{\hspace{-.5cm}\epsfbox{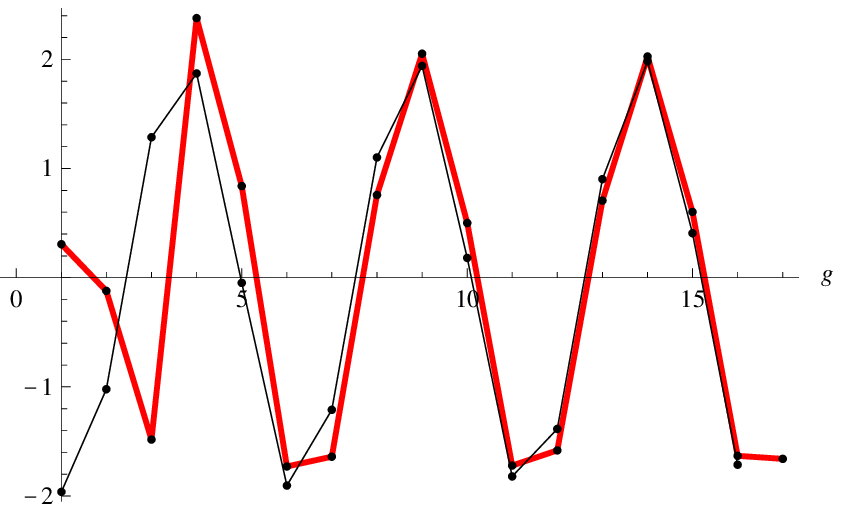}\quad\epsfbox{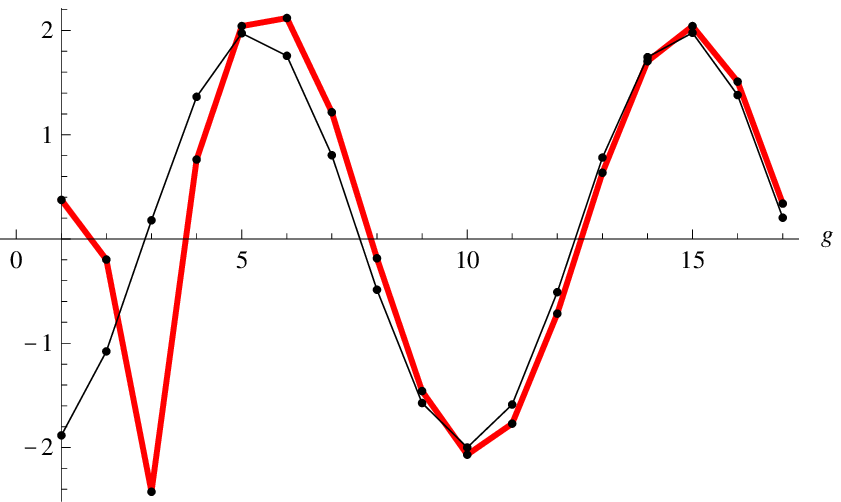}\hspace{-.5cm}}
\caption{The sequence $\pi F_g |A|^{2g-5/2}/({\Gamma(2g-{5\over 2})|\mu_1|})$ for Hurwitz theory, together with the prediction $2 \cos\left(\left(2g-{5\over 2}\right)\theta_A+\theta_{\mu_1}\right)$, at $\chi=-0.5$ (left) and $\chi=-3$ (right). The error at genus 16 is of order $3\%$.}
}
\FIGURE[ht]{\label{fig:hurwc}
    \centering
    \epsfxsize=0.5\textwidth
    \leavevmode
    \mbox{\hspace{-.5cm}\epsfbox{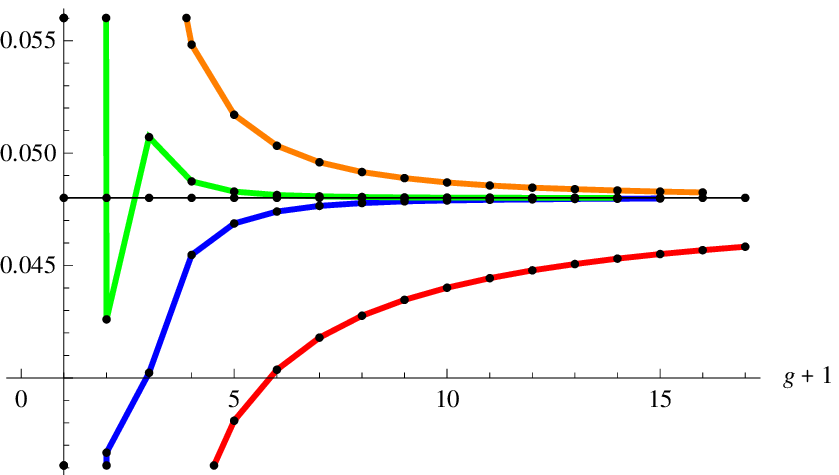}\quad\epsfbox{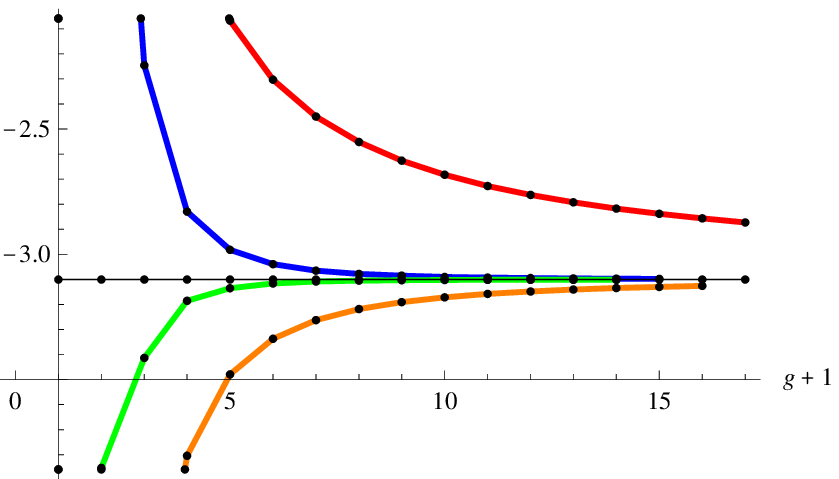}\hspace{-.5cm}}
\caption{The left figure shows ${\pi A^{2g-{5\over 2}}F_g/\Gamma(2g-{5\over 2})}$ for Hurwitz theory, and its Richardson transforms, at $\chi=0.5$. The leading asymptotics are predicted by $\mu_1$, shown as a straight line. On the right, we plot the analogous sequence \eqref{richc1}, together with the expected leading asymptotic value $\mu_2$ (straight line). The error at $g=16$ is $0.009\%$ for $\mu_1$, and $0.012\%$ for $\mu_2$.}
}
\FIGURE[ht]{\label{fig:hurwcw}
    \centering
    \epsfxsize=0.5\textwidth
    \leavevmode
    \mbox{\hspace{-.5cm}\epsfbox{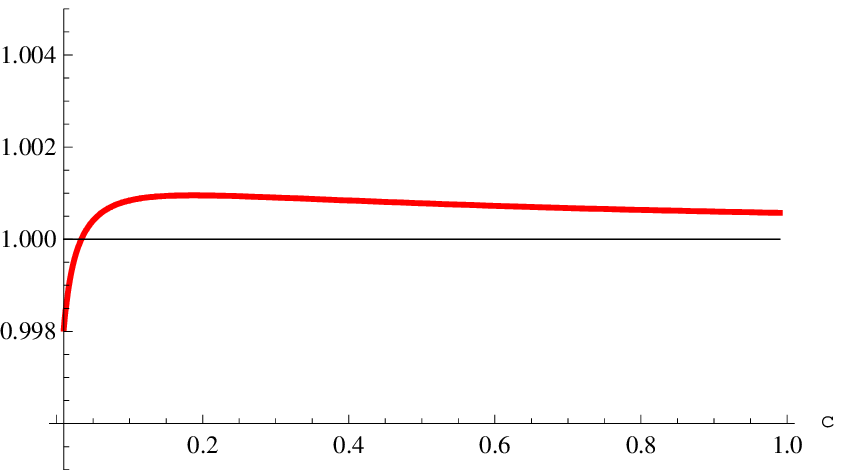}\quad\epsfbox{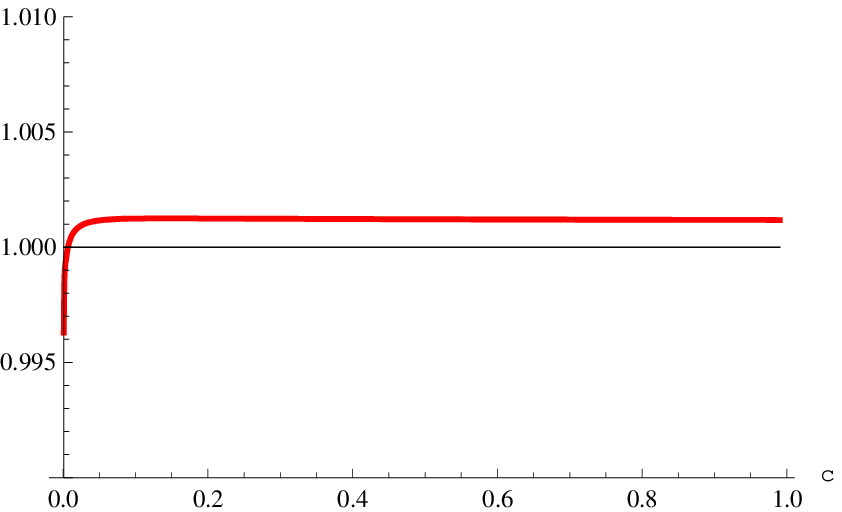}\hspace{-.5cm}}
\caption{The left figure shows $\mu_1$ as a function of $\chi$ for Hurwitz theory, as extracted from the perturbative series using the third Richardson transform of the sequence \eqref{richc0}, divided by the corresponding analytical prediction. Similarly, the second figure shows the asymptotic result for $\mu_2$ as obtained from the perturbative series using \eqref{richc1}, again divided by the corresponding analytic prediction. For $\chi>0.01$, the error is of order $0.1\%$.}
}

\sectiono{Conclusions and Outlook}

In this paper we have extended classical results on the connection between instanton effects and the large--order behavior of perturbation theory to the realm of general, one--cut matrix models and topological strings. After having performed a detailed one--instanton computation, up to two loops, we have tested this connection in both the standard quartic matrix model off--criticality and in its double--scaled limit, 2d gravity. Combining our results with the holographic matrix model descriptions of topological string theory on toric backgrounds \cite{mm,bkmp}, we have further provided a computation of nonperturbative effects in certain topological string models, and we have verified in detail that they precisely capture the large--order behavior of the perturbative amplitudes. It is important to point out that the precise agreement we find up to two loops strongly supports our claim that we have identified important nonperturbative effects in these models. This is also a strong check of the instanton/large--order correspondence proposed in this paper, as well as of the proposal of \cite{mm,bkmp} for describing topological string theories on toric backgrounds in terms of dual matrix models.

From the mathematical point of view, we have presented precise conjectures for the large--order behavior of the $F_g$ amplitudes, which are well--defined objects in all the cases we have considered. The simplest case is of course the proposed asymptotics (\ref{pred}), for the free energy of pure gravity, but the asymptotics of Hurwitz theory (\textit{i.e.}, the Hurwitz numbers) should also be of mathematical interest as they may provide new insight into this enumerative problem.

Our work raises various problems and, at the same time, suggests various venues for future research, to which we hope to return in the near future. Let us thus finish this paper by listing some of them.

\begin{itemize}

\item One model we have not studied in here is topological string theory on the resolved conifold. It is known (see, for example, \cite{mmhouches} and references therein) that this theory can be described by a Hermitian matrix model with a potential of the form $(\log \, x)^2$, which has a global minimum at $x=1$ and no saddles. In principle, the way to address the large--order behavior in such cases is to deform the model, in order to obtain an unstable potential with a calculable one--instanton amplitude. The original potential is then recovered by analytic continuation \cite{blgzj, zjbook}. However, we have not found a suitable deformation which makes it possible to find the large--order behavior for this potential. In fact, we would face the exact same problems if we were to address the large--order behavior of the perturbation series for the ground--state energy of a $(\log \, x)^2$ potential in quantum mechanics. This is a situation which, to the best of our knowledge, has also not been addressed in the literature.

\item Our work may be straightforwardly generalized to more complicated matrix models. For example, one could consider two--matrix models. Instanton effects in two--matrix models have been computed at leading order in \cite{kktwo}, and near the critical point in \cite{twoishi}, but it would be interesting to have exact results beyond leading order and off--criticality. Also, a more geometric formulation of the instanton contribution computed in this paper, along the lines of the approach in \cite{eo}, would be desirable. Although our expressions only depend on the form of the spectral curve, they are only suitable in principle for a genus--zero curve written in the form (\ref{scurve}). 

\item In this work we have restricted ourselves to the analysis of the one--instanton sector, but the nonperturbative completion of the theory involves $k$--instanton sectors corresponding to the tunneling of $k$ eigenvalues. One can easily extend the framework of \cite{lvm}, adopted in here, to compute these effects. Results for the two--instanton sector near the critical point have been obtained in \cite{st}. It would be interesting to make a detailed analysis of the multi--instanton sectors off--criticality. 

\item A possibly interesting check of our one and two--loop results for 2d gravity involves re--deriving them directly in the continuum Liouville theory, where instanton effects are described by D--brane instantons with ZZ boundary conditions \cite{m03,kms03,zz01,akk}. 

\item It would also be interesting to rigorously verify some of the proposals for the asymptotic behaviors that we have put forward, beyond the numerical tests already performed in this paper. It seems very likely that the asymptotics of pure gravity can be established via a detailed analysis of the difference equation (\ref{diffeq}). One may also envisage that the two--loop instanton computation could be checked using the methods developed in \cite{fik1, fik2, sy}. 

\item More generally, it would be rather interesting to provide a more rigorous foundation for the validity of the dispersion relation (\ref{disp}), in both cases of matrix models and string theory. We should add however that, to the best of our knowledge, this kind of relation has only been properly justified in quantum mechanical models. In most examples of field theory such a relation is simply assumed to be true, and later tested {\it a posteriori} by explicit computations \cite{zj}. In spite of this, there might be some hope to further develop this line of research in the specific case of matrix models.

\item Nonperturbative effects of order $\re^{-N}$ can be also found in models defined by sums over partitions, such as the case of two--dimensional Yang--Mills \cite{michele, jevickiram}. These effects have been used in holographic descriptions of topological string theory \cite{baby, entanglement}. It would thus be interesting to see if there is any relation between this description and the one we have proposed (and carefully tested) in terms of matrix models.

\item The topological string models we have studied in this paper are not very conventional, since they correspond to toric diagrams with intersecting lines, and this is reflected in the fact that their spectral curve is pinched. One could smooth out these models by resolving the singularity, obtaining in this way a spectral curve of genus--one with two cuts. In the context of noncritical string theory, this process is interpreted as adding ZZ branes to the background \cite{ss}. It would be very interesting to see if this leads to some geometric transition from the local curve backgrounds to other topological string backgrounds.

\item For us, the most pressing problem is to extend the analysis in this paper to more complicated topological string models. The first case to address is that of matrix models with multiple cuts. In fact, the general multi--cut model with fixed filling fractions can be regarded as a matrix model in a generic, fixed multi--instanton sector, and by studying nearby filling fractions one obtains a general framework to address multi--instanton effects in matrix models. Some aspects of this framework were discussed in \cite{bde}, albeit in a different context. Once the multi--cut case is understood in detail one could use the philosophy of this paper to extend the results to topological string theory. This would allow for computation of nonperturbative effects on new and interesting toric backgrounds, such as the case of local $\BP^2$, and would also be an important step in further strengthening our understanding of nonperturbative effects in topological string theory. It might even give precious hints for the future study of these effects in compact backgrounds.  

\item Although we have used the matrix model description of \cite{mm,bkmp} to compute nonperturbative effects in topological string theory, we have not provided a full nonperturbative definition of these models in this work. The reason is that the effects we have described in here only depend on the geometry of the spectral curve, and thus it was not necessary to write any topological string theory partition function explicitly as a matrix integral. In fact, it might happen that there is more than one way to do this, since two matrix models with different potentials and different finite $N$ partition functions might nevertheless have the same spectral curve, and therefore the same $1/N$ expansion and one--instanton amplitudes. As such, it would be important to go beyond the calculation of instanton effects and provide a full nonperturbative, holographic definition of topological string theories on toric backgrounds, in terms of matrix integrals.

\end{itemize}

\section*{Acknowledgments}
We would like to thank Vincent Bouchard, Frederik Denef, Minxin Huang, Nobuyuki Ishibashi, Albrecht Klemm, Steven Orszag, Sara Pasquetti, Daniel Robles--Llana, Stevo Stevic, J\"org Teschner, Alessandro Tomasiello and Cumrun Vafa for useful discussions and/or correspondence.



\appendix

\section{Explicit Higher--Genus Formulae}

In this appendix we present some explicit expressions for free energies at high genera, in both the quartic matrix model and Hurwitz theory. This is just a partial list of our results, as most formulae quickly become too intricate to put in print. In spite of this, we hope these explicit expressions may be of future interest (and, as far as we know, have never been computed before).

\subsection{Quartic Matrix Model}

As we have reviewed in section 4, an algorithm for computing free energies in the quartic matrix model was put forward in \cite{biz}, and we have applied it up to genus $g=10$. Here, we present a partial list of our final results. In \cite{biz}, the quartic free energies were computed up to genus two, with the result (here $t=1$)
\bea
F_0 (\alpha^2) &=& -\frac{1}{2} \log \left( \alpha^2 \right) - \frac{1}{24} \left( 1-\alpha^2 \right) \left( 9-\alpha^2 \right), \\
F_1 (\alpha^2) &=& \frac{1}{12} \log \left( 2-\alpha^2\right), \\
F_2 (\alpha^2) &=& \frac{1}{6!}\, \frac{\left( 1-\alpha^2 \right)^3}{\left( 2-\alpha^2 \right)^5}\, \left( 82+21\alpha^2-3\alpha^4 \right).
\eea
\noindent
It was further conjectured that, for genus $g \ge 2$, the general structure of the free energies should be of the form
\be
F_g (\alpha^2) = \frac{\left( 1-\alpha^2 \right)^{2g-1}}{\left( 2-\alpha^2 \right)^{5(g-1)}}\, \CP_g (\alpha^2),
\ee
\noindent
with $\CP_g (\alpha^2)$ a polynomial in $\alpha^2$ such that
\be
\CP_g (\alpha^2=1) = \frac{1}{2 \cdot 6^{2g-1}}\, \frac{(4g-3)!}{g! (g-1)!}.
\ee
\noindent
Using the exact same procedure as in \cite{biz}, we have extended the analysis up to genus ten, verifying both conjectures above. In particular, we have obtained at genus three
\be
\CP_3 (\alpha^2) = - \frac{1}{9072} \left( 17260 + \alpha^2 \left( - 32704 + 9 \alpha^2 \left( - 325 + 95 \alpha^2 - 15 \alpha^4 + \alpha^6 \right) \right)\right),
\ee
\noindent
which can be explicitly compared to another genus three calculation performed in \cite{shiro1,shiro2}, with both results in complete agreement. At genus four, we obtained
\bea
\CP_4 (\alpha^2) &=& - \frac{1}{38880} \left( - 1421392 + \alpha^2 \left( 12438536 + \alpha^2 \left( - 13719796 + 27 \alpha^2 \left( - 15694 + \right. \right. \right. \right. \nonumber \\
&& \left. \left. \left. \left. + 5810 \alpha^2 - 1456 \alpha^4 + 238 \alpha^6 - 23 \alpha^8 + \alpha^{10} \right) \right) \right) \right),
\eea
\noindent
and at genus five
\bea
\CP_5 (\alpha^2) &=& - \frac{1}{85536} \left( - 383964880 + \alpha^2 \left( - 1573981616 + \alpha^2 \left( 7592114712 + \right. \right. \right. \nonumber \\
&& + \alpha^2 \left( - 6114807776 + 81 \alpha^2 \left( - 781725 + 326811 \alpha^2 - 101961 \alpha^4 + 23535 \alpha^6 + \right. \right. \nonumber \\
&& \left. \left. \left. \left. \left. - 3915 \alpha^8 + 445 \alpha^{10} - 31 \alpha^{12} + \alpha^{14} \right) \right) \right) \right) \right).
\eea
\noindent
Finally, at genus six the free energy follows from the polynomial
\bea
\CP_6 (\alpha^2) &=& - \frac{1}{79606800} \left( 139728961867968 + \alpha^2  \left( - 369974786833952 + \right. \right. \nonumber \\
&& + \alpha^2 \left( - 955888270184512 + 3 \alpha^2 \left( 1037832523698416 + \alpha^2 \left( - 662581722466844 + \right. \right. \right. \nonumber \\
&& + 55971 \alpha^2 \left( - 39761282 + 17910398 \alpha^2 - 6371112 \alpha^4 + 1787698 \alpha^6 - 392007  \alpha^8 + \right. \nonumber \\
&& \left. \left. \left. \left. \left. \left. + 65901 \alpha^{10} - 8214 \alpha^{12} + 716 \alpha^{14} - 39 \alpha^{16} + \alpha^{18} \right) \right) \right) \right) \right) \right).
\eea
\noindent
Although we have extended this calculation up to genus ten, the expressions quickly get too messy and little illuminating, and as such we shall not display any further polynomials. Our results further allow us to conclude that the polynomial $\CP_g (\alpha^2)$ is of order $3g-4$ in $\alpha^2$. One final consistency check concerns the case of $\alpha^2=2$, corresponding to the critical point of the quartic model. In this situation it must be the case that
\be
\CP_g (\alpha^2=2) = (-1)^g\, 2^{5(g-1)}\, a_g,
\ee
\noindent
where $a_g$ are the coefficients appearing in the expansion of the double--scaled free energy (obtained from Painlev\'e I in the $1/3$ normalization; see section 4 for details). Again, our results pass the test.

\subsection{Hurwitz Theory}\label{sec:aphurw}

If we expand \eqref{todaF} in $g_H$, and use that \cite{ksw,cgmps}
\be
{\partial^2 F^{\rm H}_0 \over \partial t_H^2}=\chi=\sum_{k=1}^{\infty}{k^{k-1}\over k!}\re^{-t_H k},
\ee
\noindent
we obtain the recursion relation
\be
F^{\rm H}_g (\re^{-t_H}) = \left.{\chi\over 1-\chi} \exp \left(\sum_{l\geq 1}^{g-1}g_H^{2l} \partial_{t_H}^2F^{\rm H}_l(\re^{-t_H})+2\sum_{k\geq2}\sum_{l\geq 0}^{g-1}{1\over 2k!}\partial_{t_H} ^{2k}F^{\rm H}_l(\re^{-t_H})g_H^{2k+2l-2}\right)\right|_{g_H^{2g}},
\ee
\noindent
where we keep the coefficient of $g_H^{2g}$ in the right hand side. If we combine this recursion with the general form of Hurwitz numbers \eqref{hansatz}, we can obtain explicit expressions for the polynomials appearing in \eqref{hansatz} up to high genus. The first few are,
\be
P_2 (\chi) = \frac{\chi^3}{240}+\frac{\chi^2}{1440},
\ee
\noindent
\be
P_3 (\chi) = \frac{\chi^6}{1008}+\frac{53 \chi^5}{10080}+\frac{1741 \chi^4}{362880}+\frac{137 \chi^3}{181440}+\frac{\chi^2}{80640},
\ee
\noindent
\be
P_4 (\chi) = \frac{\chi^9}{1440}+\frac{6079 \chi^8}{604800}+\frac{42419 \chi^7}{1209600}+\frac{87739 \chi^6}{2177280}+\frac{280603 \chi^5}{17418240}+\frac{109 \chi^4}{53760}+\frac{1291 \chi^3}{21772800}+\frac{\chi^2}{7257600},
\ee
\noindent
\bea
P_5 (\chi) &=& \frac{\chi^{12}}{1056}+\frac{17387 \chi^{11}}{665280}+\frac{67289 \chi^{10}}{345600}+\frac{44696593 \chi^9}{79833600}+\frac{193701347 \chi^8}{273715200}+\frac{37315313 \chi^7}{91238400} + \nonumber \\
&& + \frac{8679559 \chi^6}{82114560}+\frac{2295119 \chi^5}{205286400}+\frac{1525901 \chi^4}{3832012800}+\frac{23 \chi^3}{7603200}+\frac{\chi^2}{958003200}.
\eea

\vfill

\eject


\bibliographystyle{plain}

\end{document}